\documentclass{ws-ijmpb}

\renewcommand{\draftnote}{ }
\renewcommand{\trimmarks}{ }
\def\be{\begin{equation}}
\def\ee{\end{equation}}
\begin{document}

\markboth{A. K. Aringazin and M. I. Mazhitov}{Stochastic models of
Lagrangian acceleration of fluid particle in developed turbulence}

\catchline{ 1}{ 2}{ 3}{ 4}{ 5}
%
%%%%%%%%%%%%%%%%%%%%%%%%%%%%%%%%%%%%%%%%%%%%%%%%%%%%%%%%%%%%%%%%%%%%

\title{STOCHASTIC MODELS OF LAGRANGIAN ACCELERATION
OF FLUID PARTICLE IN DEVELOPED TURBULENCE}

\author{A. K. ARINGAZIN$^*$ and M. I. MAZHITOV$^\dagger$}
\address{Department of Theoretical Physics, Institute for
Basic Research,\\ Eurasian National University, Astana 473021
Kazakhstan\\$^*$aringazin@mail.kz\\$^\dagger$mmi@emu.kz}

%%\author{A. K. ARINGAZIN}
%%\address{Department of Theoretical Physics, Institute for
%%Basic Research, Eurasian National University, Astana 473021
%%Kazakhstan\\aringazin@mail.kz}
%%\author{M. I. MAZHITOV}
%%\address{Department of Theoretical Physics, Institute for
%%Basic Research, Eurasian National University, Astana 473021
%%Kazakhstan\\mmi@emu.kz}

\maketitle

\begin{history}
\received{17 October 2004}
%\revised{Day Month Year}
%\accepted{(Day Month Year)}
%\comby{(xxxxxxxxxx)}
\end{history}

\begin{abstract}

\noindent Modeling statistical properties of motion of a
Lagrangian particle advected by a high-Reynolds-number flow is of
much practical interest and complement traditional studies of
turbulence made in Eulerian framework. The strong and nonlocal
character of Lagrangian particle coupling due to pressure effects
makes the main obstacle to derive turbulence statistics from the
three-dimensional Navier-Stokes equation; {{motion of a single
fluid-particle is strongly correlated to that of the other
particles.}} Recent breakthrough Lagrangian experiments with high
resolution of Kolmogorov scale have motivated growing interest to
acceleration of a fluid particle. Experimental stationary
statistics of Lagrangian acceleration conditioned on Lagrangian
velocity reveals essential dependence of the acceleration variance
upon the velocity. This is confirmed by direct numerical
simulations. Lagrangian intermittency is considerably stronger
than the Eulerian one. Statistics of Lagrangian acceleration
depends on Reynolds number. In this review we present description
of new simple models of Lagrangian acceleration that enable data
analysis and some advance in phenomenological study of the
Lagrangian single-particle dynamics. Simple Lagrangian stochastic
modeling by Langevin-type dynamical equations is one the widely
used tools. The models are aimed particularly to describe the
observed highly non-Gaussian conditional and unconditional
acceleration distributions. Stochastic one-dimensional toy models
capture main features of the observed stationary statistics of
acceleration. We review various models and focus in a more detail
on the model which has some deductive support from the
Navier-Stokes equation. Comparative analysis on the basis of the
experimental data and direct numerical simulations is made.

\end{abstract}

%%\pacs{05.20.Jj, 47.27.Jv}

\keywords{Fully developed turbulence; intermittency; turbulent
transport; Lagrangian acceleration; conditional acceleration
statistics.}

%%1
\section{Introduction}\label{Sec:Introduction}

%% THEORY
In fluid mechanics, acceleration can be defined as the substantive
derivative of the velocity,
\be\label{Eaccel}
a_i=\frac{dv_i}{dt}\equiv\partial_tv_i+v_k\partial_kv_i.
\ee
When treated in Eulerian framework, acceleration incorporates the
Eulerian local acceleration $\partial_tv_i$ and the nonlinear
advection term $v_k\partial_kv_i$ which require measurements of
the velocity $v_i$ and temporal and spatial velocity derivatives
$\partial_tv_i$ and $\partial_kv_i$ at fixed point of a flow.
Here, $\partial_k=\partial/\partial x^k$ denotes spatial
derivative in the laboratory Cartesian frame of reference,
$\partial_t =
\partial/\partial t$ is time derivative, $i,k = 1,2,3$,
and summation over repeated indices is assumed.

Using Eq.~(\ref{Eaccel}) the three-dimensional (3D) Navier-Stokes
equation for an incompressible flow can be written as
\be\label{NSaccel}
a_i=-\rho^{-1}\partial_i p + \nu\partial^2_k v_i + f_i,
\ee
where $\rho$ is the constant fluid density, $p$ is the pressure,
$\nu$ is the kinematic viscosity, $v_i$ is the velocity field, and
$f_i$ is the forcing, which usually occurs at large characteristic
spatial scale.

Measurement of time series $x_i(t)$ of the position of individual
tracer particle and using a finite-difference scheme allows one to
evaluate its Lagrangian velocity $u_i(t)$  and acceleration
$a_i(t)$ as functions of time due to the Lagrangian relations
\be\label{Laccel}
u_i(t)=\partial_tx_i(t), \quad a_i(t) =\partial^2_tx_i(t).
\ee
Here, $x_i=X_i(x_{0k},t)$ is the coordinate of an infinitesimal
fluid particle viewed as a function of the initial position
$x_{0i}\equiv x_i(0)$ of the particle and time $t\in [0,\infty)$.
With the initial data points $x_{0i}$ (Lagrangian coordinates)
running over all the fluid particles one gets a Lagrangian
description of fluid flow.

In the Eulerian framework, contributions of the viscous term and
the forcing are known to be small as compared to that of the
pressure gradient term for a certain range of scales in a locally
isotropic turbulence.
%%One expects the same picture in the Lagrangian framework, with the
%%nonlinear advection term being incorporated into the definition of
%%Lagrangian acceleration (\ref{Eaccel}) and the forces treated
%%along a particle trajectory. However,
 Direct analytical evaluation of the Lagrangian acceleration
$a_i(x_{0k},t)$ by using the 3D Navier-Stokes equation for
high-Reynolds-number (high-Re) turbulent flows is out of reach at
present. Thus one is led to estimate it theoretically in some
fashion.

Accurate evaluation of the Lagrangian velocity and acceleration in
laboratory turbulence experiments requires measurement of
positions of neutrally buoyant tracer particle by using some
tracking system to a very high accuracy. One can also use direct
measurement of the Lagrangian velocity when knowing a precise
position of tracer particle is not important. In any case, to get
Lagrangian acceleration one should have experimental access to
time scales smaller than the Kolmogorov time scale $\tau_\eta$ of
the flow. One expects that such experimental data give information
on the pressure gradient term, which is difficult to measure
experimentally.

%% EXPERIMENTS AND DNS
Growing interest in studying Lagrangian turbulence is motivated by
the recent breakthrough Lagrangian experiment by La Porta, Voth,
Crawford, Alexander, and Bodenschatz\cite{Bodenschatz} (2001), the
new data by Crawford, Mordant, Bodenschatz, and
Reynolds\cite{Bodenschatz2} (2002), Mordant, Crawford, and
Bodenschatz\cite{Mordant0303003} (2003) (optical tracking system,
$R_\lambda=690$, the measured acceleration range is $|a|/\langle
a^2\rangle^{1/2} \leq 60$, and $\tau_\eta$ is resolved), Mordant,
Delour, Leveque, Arneodo, and Pinton\cite{Mordant0206013} (2002)
(acoustic tracking system, $R_\lambda=740$, $|a|/\langle
a^2\rangle^{1/2}\leq 20$, and $\tau_\eta$ is not resolved), and
direct numerical simulations (DNS) of the 3D Navier-Stokes
equation by Gotoh, Fukayama, and Nakano\cite{GotohPF2002} (2002),
Kraichnan and Gotoh\cite{Kraichnan0305040} (2003)
($R_\lambda=380$, $|a|/\langle a^2\rangle^{1/2}\leq 150$) and
Biferale, Boffetta, Celani, Lanotte, and
Toschi\cite{Biferale0402032} (2004) ($R_\lambda=280$, $|a|/\langle
a^2\rangle^{1/2}\leq 80$, $\tau_\eta$ is resolved). Experimental
results on the 3D Lagrangian acceleration have been reported by
Mordant, Crawford, and Bodenschatz.\cite{Mordant0410070} The
classical Reynolds number Re is related to the Taylor microscale
Reynolds number due to $\mathrm{Re} = R_\lambda^2/15$. These
experiments and DNS give an important information on
single-particle dynamics and statistics, and new look to the
intermittency in high-Re fluid turbulence.

%% FIGURE 1
%%%%%%%%%%%%%%%%%%%%%%%%%%%%%%%%%%%%%%%%%%%%%%%%%%%%%%%
\begin{figure}[tbp!]
\begin{center}
\includegraphics[width=0.75\textwidth]{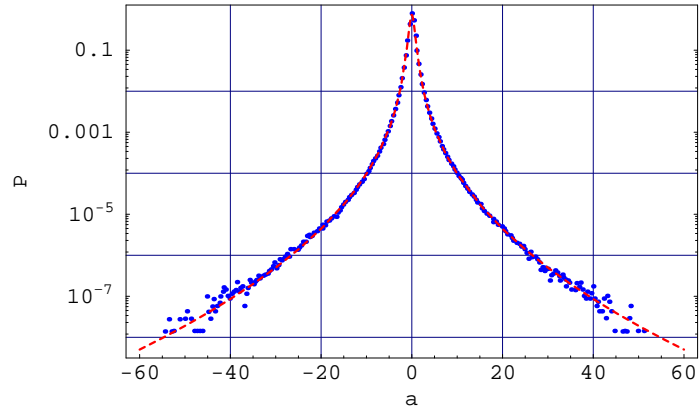}
\end{center}
\caption{ \label{Fig1} Lin-log plot of the Lagrangian acceleration
probability density function $P(a)$ for the transverse component
of acceleration. Dots: experimental data for the $R_\lambda=690$
flow by Crawford, Mordant, Bodenschatz, and Reynolds.$^2$
%%{\refcite{Bodenschatz2}}
Dashed line: the stretched
exponential fit (\ref{Pexper}). The acceleration component $a$ is
normalized to unit variance.}
\end{figure}
%%%%%%%%%%%%%%%%%%%%%%%%%%%%%%%%%%%%%%%%%%%%%%%%%%%%%%%

%% FIGURE 2
%%%%%%%%%%%%%%%%%%%%%%%%%%%%%%%%%%%%%%%%%%%%%%%%%%%%%%%
\begin{figure}[tbp!]
\begin{center}
\includegraphics[width=0.75\textwidth]{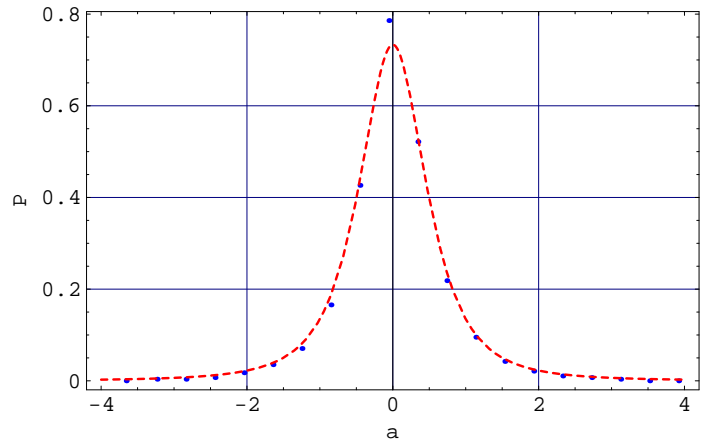}
\end{center}
\caption{ \label{Fig1b} Lin-lin plot of the central part of the
Lagrangian acceleration probability density function $P(a)$ for
the transverse component of acceleration. Same notation as in
Fig.~\ref{Fig1}.}
\end{figure}
%%%%%%%%%%%%%%%%%%%%%%%%%%%%%%%%%%%%%%%%%%%%%%%%%%%%%%%

%% FIGURE 3
%%%%%%%%%%%%%%%%%%%%%%%%%%%%%%%%%%%%%%%%%%%%%%%%%%%%%%%
\begin{figure}[tbp!]
\begin{center}
\includegraphics[width=0.75\textwidth]{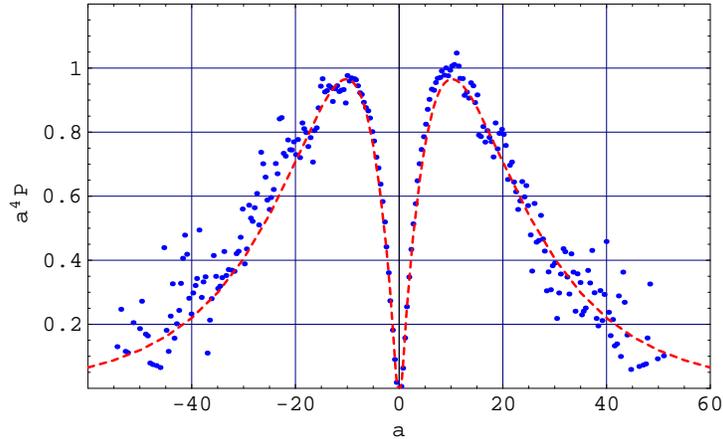}
\end{center}
\caption{ \label{Fig2} The contribution to fourth-order moment of
Lagrangian acceleration, $a^4P(a)$. Same notation as in
Fig.~\ref{Fig1}.}
\end{figure}
%%%%%%%%%%%%%%%%%%%%%%%%%%%%%%%%%%%%%%%%%%%%%%%%%%%%%%%

The experimental data on a component of Lagrangian acceleration
$a$ of polystyrene tracer particle in the $R_\lambda=690$ water
flow generated between two counter-rotating disks reveal that the
acceleration varies with time in a wild way. Statistical
description of the acceleration is then used and a huge amount of
collected data have been fitted to a good accuracy by the
probability density function of a stretched exponential
form\cite{Bodenschatz,Bodenschatz2,Mordant0303003}
\be\label{Pexper}
P(a) = C \exp\left[-\frac{a^2}{(1
+\left|{b_1a}/{b_2}\right|^{b_3})b_2^2}\right].
\ee
Here, $b_1=0.513\pm 0.003$, $b_2= 0.563\pm 0.02$, and $b_3=
1.600\pm 0.003$ are fit parameters, and $C=0.733$ is the
normalization constant.

Two coordinates $z$ and $x$, one along the large-scale symmetry
axis and the other transverse to it, were measured, while the
third coordinate $y$ was taken statistically equivalent to the
measured transverse component. Lagrangian particles are tracked in
a small central part of the flow where, in general, high degree of
statistical isotropy of small scales is expected due to Kolmogorov
1941 (K41) local isotropy hypothesis.\cite{K41} The studied
statistically stationary flow is highly anisotropic at large
scales due to the used specific stirring mechanism. This appears
to affect very small scales. Namely, one can observe small
skewness of the acceleration distribution and anisotropy of the
acceleration variance, as well as difference in distributions of
components of Lagrangian velocity.

At large acceleration magnitudes, tails of the fitted model
distribution (\ref{Pexper}) decay very slowly, asymptotically as
$\exp[-|a|^{0.4}]$, which implies finitness of the acceleration
fourth-order moment $\langle a^4\rangle =\int_{-\infty}^{\infty}
a^4P(a)da$, as confirmed by the experiment.\cite{Bodenschatz2} The
flatness factor of the distribution (\ref{Pexper}) which
characterizes widening of its tails (when compared to a Gaussian)
is $F \simeq 55.1$, which is in agreement with the experimental
value
\be\label{flatness}
F \equiv {\langle a^4\rangle}/{\langle a^2\rangle^2} =55 \pm 8.
\ee
Gaussian distribution is characterized by much smaller value
$F=3$.

The Kolmogorov time scale of the $R_\lambda=690$ flow is
$\tau_\eta=0.93$~ms. Experimental resolution of the scale is very
accurate, about 1/65 of $\tau_\eta$. Low-pass filtering with the
0.23$\tau_\eta$ width of the collected $1.7\times 10^8$ data
points was used, and the response time of optically tracked
46~$\mu$m diameter tracer particle is 0.12$\tau_\eta$. Note that
certain nonzero time scale less than the Kolmogorov time scale is
used to derive Lagrangian velocity and acceleration values. The
width of filter has a limited impact on the resulting value of
$F$, e.g. the use of 0.31$\tau_\eta$ width results in about 15\%
decrease of the experimental value of flatness $F$. Notice that
non-ideal response characteristics of tracer particle may result
in an increase of the effective integral time scale, from the
Lagrangian integral time scale $T_L$ to the Eulerian integral time
scale $T_E$ (calculated in the comoving frame), as one can show by
using Corrsin hypothesis.\cite{Corrsin1959}

The experimental data and stretched exponential fits are shown in
Figs.~\ref{Fig1}, \ref{Fig1b}  and \ref{Fig2}. One can observe
almost symmetric distributions with respect to zero acceleration
and very intermittent character of the Lagrangian acceleration.
Namely, the pronounced central peak (low accelerations) and long
tails (high accelerations)  make a highly non-Gaussian shape of
the acceleration distribution shown in Fig.~\ref{Fig1}. One
concludes that the observed fluid-particle dynamics is featured by
relatively frequent acceleration bursts, up to the measured 60
standard deviations. Such extreme events occur when the tracer
particle is captured by intense small-scale vortical structures
which are thought to be present in the turbulent flow. These
structures seem to be distributed randomly in space and time, with
large intervals between them which are characterized by
low-intensity events. As shown by Farge, Pellegrino, and
Schneider\cite{FargePRL2001} the most of the turbulent kinetic
energy is carried by vortex tubes, which are surrounded by a
background incoherent flow.

The long-standing Heisenberg-Yaglom scaling of a component of
Lagrangian acceleration,\cite{Monin-Yaglom,Heisenberg}
\be\label{HYscaling}
\langle a^2\rangle = a_0{\bar u}^{9/2}\nu^{-1/2}L^{-3/2},
\ee
was confirmed experimentally\cite{Bodenschatz} to a high accuracy,
for about seven orders of magnitude in the acceleration variance,
or two orders of the root-mean-square (rms) velocity $\bar
u=\langle u^2\rangle^{1/2}$ ($500\leq R_\lambda\leq 980$). Here,
the Lagrangian velocity $u$ is such that the average $\langle u
\rangle=0$, $a_0$ is the Kolmogorov constant, $\nu$ is the
kinematic viscosity, and $L$ is the Lagrangian integral length
scale. For $R_\lambda<500$ the Heisenberg-Yaglom scaling was found
to be broken. This signals increasing coupling of the acceleration
to large scales of the flow, and may be related specifically to
the large-scale anisotropy effect or to ``insufficient''
developing of the turbulence, or to both of them.

Very recent experimental data\cite{Mordant0410070} on the 3D
Lagrangian acceleration in turbulent flows with $R_\lambda=285,
485$, and 690 show that the three components $a_x$, $a_y$, and
$a_z$ of the acceleration are statistically dependent. For
example, the conditional variance $\langle a_y^2|a_z\rangle$
increases strongly with the magnitude of $a_z$. The acceleration
magnitude $|\bm{a}|$ was found to be characterized by the
probability density function, which is comparable to a log-normal
distribution of variance 1 at small and medium values
$|\bm{a}|/(\langle |{\bm a}|^2\rangle-\langle |{\bm
a}|\rangle^2)^{1/2}\leq 25$, and by the autocorrelation time of
about the integral time scale. The autocorrelation time of the
direction of the acceleration vector $\bm{a}$ is of the
dissipation time scale.  The observed two-time-scale character of
the stochastic dynamics is consistent with previous experimental
and DNS results.\cite{Mordant0206013} Assuming the log-normal
distribution of the magnitude $|\bm{a}|$ and statistical isotropy
of the acceleration vector $\bm{a}$ one can straightforwardly
derive distribution of each component\cite{Mordant0410070}
\be\label{logmagnitude}
P(a_i) =\frac{\exp[s^2/2]}{4m}\left[ 1-
\mbox{erf}\left(\frac{\ln[|a_i|/m] +s^2}{2s^2}\right)\right],
\ee
where erf$(x)$ is the error function, $m =\sqrt{3/e^{2s^2}}$ for a
unit variance and $s$ is a free parameter. It was shown that the
predicted distribution at $s=1$ follows experimental data points
to a good accuracy at small and medium accelerations
$|a_i|/\langle a_i^2\rangle^{1/2}\leq 25$, and overestimates them
at higher values. The origin of this departure is not clear.
However, it should be noted that for higher $R_\lambda$ tails of
the observed distributions become wider, and approach the
predicted curve.

It should be emphasized that the Lagrangian velocity components
for the studied flow follow Gaussian distribution to a good
accuracy. Theoretically, time derivative of a dynamical variable
does not necessarily follow the same statistical distribution as
that of the variable. The link between these two sharply distinct
distributions ---Lagrangian velocity and acceleration--- can be
seen from studying stationary statistics of the time increment of
a component $u$ of the velocity of individual fluid particle,
\be
\delta_\tau u(t)= u(t+\tau)-u(t).
\ee
For the time scale $\tau$ of the order of Lagrangian integral time
scale $T_L$ (large characteristic time scale of the flow implied
by simple dimensional analysis) the stationary distribution of
$\delta_\tau u(t)$ is approximately of a Gaussian form while for
$\tau$ decreasing down to the Kolmogorov time scale $\tau_\eta$
(small characteristic time scale of the flow implied by simple
dimensional analysis) the distribution continuously develops long
tails, and in a far dissipative subrange it reproduces the
acceleration distribution shown in Fig.~\ref{Fig1}. For
sufficiently small time scales $\tau<\tau_\eta$ turbulent
fluctuations are smoothed, and the increment is proportional to
the time scale, $\delta_\tau u(t)= \tau a(t)$, to a good accuracy.
The ratio between the timescales, $T_L/\tau_\eta$, is very large
for developed turbulent flows, and is characterized by Taylor
microscale Reynolds number $R_\lambda$; for the studied flow it is
of about two orders of magnitude.

High probability of extreme acceleration magnitudes, as compared
to that implied by the corresponding Gaussian distribution, is
associated with the Lagrangian turbulence intermittency, which was
found to be considerably stronger than the Eulerian one.
Equivalently, one can say that it is related to an increase of the
probability to have larger velocity increments in time with
decrease of time scale, down to the Kolmogorov one (a statistical
viewpoint). This is due to the absence of the so called sweeping
effect in the Lagrangian frame and the existence of relatively
long-lived intense vortical structures (vortex tubes) with radii
of the order of Kolmogorov length $\eta$ and total sizes extending
up to the integral length scale $L$. Recent laboratory experiments
by Mouri, Hori, and Kawashima\cite{Mouri0407111} (see also
references therein) for boundary layers with $R_\lambda =$
295--1255 confirm this picture.

In the traditional Eulerian framework, the isotropic turbulence
intermittency is understood differently, as an increase of the
probability to have larger longitudinal velocity differences
\be
\delta_l v(x)=v(x+l)-v(x)
\ee
on shorter spatial separation scales $l$, and studied through
scaling exponents $\zeta^E(p)$ of the Eulerian velocity structure
functions $\langle (\delta_l v)^p\rangle\sim\l^{\zeta^E(p)}$,
$p=1,2,3,\ldots$ (a structural viewpoint).\cite{Frisch95} The
velocity difference is taken at the same time instance. For
Lagrangian velocity structure functions one considers scaling with
respect to the time scale $\tau$, $\langle(\delta_\tau u)^p\rangle
\sim\tau^{\zeta(p)}$.

Scaling properties of the Eulerian and Lagrangian velocity
structure functions (statistical moments of $\delta_l v$ and
$\delta_\tau u$ which characterize their distributions) are
traditionally used to quantify turbulence intermittency. Here, the
intermittency exhibits itself as observed deviations from the
``normal'' scalings predicted by K41 theory. The deviations become
stronger with increasing order $p$, with the exception being
$\zeta^E(3)=1$ which corresponds to the four-fifth law by
Kolmogorov ($\zeta(2)=1$ for the Lagrangian
case).\cite{K41,Monin-Yaglom}

Under the assumption of balance between the energy injected by
driving forces, which occur presumably at large spatial scales,
and the energy dissipated by viscous processes, which are
concentrated at small scales, one can restrict consideration by a
statistically steady state, and focus on intermediate scales, the
inertial range, characteristics of which are universal to some
extent for high-Re flows. In the inertial range the energy is
transferred from large to small scales (the direct energy cascade)
and viscosity effects are not noticeable. Due to the K41
hypotheses\cite{K41} certain properties of fully developed
turbulence in the inertial range are independent on the details of
initial conditions and forcing (boundary conditions), as well as
on the details of the energy dissipation. This hypothesis is valid
only statistically, in the sense that the velocity and
acceleration (Heisenberg\cite{Heisenberg} and
Yaglom\cite{Monin-Yaglom}) are viewed as random variables. Hence,
one is interested in probability density functions and correlators
of the variables. Complete K41 scale invariance of the Eulerian
velocity difference statistics is known to be broken, with the
``anomaly'' coming from sensitivity of the inertial range of
scales to large scales of the flow. The breaking occurs also for
the predicted K41 scalings of the Lagrangian velocity structure
functions $\langle(\delta_\tau u)^p\rangle$ for the same reason.
The observed scaling exponents behave as {\em nonlinear} functions
of the order $p$ rather than linear (``normal'') ones. This is
associated in general with the so called dissipative anomaly (mean
turbulent kinetic energy dissipation rate remains finite when one
puts the viscosity parameter to zero) corresponding to a strongly
nonlinear and non-equilibrium character of a high-Re turbulent
system. Possibility to describe high-Re turbulence on the basis of
unified statistical principles, such as those successfully applied
to (quasi-)equilibrium systems, is an open problem. The strong and
nonlocal character of Lagrangian particle coupling due to pressure
effects makes the main obstacle to derive turbulence statistics
from the Navier-Stokes equation; {{motion of a single
fluid-particle is strongly correlated to that of the other
particles.}} Studying and accurate modeling Lagrangian dynamics of
a many-particle configuration,\cite{Falkovich2001} and
particularly single-particle behavior, is currently under
elaboration.

In the inertial range, which extends much for high-Re flows,
contribution of the pressure gradient to the acceleration variance
strongly dominates over that of the viscous force. The two
contributions do not correlate and the acceleration field can be
then approximated as the potential one. The DNS pressure gradient
data meet that of the experimental Lagrangian
acceleration.\cite{Bodenschatz,Bodenschatz2,Kraichnan0305040} The
effects of viscosity and intermittency are known to reduce the
effective inertial range characterized by the ascribed
scaling.\cite{AurellPRL1996}

Statistical isotropy and homogeneity of a fully developed
turbulent flow at small scales are assumed by K41 theory and
greatly simplify statistical description of turbulence. Turbulence
intermittency is related to a non-Gaussian statistical behavior
and is a more subtle matter for theoretical study. Intermittency
of the stochastic energy dissipation rate at scale $l$ is related
to intermittency of dynamics of the system that makes a link
between the Eulerian and Lagrangian aspects of intermittency. It
should be emphasized that Eulerian and Lagrangian approaches in
studying fluid flows are characterized by different theoretical
technique and implications, and compliment each other.

Modeling statistical turbulence in the Lagrangian frame is
important both for theoretical implications and applicational
studies. Simple models focus on single-particle statistical
properties and employ Langevin-type equations for one variable.
Partial justification of the use of one-dimensional models comes
from the K41 assumption on statistical isotropy of the flow at
small scales. Away from boundaries one can also use the assumption
on homogeneity of the flow and discard dependence of the
Lagrangian tracer on initial position for a sufficiently long
evolution. When comparing to the experimental or DNS data one
usually peaks up one measured component of the variable, or uses
averaging over all accessible components to get higher statistics.
Experimentally, a three-dimensional picture is difficult to
reach\cite{Mordant0410070} while DNS naturally gives a full access
to it. Some Lagrangian experiments allow very accurate resolution
of the Kolmogorov time scale with relatively short integration
time, the others allow to follow individual particle for long time
but do not resolve the Kolmogorov time scale. DNS is characterized
by a high resolution of the Kolmogorov time scale and long
integration time but at lower $R_\lambda$ achieved because of
current computational limitations.

As the experiments and performed DNS (on cubic lattices) do not
provide perfect isotropy of large-scale forcing (boundary) one
naturally expects anisotropy effects at all scales. However, these
effects are usually small at small scales (even when strong
large-scale anisotropy of a high-Re turbulent flow is present) and
thus can be ignored in the first approximation. Influence of
anisotropy effects is usually treated as a small correction. Local
isotropy of a fully developed turbulent flow
---rather strong but very fruitful condition--- is one of the main
assumptions of K41 phenomenological theory.\cite{Monin-Yaglom}
However, anisotropy effects at small scales are known to be
persistent for high Reynolds numbers. For example, persistent
anisotropy in the skewness of velocity derivatives in homogeneous
shear flows, which represent one of the simplest anisotropic flow,
is observed.\cite{ShumacherJFM2001} We remind that the forceless
3D Navier-Stokes equation is invariant under SO(3) rotational Lie
group in coordinate space, and symmetry breaking may come only due
to the forcing and/or boundary conditions. Recently developed
SO(3) decomposition theory\cite{Biferale0404014} can be used to
treat anisotropic and isotopic sectors in a rigorous way, and to
study how the isotropy recovery at small scales happens in
Navier-Stokes turbulence. The isotropy recovery is partially
justified owing to a subleading character of anisotropy found in
some exactly solvable models.

Two-time-scale stochastic dynamics in describing the Lagrangian
acceleration component jointly with the Lagrangian velocity and
position of a fluid particle was proposed long time ago by
Sawford.\cite{Sawford} The model equations are
%%\begin{eqnarray}
\be
\label{LangevinSawford}
\partial_t a = -(T_L^{-1}+\tau_\eta^{-1})a + T_L^{-1}\tau_\eta^{-1}u \\
\nonumber + \sqrt{2\sigma_u^2(T_L^{-1}+\tau_\eta^{-1})
T_L^{-1}\tau_\eta^{-1}}L(t),
\ee
%%\end{eqnarray}
\be
\partial_t u =a, \quad \partial_t x = u,
\ee
where
\be\label{TLteta}
T_L= \frac{2\sigma_u^2}{C_0{\bar\epsilon}},
 \quad
\tau_\eta=\frac{C_0\nu^{1/2}}{2a_0{\bar\epsilon}^{1/2}}
\ee
are two time scales, $T_L\gg \tau_\eta$, $L(t)$ is zero-mean
Gaussian-white noise, $C_0$ and $a_0$ are Lagrangian structure
constants, $\sigma_u^2$ is the variance of velocity distribution,
and $\bar\epsilon$ is the mean turbulent kinetic energy
dissipation rate per unit mass.

K41 hypotheses of locally isotropic character of high-Re
turbulence and similarity lead to the result that the acceleration
field is spatially isotropic, $\langle a_ia_j\rangle \simeq
\delta_{ij}$, and the stationary probability distribution of
acceleration may depend only on the parameters $\bar\epsilon$ and
$\nu$ (mean energy flux and viscosity). The second-order
Lagrangian velocity structure function $\langle \delta_\tau
u_i\delta_\tau u_j\rangle \simeq \delta_{ij}$ also should show
spatial isotropy for the inertial range of time scales. Thus, a
single-component consideration makes a sense;  $u=u_1, u_2,$ or
$u_3$ and $a=a_1, a_2,$ or $a_3$, in laboratory Cartesian frame of
reference. The constant $C_0$ enters the linear scaling of the
velocity structure function $\langle (\delta_\tau u)^2\rangle =
C_0\bar\epsilon\tau$ implied by K41 theory for the inertial range
of time scales $\tau_\eta \ll \tau \ll T_L$. Since the form of
this two-time correlator is similar to that of the displacement of
usual Brownian particle, the velocity of fluid particle in the
inertial range can be thought of as a Brownian-like motion with
the ``diffusion'' coefficient $C_0\bar\epsilon$. In other words,
the velocity is a stationary stochastic process with independent
increments. For time scales smaller than the Kolmogorov time scale
$\tau_\eta$ the predicted scaling is quite different, $\langle
(\delta_\tau u)^2\rangle = a_0\bar\epsilon^{3/2}\nu^{-1/2}\tau^2$,
and directly corresponds to Heisenberg-Yaglom scaling law
(\ref{HYscaling}); $\bar\epsilon={\bar u}^3/L$,
$\eta=(\nu^{3}/{\bar\epsilon})^{1/4}$, and $R_\lambda=
(15/{\bar\epsilon}\nu)^{1/2}{\bar u}^2$. This theory also predicts
${\bar\epsilon}/\tau$ decay of the autocorrelation of acceleration
component $\langle a(t)a(t+\tau)\rangle$, when one imposes its
independence on $\nu$, for time scales $\tau$ much bigger than
$\tau_\eta$. For the velocity autocorrelation $\langle
u(t)u(t+\tau)\rangle$ the prediction is that it decays
considerably only for $\tau$ of the order of Lagrangian integral
time scale $T_L$. The uncorrelated character of Lagrangian
acceleration then could be used to build a first approximation for
time scales within the inertial range. Note however that the
predicted decay of $\langle a(t)a(t+\tau)\rangle$ across this
range is rather slow. Qualitatively K41 relations mean that the
components of Lagrangian acceleration and velocity are associated
mainly with small and large scales of a developed turbulent flow
respectively.

Sawford model (\ref{LangevinSawford}) predicts stationary Gaussian
distributions for both the acceleration and velocity reflecting
the used uncorrelated character of fluctuations and is consistent
with K41 picture. One of the extensions of this model is due to
replacement of $\bar\epsilon$ by stochastic energy dissipation
rate $\epsilon$, and assuming that it is lognormally distributed
in correspondence to the refined Kolmogorov 1962 (K62)
approach.\cite{K62} Such extensions can be used to fit the
observed highly non-Gaussian shape of the acceleration
distribution shown in Fig.~\ref{Fig1}.

Recent Lagrangian experiments and DNS by Mordant, Delour, Leveque,
Arneodo, and Pinton\cite{Mordant0206013} and DNS by Chevillard,
Roux, Leveque, Mordant, Pinton, and
Arneodo\cite{Chevillard0310105} show that certain {\em long-time}
correlations and the occurrence of very large fluctuations at
small scales dominate the motion of a fluid particle, and this
leads to a new dynamical picture of turbulence. This requires
effective models on how to account for the specific long-time
correlations along a particle trajectory which are viewed as a key
to intermittency in turbulence.

While it is evident that the 3D Navier-Stokes equation with a
Gaussian-white random forcing belongs to a class of non-linear
stochastic dynamical equations for the velocity field with which
one can associate some generalized Fokker-Planck equations or
apply a path-integral method, it is a theoretical challenge to
make a link between the Navier-Stokes equation (\ref{NSaccel}) and
phenomenological stochastic models of Lagrangian acceleration.

Recent approach by Friedrich\cite{FriedrichPRL2003} can be traced
back to the so called probability density function method
originated by Oboukhov\cite{Oboukhov1959} and developed by
Pope\cite{Pope2000} and Sawford\cite{Sawford2001} (see also
references therein). Friedrich has shown that one can obtain
infinite chain of evolution equations for joint Lagrangian
$n$-point probability density functions and closed equation for
the associated probability density functional which stem from the
incompressible 3D Navier-Stokes equation in the Lagrangian frame;
see also work by Heppe.\cite{HeppeJFM1998} Evolution equation for
the single-particle distribution function $f(\bm v,\bm r,\bm
x_0,t)=\langle \delta(\bm v-\bm u(\bm x_0,t))\delta(\bm r-\bm
x(\bm x_0,t))\rangle$, where $\bm u(\bm x_0,t)$ and $\bm x(\bm
x_0,t)$ are Lagrangian velocity and position respectively,
includes integral of pressure gradient and dissipation operators
acting on mixed Eulerian-Lagrangian equal-time two-particle
distribution function, and so on. Particularly, he derived a
generalized Fokker-Planck equation (with memory term) for a
single-particle probability distribution of Lagrangian velocity
increments by using certain closure scheme partially justified for
high-Re homogeneous isotropic turbulence. The approach naturally
leads to consideration of acceleration covariances conditional on
Lagrangian velocity and position which correspond to a three-point
distribution function. Such a conditional dependence was dropped
in order to reduce the Fokker-Planck equation, which nevertheless
accounts for time integrated effects. This approximation means
particularly that the correlation between acceleration fields at
space-time points $(\bm r, t)$ and $(\bm r-\bm l, t')$ does not
depend on the velocity of a fluid particle at $(\bm r-\bm l, t')$,
where $\bm l = \bm v(t-t')$. K41 theory is used to derive general
form of the two-point two-time acceleration autocorrelation
function, which approximates diffusion term, for the inertial
range, whereas the drift term vanishes identically because of the
ignored conditional dependence. Power-law form for unknown
function entering the diffusion term in the Fokker-Planck equation
for modulus of velocity was used, with the exponent being treated
as a free parameter. This leads to consideration of a class of
continuous-time random walk of the velocity featured by
non-Markovian behavior, which is contrasted to Markovian treatment
(no memory effects) underlying Oboukhov model\cite{Oboukhov1959}
with Gaussian distributed Lagrangian acceleration. The resulting
equation is analytically tractable, and its solution is presented
in the form of definite integral. Timescale dependence of the free
parameter was used to fit the experimental data on statistics of
Lagrangian velocity increments in a wide range of
timescales.\cite{Mordant0103084} The introduced timescale
dependence requires a justification since this parameter was
treated constant when solving the evolution equation. The closure
scheme provides the following scaling behavior of the Lagrangian
velocity distribution: $P(u,t)=t^{-3/2}P_s(ut^{-1/2})$. Importance
of this approach is that it has deductive support from the
Navier-Stokes turbulence and directly accounts for the memory
effects.

Lagrangian acceleration statistics {\em conditional} on the same
component of Lagrangian velocity was studied experimentally by
Mordant, Crawford, and Bodenschatz.\cite{Mordant0303003} These
data add a very useful information on the Lagrangian intermittency
as well as allow one to check implications of refined stochastic
models, which describe distribution of the acceleration
conditional on velocity.

The conditional acceleration probability density function $P(a|u)$
at a set of fixed velocities $u$ ranging from 0 to 3.1 (in rms
units) was found to be of approximately the same stretched
exponential shape as that of the unconditional acceleration $P(a)$
shown in Fig.~\ref{Fig1}. Theoretically, the distribution $P(a)$
can be calculated with the use of $P(a|u)$ by integrating out $u$
in $P(a|u)$ with some (independent) distribution of $u$. The
experimental conditional acceleration variance $\langle
a^2|u\rangle$ was found to increase in a nonlinear way with the
increase of magnitude of velocity $u$. Dependence of the
acceleration variance on the velocity magnitude breaks local
homogeneity of the flow assumed by K41 theory, and is a
prerequisite to describe turbulence intermittency. One therefore
should admit influence of larger scales when describing the
small-scale dynamics by supposing that the intense structures are
characterized by both the large and small time scales in the
Lagrangian framework. The conditional mean acceleration $\langle
a|u\rangle$ was found to be nonzero and increases for higher
velocity magnitudes that reflects the large-scale anisotropy
effect of the studied flow. Recent DNS result by Biferale,
Boffetta, Celani, Devenish, Lanotte, and
Toschi\cite{Biferale0403020} based on the analysis of $3.6\times
10^9$ data points also shows an essential dependence of the
acceleration variance on magnitude of large velocities. These
findings are consistent with the understanding that the long-time
correlations along a particle trajectory dominate the motion since
Lagrangian velocity is characterized by the ``energy-containing''
scales of a turbulent flow.

The aim of the present paper is to review simple Langevin-type
single-particle modeling approach and make a comparative analysis
of different recent models of Lagrangian acceleration, on the
basis of recent Lagrangian experimental data and direct numerical
simulations of high-Re isotropic turbulence. We restrict
consideration by steady-state Lagrangian single-particle
statistics. Most of the reviewed models are one-dimensional. Such
models can shed some light to a more realistic three-dimensional
modeling. We also briefly review some recent results of
alternative approaches,
---multifractal description and multifractal random walk model of
homogeneous and isotropic turbulence--- to provide the reader with
a current view on the problem.

The layout of the paper is as follows.

In Sec.~\ref{Sec:Multifractal} we briefly review recent
multifractal approaches to the Lagrangian and Eulerian
intermittency. The formalism by Chevillard, Roux, Leveque,
Mordant, Pinton, and Arneodo\cite{Chevillard0310105} is a
Lagrangian version of Eulerian multifractal approach and describes
statistics of Lagrangian velocity increments in a wide range of
time scales, from the integral to dissipative one. Fine structure
of the range of spatial scales smaller than the inertial range has
been considered by Chevillard, Castaing, and
Leveque.\cite{Chevillard0311409} Arimitsu and
Arimitsu\cite{Arimitsu} have constructed multifractal cascade
model to derive Lagrangian acceleration distribution by making a
link between cascade picture of isotropic turbulence and Tsallis
nonextensive statistics formalism.\cite{Tsallis}

In Sec.~\ref{Sec:Langevinmodels} we review some recent
one-dimensional Langevin-type models of the Lagrangian
acceleration in developed turbulence. In Sec.~\ref{Sec:RINmodels},
we outline implications of the  models by Beck\cite{Beck,Beck4}
with the underlying $\chi$-square
(Sec.~\ref{Sec:ChisquareDistribution}) and log-normal
(Sec.~\ref{Sec:LognormalDistribution}) distributions of the model
parameter $\beta$, and the $\chi$-square Gaussian
model.\cite{Aringazin0301040} We review results of the so called
{Random Intensity of Noise} (RIN) approach\cite{Aringazin0301245}
to specify the probability density function $f(\beta)$ which is
based upon relating $\beta$ to normally distributed velocity $u$
(Sec.~\ref{Sec:GaussianDistribution}). This formalism enables to
reproduce $\chi$-square and log-normal distributions of $\beta$ as
particular cases.

A nonlinear Langevin and the associated Fokker-Planck equations
obtained by a direct requirement that the probability distribution
satisfies some model-independent scaling relation have been
recently proposed by Hnat, Chapman, and Rowlands\cite{Hnat} to
describe the measured time series of the solar wind bulk plasma
parameters. We find this result relevant to fluid turbulence since
it is based on a stochastic dynamical framework and leads to the
stationary distribution with exponentially truncated power law
tails, similar to that obtained in the above mentioned RIN models.
This model is reviewed in Sec.~\ref{Sec:HCRmodel}.

Recent second-order and third-order Langevin stochastic models of
Lagrangian acceleration developed by
Reynolds\cite{ReynoldsPF2003,ReynoldsPRL2003,ReynoldsNEXT2003,ReynoldsPF2003b}
are reviewed in Sec.~\ref{Sec:Reynoldsmodels}. The second-order
model generalizes Sawford stochastic model (\ref{LangevinSawford})
while the third-order model introduces hyper-acceleration
(substantive derivative of the acceleration) and the associated
time scale. When neglecting third-order processes one recovers a
second-order model. {{Reynolds-number effects are incorporated
into the second-order model, which is applicable at large time
scale.}} Such a modeling of accelerations in homogeneous
anisotropic turbulence has been recently made by Reynolds, Yeo and
Lee.\cite{ReynoldsPRE2004} Reynolds and
Veneziani\cite{ReynoldsPLA2004} have shown importance of
trajectory-rotations and that non-zero mean rotations are
associated with suppressed rates of turbulent dispersion and
oscillatory Lagrangian velocity autocorrelation functions.
Particularly, due to the developed extended second-order model,
non-zero conditional mean acceleration endows trajectories with a
preferred sense of rotation.

The Navier-Stokes equation based approach to describe statistical
properties of small-scale velocity increments, both in the
Eulerian and Lagrangian frames, was developed in much detail by
Laval, Dubrulle, and Nazarenko;\cite{Laval0101036} see also recent
work by Laval, Dubrulle, and McWilliams.\cite{Laval2} This
approach introduces nonlocal interactions between well separated
scales, the so called elongated triads, and is referred to as the
Rapid Distortion Theory (RDT) approach. This approach is
contrasted with Gledzer-Ohkitani-Yamada (GOY) shell model,
 %% see references in:
 %%Growth of Noninfinitesimal Perturbations in Turbulence
 %%E. Aurell, G. Boffetta,  A. Crisanti, G. Paladin, A. Vulpiani
 %%PRL 77, 1262 (1996).
 %%file: Growth of Noninfinitesimal Perturbations in Turbulence PRL 1996.pdf
in which interactions of a shell of wave numbers with only its
nearest and next-nearest shells are taken into account. In
Sec.~\ref{Sec:LDNmodel} we outline results of this approach and
focus on the proposed one-dimensional Langevin-type model of
Lagrangian small-scale turbulence to which we refer as
Laval-Dubrulle-Nazarenko (LDN) model. This model includes
Gaussian-white additive and multiplicative noises with constant
intensities, while local interactions are accounted for by
introducing a turbulent viscosity. {{LDN-type model for Lagrangian
acceleration exploits such a simple form of the noises, which
represent effects of stochastic distortion produced by large
scales.}}

In Sec.~\ref{Sec:Comparison} we represent in some detail
qualitative and quantitative comparative analysis of the
one-dimensional LDN-type model at zero correlation between the
noises and simple RIN models.\cite{Aringazin0305186}

In Sec.~\ref{Sec:ConditionalProbability} we review very recent
models of the conditional acceleration statistics by Sawford,
Yeung, Borgas, Vedula, La Porta, Crawford, and
Bodenschatz,\cite{SawfordPF2003} Reynolds,\cite{ReynoldsNEXT2003}
and Biferale {\it et al.}\cite{Biferale0403020} We present our
study\cite{Aringazin0305186} on the conditional probability
density function $P(a|u)$ where the one-dimensional LDN-type model
with mutually $\delta$-correlated Gaussian-white additive and
multiplicative noises is taken as a constitutive model and certain
model parameters are assumed to depend on the amplitude of
Lagrangian velocity $u$. We also present results of a complete
quantitative description of the available experimental
data\cite{Bodenschatz,Bodenschatz2,Mordant0303003} on conditional
and unconditional acceleration statistics within the framework of
a single LDN-type model with a single set of fit
parameters.\cite{Aringazin0312415}

In Sec.~\ref{Sec:MRW} we briefly review recent results of the
application of multifractal random walk theory by Muzy and
Bacry\cite{Muzy0206202,Bacry0207094} to developed turbulence. This
approach allows one to go beyond modeling of the Lagrangian
velocity of fluid particle by Gaussian process to include Poisson
process, and to use Kolmogorov-Levy-Khinchin theory of stochastic
processes with independent increments.

%% 2
\section{Multifractal approaches}
\label{Sec:Multifractal}

Recently, Chevillard, Roux, Leveque, Mordant, Pinton, and
Arneodo\cite{Chevillard0310105} have constructed an appropriately
recasted multifractal approach, which is widely used in Eulerian
studies of turbulence, to describe statistics of Lagrangian
velocity increments in a wide range of timescales, from the
integral to dissipative one. The resulting theoretical
distribution reproduces continuous widening of the velocity
increment probability density function (PDF) with the decrease of
time scale, from a Gaussian-shaped to the stretched exponential,
as observed in Lagrangian experiments carried out at
Cornell\cite{Bodenschatz,Bodenschatz2,Mordant0303003} and
ENS-Lyon,\cite{Mordant0206013,Mordant0103084} and DNS of the 3D
Navier-Stokes equation. Two global parameters (Reynolds number and
Lagrangian integral time scale) and two local parameters
(smoothing parameter and intermittency parameter) with a parabolic
singularity spectrum were used to cover the data in the entire
range of time scales.

At dissipative time scale the obtained PDF fits the experimental
data on Lagrangian acceleration to a good accuracy. The cumulant
analysis made in this approach provides an understanding of the
observed departures from the scaling when going from the integral
to dissipative time scale. The used parabolic singularity spectrum
$D(h)$ is a hallmark of the log-normal (Kolmogorov 1962)
statistics and reproduces well the left-hand side (corresponding
to intense velocity increments) of the observed curve, which is
centered at 0.58 ($>1/2$), but increasingly deviates at the
right-hand side (rhs) of it (corresponding to weak velocity
increments). Another widely used statistics, the log-Poisson one,
was shown to imply departure from the Lagrangian observations in
the same manner. The acceleration statistics conditional on
velocity was not considered in this work.

The basic assumption of the multifractal approach is to relate
Lagrangian velocity increments at different time scales to each
other,\cite{Chevillard0310105}
\be
u(t+\tau)-u(t) = \beta(\tau/T)(u(t+T)-u(t)),
\ee
by using independent random function $\beta(\tau/T)$, where the
time scale $\tau$ is such that $\tau<T$ and $T$ is fixed at the
order of Lagrangian integral time scale $T_L$. This relation is
understood as a statistical law. When considering the function
$\beta(\tau/T)$ to be deterministic, one ends up with a
monofractal (monoscale, or self-similar) picture, well-known
example of which is Brownian motion. Random character of
$\beta(\tau/T)$ leads to a multiscale
behavior\cite{Yaglom1966,Mandelbrot1974} of the stochastic
velocity, for which scaling properties of structure functions can
be readily derived.\cite{Novikov1994} The scaling exponent
$\zeta(p)$ of the $p$-order structure function is linear for
monofractal processes and nonlinear in $p$ for multifractal
processes. The simplest example of multifractal process is given
by log-normal distribution of $\beta$, for which case the
resulting scaling exponent is a second-order polynomial in $p$
(parabola).

The multifractal approach has been managed to describe both the
inertial and {\em dissipative} range of time scales. Thus the
effect of dissipation has been accounted. The condition that the
local Reynolds number at some time scale between the inertial and
dissipative ranges is of the order of unity was used, and the
Lagrangian integral time scale and Reynolds number, which are
available from the experimental data, are explicitly incorporated
into the formalism. The remaining two free parameters were used
for fitting. The obtained PDF of the Lagrangian velocity
increments is symmetric and includes both the distinct regimes in
a unified way by using Batchelor's interpolation formula, which
contains the smoothing parameter. The other free parameter is the
Lagrangian intermittency parameter measuring second-order
nonlinearity of the scaling exponent of the Lagrangian velocity
structure function. Gaussian shape for the PDF of velocity
increments at the integral time scale was used, and the Kolmogorov
scaling of the second-order Lagrangian velocity structure function
$\langle (\delta_\tau u)^2\rangle \simeq \tau$ was assumed for the
inertial range. The results of this approach are in a good
agreement with various sets of experimental and DNS data.

In a very recent paper Chevillard, Castaing, and
Leveque\cite{Chevillard0311409} have considered a fine structure
of the range of scales smaller than the inertial range, within the
framework of Eulerian multifractal approach. Below, we briefly
review and discuss results of this work as it concerns very small
scales with which fluid-particle acceleration is generally
associated. Indeed, Heisenberg-Yaglom scaling law
(\ref{HYscaling}) shows that the acceleration essentially depends
on the viscosity $\nu$ so that it is associated with very small
scales $l<\eta$ of a turbulent flow for which viscous effects are
known to dominate over inertial effects.

They introduced the so called far-dissipative range of spatial
scales, $l<\eta_-$, and the near-dissipative range, $l\in
[\eta_-,\eta_+]$, and found that the Eulerian intermittency grows
faster across the scales in the near-dissipative range as compared
with that in the inertial range, $l>\eta_+$, with decreasing
separation length $l$ of the longitudinal velocity difference; the
Kolmogorov length scale $\eta$ is such that $\eta_-<\eta<\eta_+$.
This observed phenomenon has been qualitatively described and
attributed to the reinforcement of contrast between intense and
quiescent motions due to the gradually increasing scale-dependent
viscous cutoff effect when going to smaller scales, from $\eta_+$
to $\eta_-$. As the typical scalings in the inertial and
far-dissipative ranges are known one can compute relationship
between the values of cumulants at the endpoints of the
near-dissipative range. The so called 9/4 amplification law for
the intermittency has been established: the parameter measuring
intermittency increases more rapidly during the crossover from the
inertial range to the far-dissipative range as compared to that in
the inertial range. Highly remarkably, this result has been found
to be independent on the Reynolds number.

In general, the far-dissipative range is characterized by a strong
viscous damping and eventual saturation of the intermittency with
decreasing scale which reaches some highest possible value for a
given turbulent flow. Reynolds number dependence of the left and
right endpoints $\eta_-$ and $\eta_+$ has been established and
described: the near-dissipative range of scales becomes wider for
higher Reynolds numbers exhibiting approximately
$\sqrt{\ln\mathrm{Re}}$ dependence but the inertial range grows
faster, approximately as $\ln\mathrm{Re}$. Scaling properties of
Eulerian velocity structure functions were found to be {\em
non-universal} in the near-dissipative range of scales as they
depend on the Reynolds number and the order of structure function.
The experiment and DNS were made for Re $=11750$ and Re $=1070$
flows respectively\cite{Chevillard0311409} for which pronounced
near-dissipation range is observed. For sufficiently high Reynolds
numbers the near-dissipative range can be taken negligible when
compared to the inertial range, and the Kolmogorov length
$\eta\simeq\eta_-\simeq\eta_+$ becomes the only scale
characterizing dissipative range that appears to be in accord to
the original K41 theory.

While it is evident that viscous effects are ultimately
responsible for such a behavior of the intermittency parameter in
the near-dissipative range and that the Kolmogorov scale can be
defined by setting local Reynolds number to unity, it is a rather
subtle matter to relate corrections to intermittency coming from
viscous cutoff scales with Navier-Stokes turbulence dynamics.
Below we present a tentative picture.

In general, one naturally expects a specific crossover from the
mechanism of downscale energy-transfer operating within the
inertial range (the direct energy cascade with negligible effect
of viscosity) to the mechanism of strong viscous dissipation of
energy at very small scales of the flow. These two mechanisms are
evidently of a quite different character, and matching one to the
other requires specific intermittent structures which span the
near-dissipative range.

Possible downscale phenomenological picture is that small scales
become much more intense as one riches the Kolmogorov length scale
$\eta$, which is a characteristic size of very intense vortical
structures responsible for intermittent bursts in this range. The
viscous damping tends to weaken and destroy such intense vortices
and their correlations, and terminate formation of smaller-radius
vortex tubes for a given Reynolds number. For local Reynolds
numbers less than unity one therefore expects no cascade picture
except for a single (the smallest-radius) vortex tube, while for
that bigger than unity cluster of highly correlated vortex tubes
is likely to be present. The 9/4 amplification law might be due to
relative under-population and/or more coherent character of such
intense vortical structures as compared to that of vortical
structures in the inertial range.

Vortex tubes are viewed as the most elementary structures in
turbulence.\cite{Pullin1998} Recent rough-wall boundary-layer
experimental analysis by Mouri, Hori, and
Kawashima\cite{Mouri0407111} of spatial distribution of the
small-scale vortex tubes with characteristic radii of
6.1$\eta$--7.4$\eta$ shows that the probability to find small
separations between the tubes is considerably higher than that of
large separations. The experimental data are due to
one-dimensional cut. The vortex tubes have been identified using
enhancements of the velocity increment above certain threshold,
which eliminates detection of weak tubes and other low-intensity
structures. The data give access down to fractions of Taylor
microscale $\lambda=[2\langle v^2\rangle/\langle(\partial_x
v)^2\rangle]^{1/2}$ of the studied wind tunnel flows. The length
parameter $\lambda$ increases from 34$\eta$ to 70$\eta$ for the
flows with $R_\lambda$ from 295 to 1255, while Kolmogorov scale
$\eta$ decreases from 0.06~cm to 0.01~cm respectively;
$R_\lambda=\langle v^2\rangle^{1/2}\lambda/\nu$. The radius of
vortex tube is identified by the position of maximum of the
obtained Burgers-like antisymmetrized velocity profile with
respect to zero. The finding is that large separations are
distributed in a random and independent way due to the observed
exponential tail of the distribution whereas smaller separations,
below few Taylor lengths, occur increasingly frequent.
Superposition of two exponential functions was used as a model
which fits the experimental distribution at large and middle
separations (from 25$\lambda$ to 5$\lambda$) to a good accuracy
but increasingly underestimates it at smaller separations (less
than $5\lambda$). This indicates that the vortex tubes tend to
cluster together and correlate to each other at the middle and
small separation scales. Shapes of the experimental spatial
distributions, argument of which is expressed in Taylor microscale
units, are found approximately the same for a wide range of
Reynolds numbers. This can be treated as a universal character of
the spatial distribution of vortex tubes of the same small
dimensionless radius (in Kolmogorov length units) in high-Re
turbulent flows, $R_\lambda >400$. Reynolds-number dependence of
the tube parameters was studied as well: the radius scales
linearly with Kolmogorov length, $R_0 \sim \eta$, and the maximum
circulation velocity scales linearly with the rms velocity,
$V_0\sim \langle v^2\rangle^{1/2}$, for $400<R_\lambda<1255$; the
linear scaling of the velocity with respect to the Kolmogorov
velocity $u_K=(\nu{\bar\epsilon})^{1/4}$ which is acceptable
within the framework of K41 theory is not observed.

As the tendency is that small-radius vortex tubes form clusters
and correlate to each other at small separations (which
incorporate the inertial range of scales), one expects more sparse
character of vortical structures at smaller scales. Existence of
vortex tubes with smaller radii is supported by the observation
that the Reynolds number Re$_0=V_0R_0/\nu$ characterizing
circulations of the vortex tubes scales as $R_\lambda^{1/2}$ (i.e.
not constant), with the value increasing from Re$_0 = 32$ to 62
for $R_\lambda = 295$ to 1255. Despite that high Re$_0$ may result
in less stable character of the vortex tubes they have a rather
long lifetime, which is of the order of large-eddy turnover time
$L/\langle v^2\rangle^{1/2}$. This effect may be due to the
clustering.

Since one of the characteristic parameters of the small-scale
vortical structures essentially depends on $R_\lambda$ it is then
natural to observe Reynolds-number dependence of the Lagrangian
acceleration statistics;\cite{Bodenschatz} the level of
intermittency increases with Reynolds number.

%%Also, higher circulation velocity of the tubes for higher
%%$R_\lambda$ could explain the observed essential dependence of the
%%Lagrangian acceleration variance on Lagrangian
%%velocity.\cite{Mordant0303003,Biferale0403020}

We note that for vortex tubes with small Re$_0\leq 1$ one expects
no clustering. High probability to find the same-radius vortex
tubes with Re$_0>1$ separated by small distance suggests that they
form coupled pairs, or multipole clusters consisting of several
vortex elements in a more general case. Conservation of the
enstrophy moment for a vortex pair can be used to find minimal
distance between centers of vortex tubes which is estimated to be
of about 3.1--3.7 radius of the tube.\cite{Zhanabaev2002} Hence
the probability of separations smaller than about 3 radius
($\approx 21\eta$ for the studied tubes) should tend to zero. The
experimental distribution\cite{Mouri0407111} is currently
available for separations down to few radii for which a tendency
to reach maximum and drop with decreasing separation is indeed
observed. Accurate resolution would allow to identify the relative
separation corresponding to maximum probability and study its
Reynolds-number dependence. This could help to investigate the
vortex clusters which appear to be typical small-scale objects in
high-Re turbulent flows. The vortical structures are advected by a
noisy background flow, such as the eddy-noise
forcing\cite{Frisch95} and the energy-equipartition based random
forcing.\cite{FargePRL2001}

It is important to note that just higher mean intensity of small
scales does not guarantee increase of intermittency since the
latter is associated mainly with (nonlocal) interactions of the
small scales with much larger scales, as argued recently by Laval,
Dubrulle, and Nazarenko.\cite{Laval0101036}

The specific increase of intermittency in the near-dissipative
range indicates just a noticeable character rather than the
overall essential gradual increase of the effective viscous
damping across the near-dissipative range. This is contrasted to
what happens in the far-dissipative range which corresponds mainly
to the interior and surrounding of small-scale intense vortex
tubes where strong viscous dissipation occurs. However, one should
keep in mind that some other parameters control intermittency as
well when one treats turbulence intermittency as a nonlocal
phenomenon. For example, (a) higher intensity of stochastic
large-scale strain coupled to small scales and (b) stronger
large-scale correlations across the inertial range produce
considerable increase of the intermittency at the small scales.

It seems that more sparse character of coherent structures at
small scales, rather than some more or less intense local
interactions, could be directly associated with the rapid increase
of intermittency in the near-dissipative range. The role of
viscous damping is essential here, but it is likely restricted to
the strong damping effect at the smaller-scale end of the
near-dissipative range. In general, this could be viewed as a
phenomenon related to maintaining of the mean energy flux
downscale to the far-dissipative range. It seems that the local
Reynolds number varies much due to high inhomogeneity of the flow
at the small scales so that more refined tools as compared to the
usual Fourier transform are required to capture details of the
small scales.

%%One can also speculate that the small-scale coherent structures in
%%the near-dissipative range are reorganized in such a way as that
%%they would favor upscale energy transfer towards the inertial
%%range due to extremely big fluctuations (of course the mean energy
%%flux is still downscale).

More detailed analysis and interpretation of experiments resolving
Kolmogorov scale are required for the near-dissipative range which
is currently much less understood than the inertial range. In
particular, whether the flatness factor of the distribution of
Lagrangian velocity increments exhibits a pronounced range of
timescales similar to the near-dissipative range of spatial scales
is still an open question.\cite{Mordant0303003,Aringazin0305459}

In a very recent paper Biferale, Boffetta, Celani, Lanotte, and
Toschi\cite{Biferale0402032} have presented interesting results of
DNS of Lagrangian transport in homogeneous and isotropic
turbulence with $R_\lambda$ up to 280, a very accurate resolution
of dissipative scale, and an integration time of about Lagrangian
integral time scale. This is contrasted to the experimental
optical tracking\cite{Bodenschatz,Bodenschatz2} which gives access
to the resolution of about 1/65 of the Kolmogorov time scale and
integration time of about few Kolmogorov time scales of the
$R_\lambda=690$ flow, and to the experimental acoustic
tracking\cite{Mordant0206013} which enables the integration time
of the order of Lagrangian integral time scale but no access to
time scales less than the Kolmogorov time scale of the
$R_\lambda=740$ flow. While the value $R_\lambda=280$ is not as
high as those in the experiments, an additional advantage of the
DNS data is that it gives access to a full three-dimensional
picture of the flow and high statistics is reached by ``seeding''
and tracking simultaneously millions of Lagrangian tracers.

In the subsequent work, Biferale, Boffetta, Celani, Devenish,
Lanotte, and Toschi\cite{Biferale0403020} have shown how the
multifractal formalism offers an alternative approach which is
rooted in the phenomenology of turbulence. The Lagrangian
statistics was derived from the Eulerian statistics without
introducing {\it ad hoc} hypotheses. She-Leveque empirical formula
for the Eulerian scaling exponents has been used and time scale is
related to the length scale by using the assumption that Eulerian
velocity differences are proportional to Lagrangian velocity
increments, $\delta_l v \simeq \delta_\tau u$. Although the
formalism is not capable to account for small acceleration values
(typical situation for the multifractal approach), the obtained
acceleration PDF captures the DNS data to a good accuracy in the
tails, with acceleration values ranging from about $|a|/\langle
a^2\rangle^{1/2}=1$ up to 80. Alas, one can observe an
overestimation in this range which can be clearly seen from the
predicted contribution to fourth-order moment, $a^4P(a)$, as
compared to the DNS data. High degree of isotropy of the simulated
stationary flow suggests statistical equivalence of Cartesian
components of acceleration aligned to fixed directions, and the
resulting DNS acceleration distribution obtained by averaging over
the components has been found, as expected, with no observable
asymmetry with respect to $a\to -a$.

Also, acceleration variance conditional on the velocity has been
derived\cite{Biferale0403020} within the same multifractal
approach and compared to the DNS data. We will consider this issue
below in Sec.~\ref{Sec:ConditionalProbability}.

Recent multifractal cascade model by Arimitsu and
Arimitsu\cite{Arimitsu} implies Lagrangian acceleration
distribution, which fits DNS acceleration
data\cite{GotohPF2002,Kraichnan0305040} to a very good accuracy.
The model is based on the analysis of scale invariance of the 3D
Navier-Stokes equation for high Reynolds numbers, and on the
assumption that singularities due to the invariance distribute
themselves multifractally in physical space. The guiding principle
is an extremum of Tsallis nonextensive entropy\cite{Tsallis} under
certain constraints from which distribution function $P(\alpha)$
of singularity exponent $\alpha$ is obtained. Basically, two fit
parameters determine the resulting distribution: the total number
of ``steps'' in turbulent cascade, $n$, and the intermittency
exponent $\mu$. The ascribed ``eddy size'' decrement factor is 2.
Each step assumes statistical independence of the corresponding
flow modes within the multifractal scaling range. The acceleration
statistics is obtained from the scaling $\delta p_m/\delta
p_0=(l_m/l_0)^{2\alpha/3}$, where $\delta p_m$ is the pressure
difference across the separation length $l_m=2^{-m}l_0$ and $l_0$
is the turbulence integral scale. At the step $m=n$ the cascade is
terminated and one expects good approximation for the pressure
gradient. Minimum and maximum values of $\alpha$ for the
singularity spectrum $f(\alpha)$ are related to Tsallis entropic
index $q$ by
\be
(1-q)^{-1} = 1/\alpha_- - 1/\alpha_+,
\ee
$f(\alpha_{\pm})=0$. The representation for spectrum corresponding
to the cascade model of isotropic turbulence is found as follows:
\be
f(\alpha) =  1+(1-q)^{-1}\log_2[1-
(\alpha-\alpha_0)^2/(\Delta\alpha)^2],
\ee
where $(\Delta\alpha)^2 = 2X/[(1-q)\ln 2]$ and $q$, $X$, and
$\alpha_0$ are determined from $\mu$;
$\alpha_{\pm}=\alpha_0\pm\Delta\alpha$. Energy conservation and
definition of the intermittency exponent were used.

It was argued that the acceleration distribution should include
two parts: one coming from the above multifractal analysis and the
other corresponding to contribution of dissipative term (the so
called ``thermal fluctuations'' and/or measurement errors). The
first part determines shape of the tails whereas the second part
makes the core of distribution (small acceleration magnitudes)
with another parameter $q'$ entering model Tsallis distribution.
Very good fit to the DNS acceleration data is obtained for the
values $n=18$, $\mu=0.240$ $(q=0.391)$, and $q'=1.7$. However, as
pointed out by Kraichnan and Gotoh,\cite{Kraichnan0305040} the
total number of steps for the simulated $R_\lambda=380$ flow
should not exceed $n=9$ to provide consistent treatment of the
cascade, and that there is no way to fit the tails at $n=9$ by
tuning the value of $\mu$.

%% 3
\section{One-dimensional Langevin toy models
of Lagrangian acceleration in turbulence}
\label{Sec:Langevinmodels}

In this Section, we outline results of some recent one-dimensional
Langevin-type models of Lagrangian fluid particle acceleration in
a developed turbulent flow.

Modeling of the Lagrangian acceleration dynamics can be naturally
made by employing Langevin-type equation, which contains time
derivative of the acceleration, so that random forces responsible
for the time evolution of acceleration of a fluid particle are
related to the time derivative of the rhs of Eq.~(\ref{NSaccel})
treated in the Lagrangian frame.

Various one-dimensional models were suggested recently to describe
Lagrangian acceleration statistics among which we mention the
$\chi$-square model\cite{Beck,Beck2} and the log-normal
model\cite{Beck4} by Beck which are based on the Tsallis
nonextensive statistics\cite{Tsallis} inspired
approach,\cite{Johal,Wilk,Aringazin0204359,Beck3} the second-order
and third-order models by
Reynolds\cite{ReynoldsPF2003,ReynoldsPRL2003,ReynoldsNEXT2003,ReynoldsPF2003b}
which extends the model by Sawford,\cite{Sawford} the
$\chi$-square Gaussian
model,\cite{Aringazin0301040,Aringazin0212642} and the model with
underlying normally distributed Lagrangian velocity
fluctuations.\cite{Aringazin0301245}

It is worthwhile to note that the idea to use stochastic averaging
over random variance of intermittent variable was used long time
ago by Castaing, Gagne, and Hopfinger.\cite{Castaing} Their
propagator approach is related to the so called Markovian
description by Renner, Peinke, and Friedrich\cite{Friedrich} as
shown by Donkov, Donkov, and Grancharova;\cite{Donkov} see also
work by Amblard and Brossier.\cite{Amblard}

Review and critical analysis of the applications of various recent
nonextensive statistics based models to the turbulence have been
made by Gotoh and Kraichnan.\cite{Kraichnan0305040} An emphasis
was made that some models lack justification of a fit from
turbulence dynamics although being able to reproduce experimental
and DNS data to more or less accuracy. Deductive support from the
3D Navier-Stokes equation was stressed to be essential for the
fitting procedure to be considered meaningful; see also
Ref.~\refcite{Aringazin0305186} for a review. Also, Zanette and
Montemurro\cite{Zanette2004} have argued recently that the
connection between specific non-thermodynamical processes and
non-extensive mechanisms is generally not well defined.

%% 3.1
\subsection{Simple RIN models}
\label{Sec:RINmodels}

Tsallis nonextensive statistics\cite{Tsallis} inspired
formalism\cite{Johal,Aringazin0204359,Beck3} was recently used by
Beck\cite{Beck,Beck4} to describe Lagrangian statistical
properties of developed turbulence; see also
Refs.~\refcite{ReynoldsPF2003,Beck2,Wilk}. In recent
papers,\cite{Aringazin0301040,Aringazin0301245,Aringazin0212642}
we have made some refinements of this approach.

In this approach, PDF of a component of acceleration of
infinitesimal fluid particle in developed turbulent flow is found
due to the equation
\be\label{P}
P(a) = \int_{0}^{\infty}\! d\beta\ P(a|\beta)f(\beta),
\ee
where $P(a|\beta)$ is PDF of $a$ conditional on $\beta$. This
distribution is associated with a surrogate dynamical equation,
the one-dimensional Langevin equation for the Lagrangian
acceleration $a$,
\be\label{Langevin}
\partial_t a = \gamma F(a) + \sigma L(t).
\ee
Here, $\partial_t$ denotes time derivative, $F(a)$ is the
deterministic drift force, $\gamma$ is the drift coefficient,
$\sigma^2$ measures intensity of the noise, a strength of the
additive stochastic force, and $L(t)$ is Gaussian-white noise with
zero mean,
\be\label{Lnoise}
\langle L(t)L(t')\rangle=2\delta(t-t'), \quad
 \langle L(t)\rangle=0.
\ee
The averaging is made over ensemble realizations. Short-time
correlated force, which is approximated here by $L(t)$, is assumed
to come as a combined dynamical effect of the flow modes. This
force can be viewed as a ``background'' stochastic force which
acts along a particle trajectory.

For constant model parameters $\gamma$ and $\sigma$, this usual
Langevin model ensures that the stochastic process $a(t)$ defined
by Eq.~(\ref{Langevin}) is Markovian. The PDF $P(a|\beta)$ of the
acceleration at fixed
%%\be\label{beta0}
$\beta =\gamma/\sigma^2$
%%\ee
can be found as a stationary solution
of the corresponding Fokker-Planck equation
\be\label{FPconstant}
\partial_t P(a,t)
 = \partial_a[-\gamma F(a) + \sigma^2\partial_a]P(a,t),
\ee
where $\partial_a =\partial/\partial a$. This equation can be
derived from the Langevin equation (\ref{Langevin}) using the
noise (\ref{Lnoise}) either in Stratonovich or Ito
interpretations. Particularly, for a linear drift force $F(a)=-a$,
the stationary PDF, i.e. $\partial_t P(a,t)=0$, is of a Gaussian
form,
\be\label{PGauss}
P(a|\beta)= C(\beta)\exp[-\beta a^2/2],
\ee
where $C(\beta)=\sqrt{\beta/(2\pi)}$ is a normalization constant
and $a\in (-\infty,\infty)$.

With constant $\beta$, the Gaussian PDF (\ref{PGauss}) corresponds
to the non-intermittent K41 picture of fully developed turbulence,
and formally agrees with the experimental statistics of components
of velocity increments in time for the integral time scale.
However, it fails to describe observed Reynolds-number-dependent
stretched exponential tails of the experimental acceleration PDF
shown in Fig.~\ref{Fig1} which correspond to anomalously high
probabilities for the tracer particle to have extremely high
accelerations, bursts with dozens of {rms} acceleration.

The function $f(\beta)$ entering Eq.~(\ref{P}) is a PDF arising
from the assumption that $\beta$ is a random parameter with
prescribed external statistics. This is the main point of the
approach, which extends the usual Gaussian picture. Evidently, the
characteristic time of variation of the parameter $\beta$ should
be sufficiently large to justify approximation that the resulting
PDF (\ref{PGauss}) is used with this independent randomized
parameter. Two well separated time scales in the Lagrangian
velocity increment autocorrelation have been established both by
experiments and DNS.\cite{Mordant0206013} The large time scale has
been found of the order of the Lagrangian integral time scale and
corresponds to a magnitude part of Lagrangian velocity increments.

%%that is in accord to assumption that the intensity of noise along
%%a trajectory is longtime fluctuating.

The model (\ref{Langevin}) belongs to a class of stochastic
single-particle models of Lagrangian turbulence and deals with an
evolution of the acceleration in time which in accord to the
Navier-Stokes equation is driven by time derivative of the rhs of
Eq.~(\ref{NSaccel}). This type of modeling corresponds to the
well-known universality\cite{K41,Heisenberg,Yaglom1966} in
statistically homogeneous and isotropic turbulence. This is in an
agreement with the observed temporally irregular character of the
Lagrangian velocity and acceleration of a tracer particle in
high-Re turbulent flows.

In a physical context, an essential fluid-particle dynamics in the
developed turbulent flow is described here in terms of a
generalized Brownian-like motion, a stochastic particle approach,
taking the particle acceleration (\ref{Laccel}) as the dynamical
variable. Such models are generally based upon a hierarchy of
characteristic time scales in the system and naturally employ a
one-point statistical description using Langevin-type equation (a
stochastic differential equation of first order) for the dynamical
variable, or the associated Fokker-Planck equation (a partial
differential equation) for one-point probability density function
of the variable.

With the choice of $\delta$-correlated noises such Langevin-type
models fall into the class of Markovian models (no memory effects)
allowing well established Fokker-Planck approximation.
Consideration of finite-time correlated noises and the associated
memory effects requires a deeper analysis which should be made
separately in each particular case. The evolution equations are
formulated and solved in the Lagrangian framework, in a purely
temporal treatment, with fluctuations being treated along a
particle trajectory. Fokker-Planck equation with memory term for
joint Lagrangian single-particle PDF of velocity increments has
been studied recently by Friedrich.\cite{FriedrichPRL2003}

Approximation of a short-time correlated noise by the zero-time
correlated one is usually made due to the timescale hierarchy
emerging from the general physical analysis of the system and
experimental data. Under the stationarity condition, one can try
to solve the Fokker-Planck equation to find stationary PDF of the
acceleration, $P(a)$. This function as well as the associated
statistical moments can then be compared with the experimental
data on acceleration statistics. The Fokker-Planck approximation
allows one to make a link between the dynamics and the statistical
approach. In the case when stationary probability distribution can
be found exactly one can make a further analysis without a
dynamical reference, yet having a possibility to extract
stationary time correlators.

In contrast to the usual Brownian-like motion, the fluid-particle
acceleration does not merely follow a random walk with a complete
self-similarity at all scales. It was found to reveal a different,
multiscale self-similarity, which can be seen from wide tails of a
non-Gaussian distribution of the experimental data shown in
Fig.~\ref{Fig1}. This requires a consideration of some specific
Langevin-type equations, which may include nonlinear terms, {e.g.}
to account for turbulent viscosity effects, and an extension of
the usual properties of model forces, additive and multiplicative
noises.

Specifically, the class of models represented by
Eqs.~(\ref{P})-(\ref{PGauss}) is featured by consideration of the
acceleration evolution driven by the ``forces'' characterized by
{\it fluctuating} drift coefficient $\gamma$ (or fluctuating
intensity of multiplicative noise in a more general case) and/or
{\it fluctuating} intensity $\sigma^2$ of the additive noise. This
was found to imply stationary distributions of the acceleration
(or velocity increments in time, for finite time lags) of a
non-Gaussian form with wide tails which are a classical signature
of the turbulence intermittency. Earlier work on such type of
models are due to Castaing, Gagne, and Hopfinger,\cite{Castaing}
referred to as the Castaing model, in which a log-normal
distribution of fluctuating variance of intermittent variable was
used without reference to a stochastic dynamical equation.

The difference from the well known class of stochastic models with
Gaussian-white additive and {multiplicative} noises which are also
known to imply stationary distributions with wide tails is that
one supposes that {\em intensities} of the noises are not constant
but fluctuate at a large time scale. We refer to the models with
such {\em random intensities of noises} as RIN models.

This class of models introduces a two-time-scale dynamics, one
associated with $\delta$-correlation of noises (modeling the
smallest time scale under consideration) and the other associated
with variations of intensities of the noises, their possible
coupling to each other, and other parameters assumed to occur at
large time scales, up to the Lagrangian integral time scale. From
a general point of view, one can assume a hierarchy of a number of
characteristic time scales.\cite{Biferale0402032,Biferale0403020}
However, as a first step one simplifies the consideration in order
to make it more analytically tractable, in accord to the presence
of two characteristic time scales in the Kolmogorov 1941 picture
of fully developed turbulence.

In the approximation of two time-scales, one can start with a
Langevin-type equation, derive the associated Fokker-Planck
equation in Stratonovich or Ito formulations, and try to find a
stationary solution of the Fokker-Planck equation, in which slowly
fluctuating parameters are taken to be fixed. As the next step,
one evaluates stochastic expectation of the resulting {\em
conditional} PDF over the parameters with some distributions
assigned to them. By this way one can obtain a stationary marginal
PDF as the main prediction of the model.

The dynamical model (\ref{Langevin}) represents a particular
simple one-dimensional RIN model characterized by the presence of
additive noise (a short time scale) and the fluctuating composite
parameter $\beta=\gamma/\sigma^2$ (a long time scale), where
$\gamma$ is simply kinetic coefficient (a multiplicative noise is
not present explicitly) and $\sigma^2$ is the additive noise
intensity. This model is, of course, far from being a full model
of the essential dynamics of fluid particle in the developed
turbulence regime. It is a theoretical challenge to make a link
between the Navier-Stokes equation and surrogate one-dimensional
Langevin models for acceleration such as that defined by
Eq.~(\ref{Langevin}).

The averaging (\ref{P}) of the Gaussian distribution
(\ref{PGauss}) over randomly distributed positive $\beta$, an
evaluation of the stochastic expectation, was found to be a simple
{\it ad hoc} procedure to obtain observable predictions, with one
free parameter, which meets experimental statistical data on the
acceleration of tracer particle. One can think of this as the
averaging over a large time span for one tracer particle, or as
the averaging over an ensemble of tracer particles, moving in the
three-dimensional flow characterized by domains with different
values of $\beta$ randomly distributed in space.

In a physical context one would like to know processes underlying
the random character of the model parameter $\beta$. Due to the
definition the random character of $\beta$ is attributed to a
random character of the drift coefficient $\gamma$ and/or the
additive noise intensity $\sigma^2$. In contrast to the usual
Brownian motion, in which medium is treated thermodynamically in
an equilibrium state and therefore parameters entering dynamical
equation are taken constant, the system under consideration is
characterized by extreme fluctuations and presence of coherent
structures that naturally suggest fluctuating character of some of
the model parameters.

The distribution of $\beta$ is not fixed uniquely by the theory so
that a judicious choice of $f(\beta)$ makes a problem in the RIN
model (\ref{P})-(\ref{PGauss}). Below we briefly consider three
specific models characterized by different prescriptions for
distribution of the parameter $\beta$, and compare them to the
experimental data.

%% 3.1.1
\subsubsection{The underlying $\chi$-square distribution}
\label{Sec:ChisquareDistribution}

With the $\Gamma$ ($\chi$-square) distribution of $\beta$ of order
$n$ ($n=1,2,3,\dots$),
\be\label{chi2}
f(\beta) =\frac{1}{\Gamma(n/2)}
 \left(\frac{n}{2\beta_0}\right)^{{n}/{2}}
 \beta^{{n}/{2}-1}\exp\left[-\frac{n\beta}{2\beta_0}\right],
\ee
the resulting marginal probability density function (\ref{P}) with
$P(a|\beta)$ given by the Gaussian (\ref{PGauss}) is found in the
form~(cf. Ref.~\refcite{Beck})
\be\label{Pgamma}
P(a)=\frac{C}{(a^2+{n}/{\beta_0})^{(n+1)/2}},
\ee
where
%%\be
%%C= \frac{(n/\beta_0)^{n/2}\Gamma(\frac{n+1}{2})}
%%{\sqrt{\pi}\Gamma(\frac{n}{2})}
%%\ee
$C$ is normalization constant. With $n=3$  ($\beta_0=3$ for a unit
variance) one obtains the normalized marginal distribution in the
following simple form:
\be\label{Pchi2}
P(a)=\frac{2}{\pi(a^2+1)^2},
\ee
This is the prediction of the $\chi$-square model with the Tsallis
entropic index taken to be $q=(n+3)/(n+1)=3/2$ due to the
theoretical argument that the number of independent random
variables at Kolmogorov scale is $n=3$ for the three-dimensional
flow.\cite{Beck} One can see that the resulting marginal
distribution is characterized by power-law tails that {\it a
priori} lead to divergent fourth-order moment $\langle a^4\rangle
= \int_{-\infty}^{\infty}a^4P(a)da$.

A Gaussian truncation of the power-law tails naturally arises
under the assumption that the parameter $\beta$ contains a
non-fluctuating part, which can be separated out as follows:
$\beta/2 \to a_c^{-2} +\beta/2$.\cite{Aringazin0301040} Here,
$a_c$ is a free parameter measuring the non-fluctuating part. This
leads to essentially modified marginal distribution
\be\label{PAringazin}
P(a) = \frac{C\exp[-a^2/a_c^2]}{(a^2+{n}/{\beta_0})^{(n+1)/2}},
\ee
where $C$ is normalization constant and $a_c$ can be used for a
fitting.
%%Taking the theoretical value $n=3$ and $\beta_0=3$ as in the above
%%case, one obtains that
%%\be
%%C= \frac{2a_c^2}{\pi
%%(a_c^2-2)\exp[a_c^{-2}][1-\mathrm{erf}(a_c^{-1})]+2\sqrt{\pi}a_c},
%%\ee
%%where $\mathrm{erf}(x)$ denotes the error function.
This distribution at the fitted value $a_c=39.0$ is in a good
agreement with the experimental probability density
function.\cite{Bodenschatz,Bodenschatz2}

Note that at $a_c \to \infty$ (no constant part) the model
(\ref{PAringazin}) covers the model (\ref{Pgamma}). Within the
framework of Tsallis nonextensive statistics, the parameter $q-1$
measures a variance of fluctuations. For $q\to 1$ (no
fluctuations), Eq.~(\ref{PAringazin}) reduces to a Gaussian
distribution, which meets the experimental data for temporal
velocity increments at the integral time scale.

%% FIGURE 4
%%%%%%%%%%%%%%%%%%%%%%%%%%%%%%%%%%%%%%%%%%%%%%%%%%%%%%%
\begin{figure}[tbp!]
\begin{center}
\includegraphics[width=0.75\textwidth]{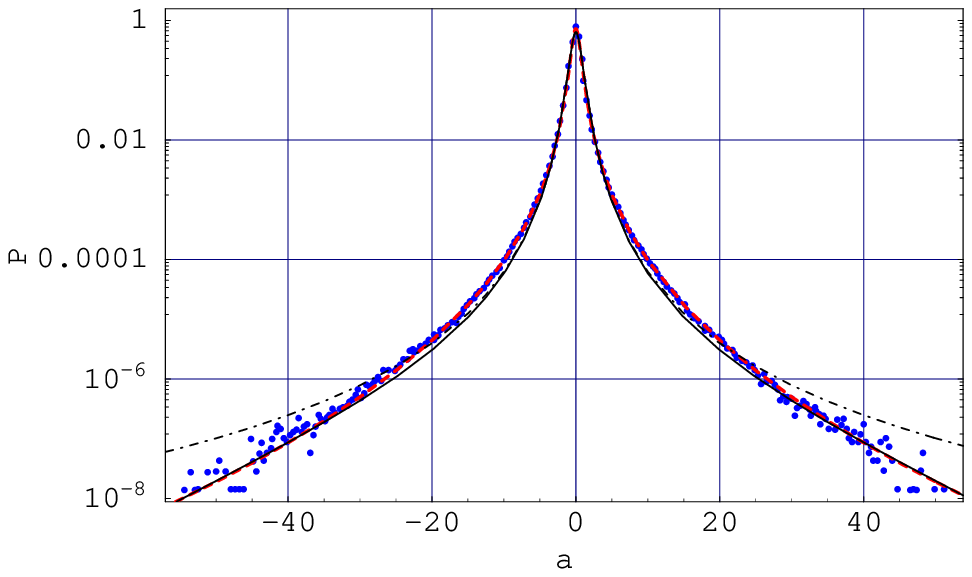}
\includegraphics[width=0.75\textwidth]{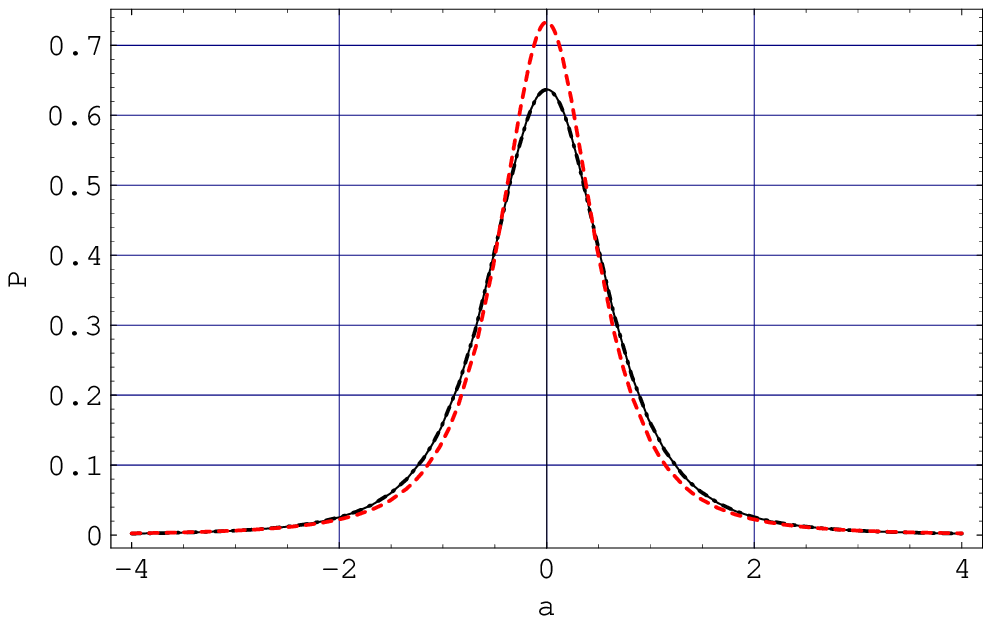}
\end{center}
\caption{ \label{Fig3} Lagrangian acceleration probability density
function $P(a)$. Dots: experimental data for the $R_\lambda=690$
flow by Crawford, Mordant, Bodenschatz, and Reynolds.$^2$
%%\cite{Bodenschatz2}
Dashed line: the stretched exponential fit (\ref{Pexper}).
%%$b_1=0.513$, $b_2= 0.563$, $b_3= 1.600$, $C=0.733$.
Dot-dashed line: Beck $\chi$-square model (\ref{Pchi2}), $q=3/2$.
Solid line: the $\chi$-square Gaussian model (\ref{PAringazin}),
$a_c=39.0$, $C=0.637$. $a$ denotes acceleration normalized to unit
variance.}
\end{figure}
%%%%%%%%%%%%%%%%%%%%%%%%%%%%%%%%%%%%%%%%%%%%%%%%%%%%%%%

%% FIGURE 5
%%%%%%%%%%%%%%%%%%%%%%%%%%%%%%%%%%%%%%%%%%%%%%%%%%%%%%%
\begin{figure}[tbp!]
\begin{center}
\includegraphics[width=0.75\textwidth]{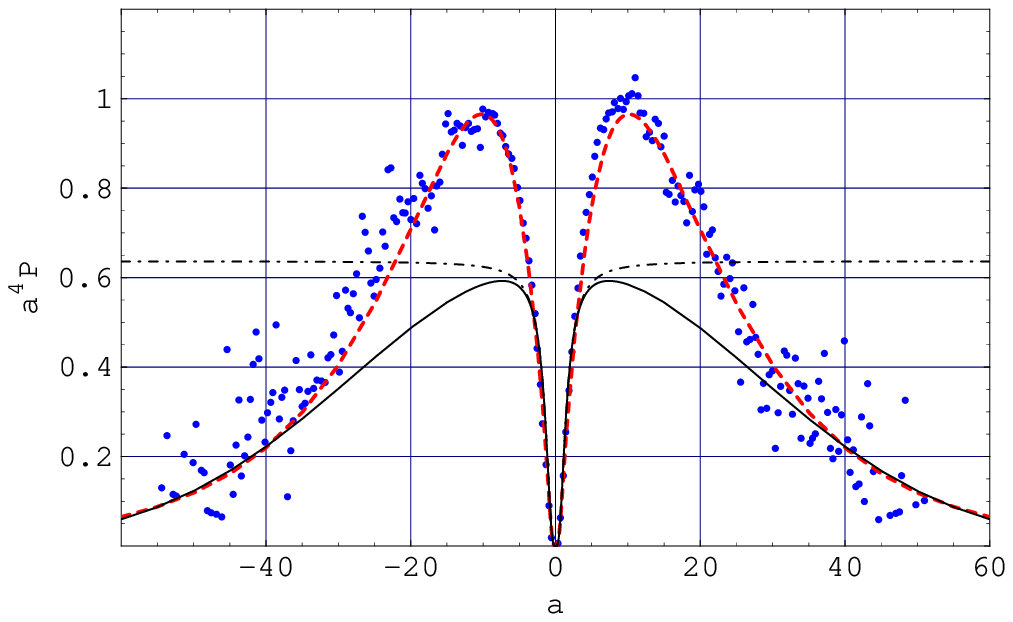}
\includegraphics[width=0.75\textwidth]{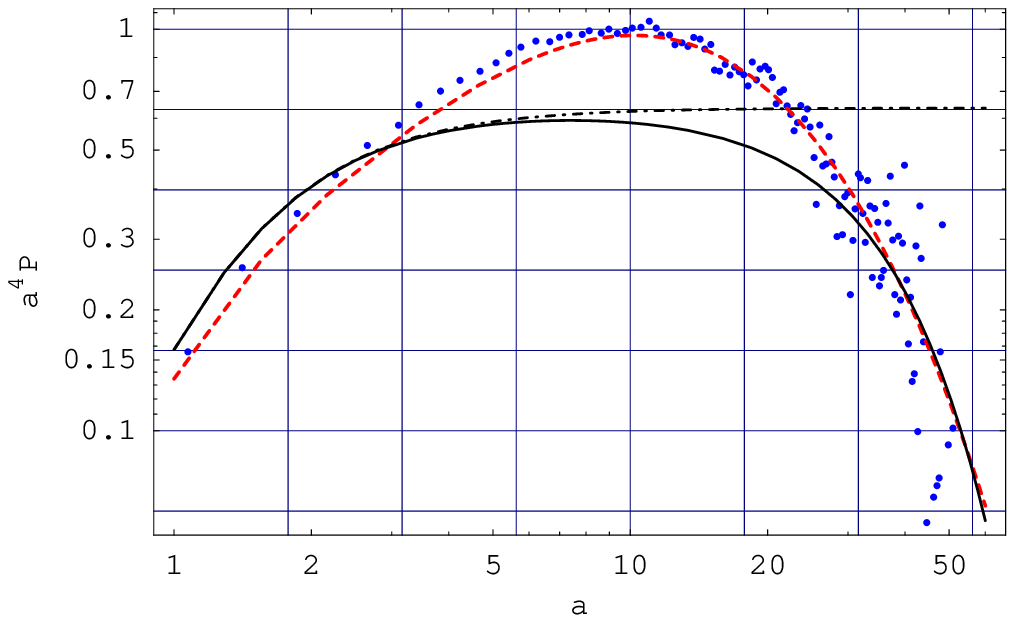}
\end{center}
\caption{ \label{Fig4} The contribution to fourth-order moment,
$a^4P(a)$. Top panel: a linear plot, bottom panel: a log-log plot.
Same notation as in Fig.~\ref{Fig3}.}
\end{figure}
%%%%%%%%%%%%%%%%%%%%%%%%%%%%%%%%%%%%%%%%%%%%%%%%%%%%%%%

A comparison of the $\chi$-square model (\ref{Pchi2}) and
$\chi$-square Gaussian model (\ref{PAringazin}) with the
experimental data is shown in Figs.~\ref{Fig3} and \ref{Fig4}. One
can see that both the distributions follow the experimental $P(a)$
to a good accuracy, although the tails of the $\chi$-square model
distribution depart from the experimental curve at large $|a|$. A
major difference is seen from the contribution to fourth-order
moment, $a^4P(a)$, shown in Fig.~\ref{Fig4}. The $\chi$-square
model yields a qualitatively unsatisfactory behavior indicating a
divergency of the predicted fourth-order moment $\langle
a^4\rangle$. In contrast, the $\chi$-square Gaussian model is in a
good qualitative agreement with the data, reproducing them well at
small and large acceleration values although quantitatively it
deviates at intermediate acceleration values and gives the
flatness value $F \simeq 46.1$ for $a_c=39.0$, as compared to the
flatness value given by Eq.~(\ref{flatness}).

%% 3.1.2
\subsubsection{The underlying log-normal distribution}
\label{Sec:LognormalDistribution}

%% FIGURE 6
%%%%%%%%%%%%%%%%%%%%%%%%%%%%%%%%%%%%%%%%%%%%%%%%%%%%%%%
\begin{figure}[tbp!]
\begin{center}
\includegraphics[width=0.75\textwidth]{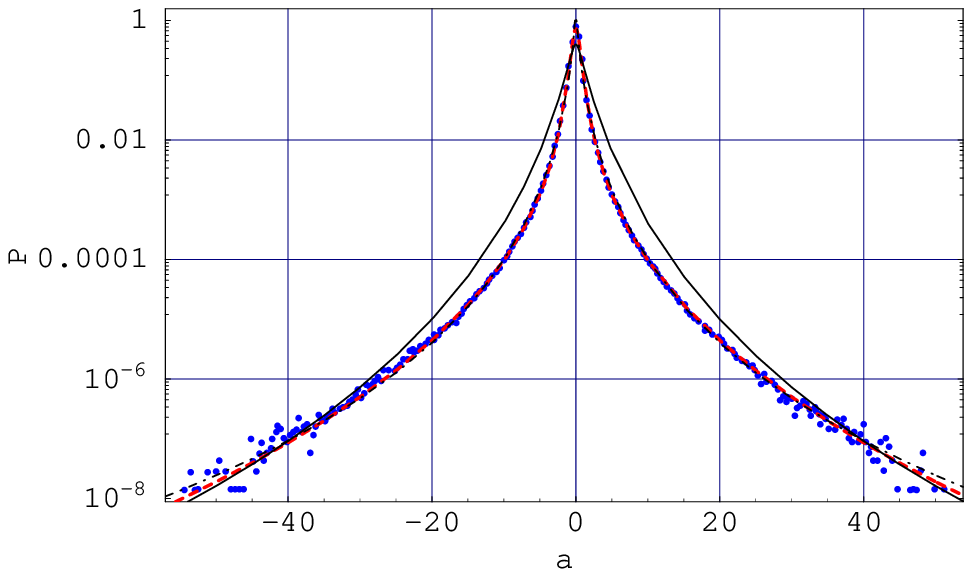}
\includegraphics[width=0.75\textwidth]{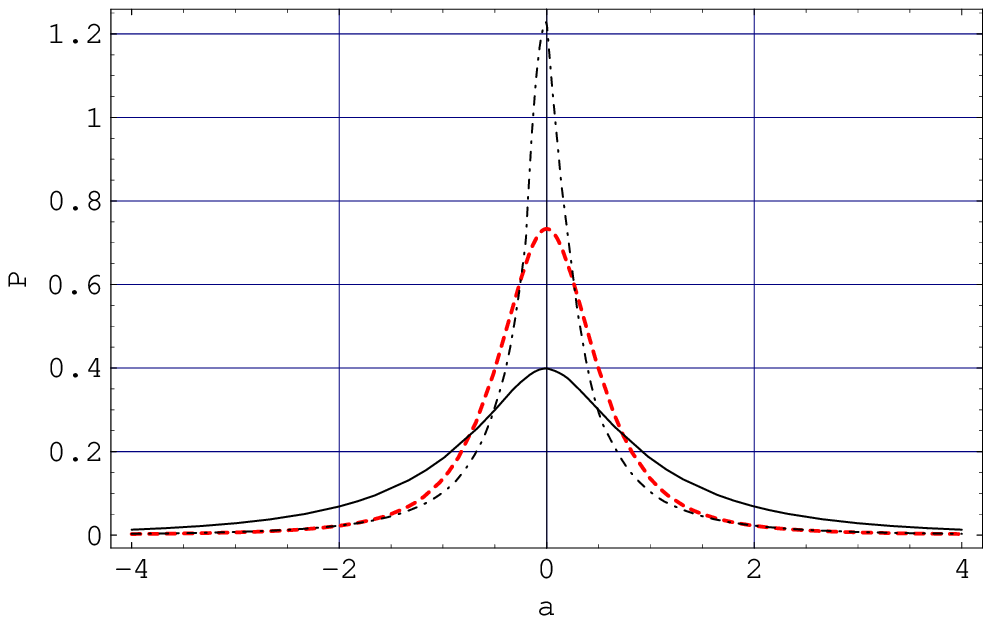}
\end{center}
\caption{ \label{Fig5} Lagrangian acceleration probability density
function $P(a)$. Dots: experimental data for the $R_\lambda=690$
flow by Crawford, Mordant, Bodenschatz, and Reynolds.$^2$
%%\cite{Bodenschatz2}.
Dashed line: the stretched exponential fit (\ref{Pexper}).
%%$b_1=0.513$, $b_2= 0.563$, $b_3= 1.600$, $C=0.733$.
Dot-dashed line: Beck log-normal model (\ref{PBeck}), $s=3.0$.
Solid line: Castaing log-normal model (\ref{PCastaing}),
$s_0=0.625$. $a$ denotes acceleration normalized to unit
variance.}
\end{figure}
%%%%%%%%%%%%%%%%%%%%%%%%%%%%%%%%%%%%%%%%%%%%%%%%%%%%%%%

%% FIGURE 7
%%%%%%%%%%%%%%%%%%%%%%%%%%%%%%%%%%%%%%%%%%%%%%%%%%%%%%%
\begin{figure}[tbp!]
\begin{center}
\includegraphics[width=0.75\textwidth]{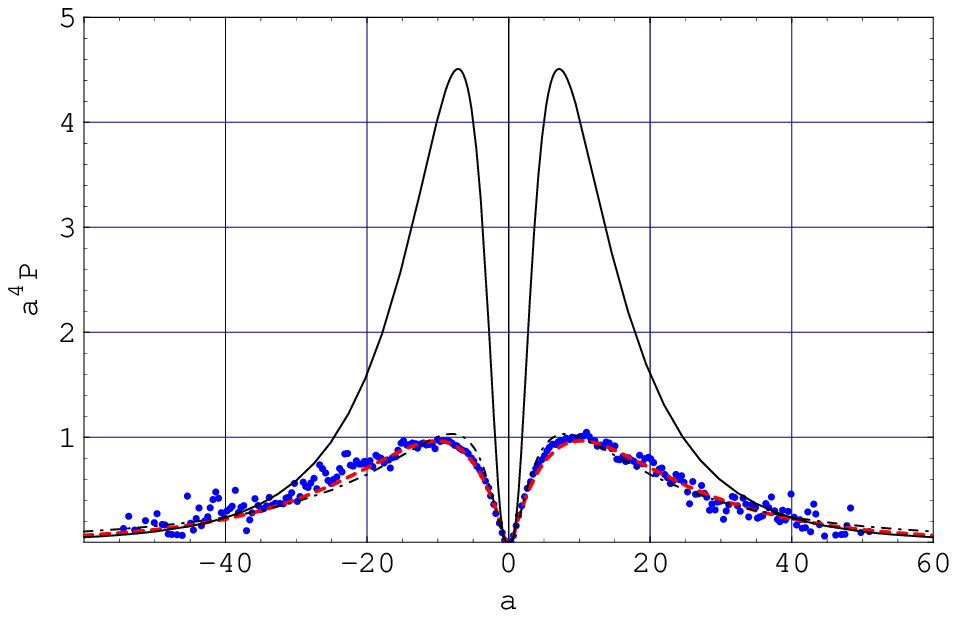}
\includegraphics[width=0.75\textwidth]{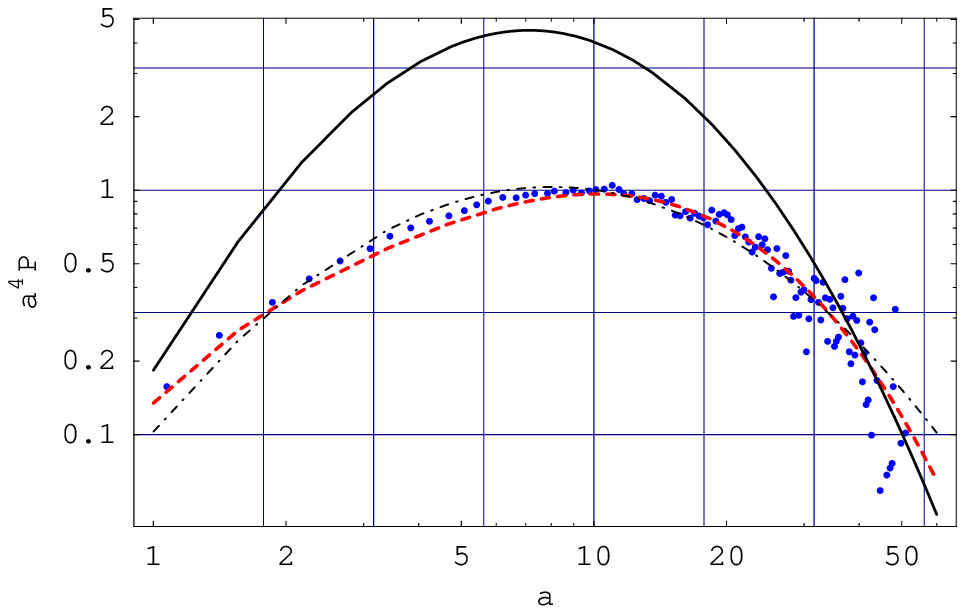}
\end{center}
\caption{ \label{Fig6} The contribution to fourth-order moment,
$a^4P(a)$. Top panel: a linear plot, bottom panel: a log-log plot.
Same notation as in Fig.~\ref{Fig5}.}
\end{figure}
%%%%%%%%%%%%%%%%%%%%%%%%%%%%%%%%%%%%%%%%%%%%%%%%%%%%%%%

With the log-normal distribution of $\beta$,
\be\label{lognormal}
f(\beta) =
\frac{1}{\sqrt{2\pi}s\beta}
\exp\left[-\frac{(\ln\frac{\beta}{m})^2}{2s^2}\right],
\ee
the resulting marginal PDF (\ref{P}) with $P(a|\beta)$ given by
the Gaussian (\ref{PGauss}) was recently proposed to
be\cite{Beck4}
\be\label{PBeck}
P(a) = \frac{1}{2\pi s}\int_{0}^{\infty}\!\!\! d\beta\
\beta^{-1/2}
\exp\left[-\frac{(\ln\frac{\beta}{m})^2}{2s^2}\right] e^{-\beta
a^2/2},
\ee
where the only free parameter $s$ can be used for a fitting, or
derived from theoretical arguments, $s^2=3$ ($m=\exp[s^2/2]$ to
provide unit variance). This distribution is shown in
Fig.~\ref{Fig5} and was found to be in a good agreement with the
Lagrangian experimental data by La Porta {\it et
al.},\cite{Bodenschatz} the new data by Crawford {\it et
al.},\cite{Bodenschatz2} Mordant {\it et
al.},\cite{Mordant0206013} and DNS of the Navier-Stokes equation
by Kraichnan and Gotoh.\cite{Kraichnan0305040}

However, the central part of the distribution shown in
Fig.~\ref{Fig5} reveals much greater inaccuracy of the log-normal
model ($P(0)\simeq 1.23$) as compared with that of both the
$\chi$-square and $\chi$-square Gaussian models ($P(0)\simeq
0.65$) which are almost not distinguishable in the region
$|a|/\langle a^2\rangle^{1/2}\leq 4$ (Fig.~\ref{Fig3}); see also
recent work by Gotoh and Kraichnan.\cite{Kraichnan0305040} This is
the main failing of the log-normal model (\ref{PBeck}) for
$s^2=3.0$ although the predicted distribution follows the measured
low-probability tails, which are related to acceleration bursts,
to a good accuracy. The core of the experimental curve
(\ref{Pexper}) ($P(0)\simeq 0.73$) contains most weight of the
experimental distribution and is the most accurate part of it,
with the relative uncertainty of about 3\% for $|a|/\langle
a^2\rangle^{1/2}<10$ and more than 40\% for $|a|/\langle
a^2\rangle^{1/2}>40$.\cite{Mordant0303003}

The distribution (\ref{PBeck}) is characterized by a bit bigger
flatness value, $F=3\exp[s^2]\simeq 60.3$ for $s^2=3$, as compared
to the flatness value (\ref{flatness}) which is nevertheless
acceptable from the experimental point of view. The peaks of the
contribution to fourth-order moment shown in Fig.~\ref{Fig6} do
not match that of the experimental curve for the $R_\lambda=690$
flow.

%%We note that the best fit is achieved for $s^2$ close to the
%%theoretical value 3 but this does not significantly improve
%%overlapping of the peaks with the data points.

One naturally expects that a better correspondence to the
experiment could be achieved by an accounting for small scale
interactions via turbulent viscosity [certain nonlinearity in the
first term of the rhs of Eq.~(\ref{Langevin})] as it implies a
damping of large events, {i.e.} less pronounced enhancement of the
tails of $P(a)$.

It should be noted that the idea to describe turbulence
intermittency via averaging of the Gaussian distribution over
lognormally distributed variance of some intermittent variable was
proposed long time ago by Castaing, Gagne, and
Hopfinger,\cite{Castaing}
\be\label{PCastaing}
P(x) =\frac{1}{2\pi s_0} \int_0^\infty\!\!\!
d\theta\,\theta^{-2}\exp
\left[-\frac{(\ln\frac{\theta}{m_0})^2}{2s_0^2}\right]
e^{-x^2/(2\theta^2)},
\ee
where $x$ is a variable under study. Below, we apply this model to
the Lagrangian acceleration, $x=a$.

In technical terms, the difference from the Castaing log-normal
model is that in Eq.~(\ref{PBeck}) the {\em inverse} square of the
variance, $\beta=\theta^{-2}$, is taken to be lognormally
distributed. In essence, the models (\ref{PBeck}) and
(\ref{PCastaing}) are of the same type, with different parameters
assumed to be fluctuating at large time scale and hence different
resulting marginal distributions.

One can check that the change of variable, $\theta=\beta^{-1/2}$,
in Eq.~(\ref{PCastaing}) leads to the density function different
from that given by Eq.~(\ref{PBeck}),
\be\label{PCastaing2}
P(x) \simeq \int_0^\infty\!\!\! d\beta\,\beta^{-3/2}\exp
\left[-\frac{(\ln\frac{\beta}{m_1})^2}{2s_1^2}\right] e^{-\beta
x^2/2},
\ee
where we have denoted $m_1=m_0^{-2}$ and $s_1=2s_0$. Therefore,
the distributions (\ref{PBeck}) and (\ref{PCastaing}) are indeed
not equivalent to each other, both being of a stretched
exponential form.

As to a comparison of the fits, we found that the fit of the
Castaing log-normal model (\ref{PCastaing}) for the acceleration,
with the fitted value $s_0=0.625$ ($m_0=\exp[s_0^2/2]$ for unit
variance), is of a considerably lesser quality as one can see from
Fig.~\ref{Fig5} and, more clearly, from Fig.~\ref{Fig6}. Positions
of the peaks of $a^4P(a)$ are approximately the same for both the
models, namely, $|a|/\langle a^2\rangle^{1/2}\simeq 8.0$ as
compared to $|a|/\langle a^2\rangle^{1/2} \simeq 10.2$ for the
experimental curve.

Notice that the existence and positions of the peaks of $a^4P(a)$
reflect some characteristic property of the Lagrangian particle
dynamics. The value $|a|/\langle a^2\rangle^{1/2} \simeq 10.2$
possibly separates different mechanisms underlying stochastic
motion of the particle. One therefore may expect multiple
autocorrelation time scales for $a(t)$.

We conclude this subsection with the following remark. The
Langevin model of the type (\ref{Langevin}), Fokker-Planck
approximation of the type (\ref{FPconstant}), and the underlying
log-normal distribution (\ref{lognormal}) within the Castaing
approach were recently used by Hnat, Chapman, and
Rowlands\cite{Hnat} to describe intermittency and scaling of the
solar wind bulk plasma parameters. This model will be reviewed in
Sec.~\ref{Sec:HCRmodel} below.

%% 3.1.3
\subsubsection{The underlying Gaussian distribution of velocity}
\label{Sec:GaussianDistribution}

The problem of selecting appropriate distribution of the parameter
$\beta$ among possible ones was recently addressed in
Ref.~\refcite{Aringazin0301245}. A specific model based on the
assumption that the Lagrangian velocity $u$ follows normal
distribution with zero mean and variance $\sigma_u^2$ was
developed. The result is that a class of underlying distributions
of $\beta$ can be encoded in the function $\beta=\beta(u)$, and
the marginal distribution is found to be
\be\label{Pnormal}
P(a) \!=\! C(s)\!\!\int_{0}^{\infty}\!\!\!\!\!\! d\beta\,
P(a|\beta)
\exp\!\left[{-\frac{[u(\beta)]^2}{2\sigma_u^2}}\right]\!
\left|\frac{du}{d\beta}\right|,
\ee
where $u(\beta)$ is the inverse function. Note that only an {\em
absolute} value of $u$ contributes to this probability
distribution. Particularly, the exponential dependence
\be
\beta(u) = \exp[\pm u],
\ee
features the log-normal distribution of $\beta$ so that
Eq.~(\ref{Pnormal}) leads to Eq.~(\ref{PBeck}) used in
Ref.~\refcite{Beck4}, while the $\chi$-square distribution of
order 1 is recovered when one takes
\be
\beta(u) = u^2.
\ee

In general, this model is relevant when $\beta(u)$ is a monotonic
Borel function of the stochastic variable $u$ mapping $
(-\infty,\infty) \ni u$ to $[0, \infty) \ni\beta$, and allows one
to rule out some {\it ad hoc} distributions of $\beta$ as well as
to make appropriate generalizations of both $\chi^2$ and
log-normal distributions of the parameter $\beta$.

The possible dynamical foundation of the above model is as
follows. The stationary distribution (\ref{Pnormal}) with
$\beta(u)=\exp[\pm u]$ can be associated with the Langevin
equation of the form\cite{Aringazin0301245}
\be\label{LangevinExp}
\partial_t a = \gamma F(a) + e^{\omega}L(t),
\ee
where we denote $\omega=\mp u/2$, $u$ follows Gaussian
distribution with zero mean, and we take $\gamma$ = const to
simplify the consideration. Here, we adopt a viewpoint that
statistical properties of the acceleration $a$ are associated with
velocity statistics due to the Heisenberg-Yaglom scaling
(\ref{HYscaling}). Notice also that in Sawford model the Langevin
equation (\ref{LangevinSawford}) for $a$ includes velocity $u$ and
its variance.

Below, we outline a relationship of the model (\ref{LangevinExp})
to some recent approaches in studying the intermittency.

(i) The form of the last term in Eq.~(\ref{LangevinExp}), in which
$\omega$ can be viewed as a Gaussian process $\omega=\omega(t)$
independent of the white noise $L(t)$, strikingly resembles that
involved in the recently developed log-infinitely divisible
multifractal random walks model by Muzy and
Bacry,\cite{Muzy0206202,Bacry0207094} a continuous extension of
discrete cascades.\cite{Mandelbrot1967} This model will be
reviewed in Sec.~\ref{Sec:MRW} below.

(ii) The driving force amplitude of the form $e^{\omega(t)}$, with
the ultraslow decaying correlation function of $\omega$,
\be
\label{ultraslow}
 \langle\omega(t)\omega(t+\tau)\rangle =
-\lambda_0^2\ln[\tau/T],
\ee
$\tau < T$, in the Langevin-type equation has been recently
considered by Mordant {\it et al.}\cite{Mordant0206013} The
results of this model have been found in a very good agreement
with the experimentally observed very slow decay of the
autocorrelation of Lagrangian velocity increment magnitudes
$|\delta_\tau u_i|$ for each of the two measured component. Also,
very slow decay was observed for the cross correlation of
magnitudes of components. Both the dynamical correlations were
found to vanish only for time scale of about $T_L$, while the
autocorrelations of full signed $\delta_\tau u_i$ decay rapidly
(the autocorrelation functions cross zero at about $0.06T_L$) and
the cross correlation function is approximately zero. The used
time scale $\tau=0.03T_L$. It was emphasized that the parameter
$\lambda_0$ enters both the model autocorrelation functions of
velocity increments and Lagrangian velocity structure functions.

This may correspond to the slow $\bar\epsilon/\tau$ decay of the
Lagrangian acceleration autocorrelation predicted by K41 theory.
We also note that the fitted value of the above intermittency
parameter $\lambda_0$ ($\lambda_0^2=0.115\pm 0.01$) is very close
to $1/s^2=1/3$, with the value $s^2=3$ in Eq.~(\ref{PBeck})
interpreted as the number of independent random variables in
three-dimensional space at the Kolmogorov scale. If this is not
due to a coincidence, the intermittency parameter $\lambda_0$
approaches simply the inverse of effective space dimension,
%%\be
$\lambda_0 = 1/d$,
%%\ee
$d=3$, for high-Re turbulent flows. The above connection to the
MRW model and very slow decay of correlations of absolute values
of acceleration components indicate relevance of the specific
representation (\ref{LangevinExp}), with very slow varying $|u|$,
in the description of intermittency. In fact, due to the
experiments\cite{Mordant0103084} the Lagrangian velocity
autocorrelation function $\langle u(t)u(t+\tau)\rangle/\langle
u^2\rangle$ decays almost exponentially but very slowly, to vanish
only for $\tau > 3T_L$, where the integral time scale
$T_L=2.2\times 10^{-2}$~s is two orders of magnitude bigger than
the Kolmogorov time scale $\tau_\eta = 2.0\times 10^{-4}$~s;
$R_\lambda = 740$ and the mean velocity is about 10\% of the {rms}
velocity.

(iii) Due to the well-known Kolmogorov power-law relationship
between the mean energy dissipation rate $\bar\epsilon$ and the
rms velocity $\bar u$, $\bar\epsilon \simeq\bar u^3/L$, the
representation (\ref{LangevinExp}) can be thought of as the result
of using the relation $\ln\beta \simeq\ln\epsilon$ (statistical
law), with $\ln\epsilon$ being normally distributed due to K62
theory.\cite{Frisch95} Here, $\epsilon$ denotes {\em stochastic}
energy dissipation rate per unit mass. From this point of view,
one can identify
\be\label{omega}
\omega = g\ln\epsilon,
\ee
where $g$ is a constant. This means that stochastic dynamics of
logarithm of the energy dissipation rate is independent, and it
influences the acceleration dynamics specifically through the
intensity of driving stochastic force entering
Eq.~(\ref{LangevinExp}). Stationary normal distribution of
$\omega$ can be in turn derived from the Fokker-Planck equation
associated with the Langevin equation of a linear form,
%%\be\label{Langevinomega}
$\partial_t \omega = g_0 + g_1 \omega + g_2 L(t)$,
%%\ee
where $g_i$ are constants. This equation is in an agreement with
the recent results of Eulerian (hotwire anemometer) study of the
interaction between velocity increments and normalized energy
dissipation rate by Renner, Peinke, and Friedrich.\cite{Friedrich}
Particularly, they found that an exponential dependence of the
diffusion coefficient on the logarithmic energy dissipation in the
Fokker-Planck equation for the velocity increments in space is in
a very good agreement with the experimental data. Notice that this
equation does not imply a logarithmic decay of the Lagrangian
correlation function $\langle\omega(t)\omega(t+\tau)\rangle$
proposed by Mordant {\it et al.}\cite{Mordant0206013} This may be
attributed to the well-known difference between the Eulerian
(fixed probe) and Lagrangian (trajectory) frameworks.

(iv) For the choice $\beta(u)=\exp[u]$ corresponding to the
log-normal distribution of $\beta$, using Eq.~(\ref{LangevinExp})
one can derive the stationary probability density function of the
form\cite{Aringazin0301245}
\be\label{gu}
P(a)= \int_{-\infty}^{\infty}\!\! du\ C(u)\exp\{\ln[g(u)]-e^{u}
a^2/2\},
\ee
where $g(u)$ is PDF of $u$. Hence the joint PDF can be written
\be\label{PDFfluct}
P(a,u)= C(u)\exp\{\ln[g(u)]-e^{u}a^2/2\}.
\ee
Such a form of the distribution, containing specifically the
double exponent, resembles the ``universal" distribution of
fluctuations (Gumbel function),
\be\label{PChapman}
P(x)= c_0\exp[c_1(y-e^{y})], \quad y\equiv c_2(x-c_3),
\ee
where $c_i$ are constant, recently considered by Chapman,
Rowlands, and Watkins\cite{Chapman} (see also references therein)
following the work by Portelli, Holdsworth, and
Pinton\cite{Portelli}. They used an apparently different approach
(not related to a Langevin-type equation) based on the
multifractal-type energy cascade and $\chi^2$ or log-normal (K62
theory) underlying distribution of the energy dissipation rate at
fixed level. They pointed out a good agreement of such $P(x)$ with
experimental data, where $x$ denotes a fluctuating entity observed
in a variety of model correlated systems, such as turbulence,
forest fires, and sandpiles. The result of this approach meets
ours and we consider it as an alternative way to derive the
characteristic probability measure of fluctuations; with $g(u)$
taken to be a $\chi^2$ (respectively, Gaussian) density function,
one obtains, up to a pre-exponential factor and constants,
$P(u)\simeq \exp[-u-\exp(u)]$ (respectively, $P(u)\simeq
\exp[-u^2-\exp(u)]$). Thus we conclude that the ``universal''
distribution (\ref{PChapman}) can be derived also within the
general framework proposed in Ref.~\refcite{Aringazin0301245} that
reflects a universal character of the underlying $\chi^2$
distribution pointed out by Beck and Cohen.\cite{Beck3}

Strong correlation between bursts of Eulerian velocity magnitude
and bursts of energy dissipation rate has been indicated in the
recent work by Pearson {\it et al.}\cite{Pearson0404114} which
reports results of DNS of statistically stationary and isotropic
slightly compressible $R_\lambda\simeq 219$ flow. Unlike to
results of some pseudo-spectral methods, highly fluctuating
character of time series of $\epsilon$, $L$, $u$, and $R_\lambda$,
where $L$ and $u$ are characteristic large length and velocity
scales respectively, has been encountered. A correct procedure for
determination of $R_\lambda$ dependence of $C_\epsilon =
\bar\epsilon L/{\bar u}^3$ is based on accounting for time lapse
between the bursts and for the averaging over entire time series
for which stationarity condition is fulfilled. This leads to
elimination of the scatter in determining the value of
$C_\epsilon$ from various recent DNS and experimental data. The
value $C_\epsilon \simeq 0.5$ has been established as the high-Re
asymptotic value, which is in agreement with K41 assumption on
viscosity independence of the mean turbulent kinetic energy
dissipation rate for stationary isotropic turbulence.

In summary, we have presented a class of models
(\ref{Pnormal})-(\ref{LangevinExp}) using the basic assumption
that the parameter $\beta$ depends on normally distributed
velocity fluctuations. This class has been found to incorporate
the previous RIN models in a unified way, with the dependence
$\beta(u)$ required to be a (monotonic) Borel function of the
stochastic variable $u$.

Finally, we note that although successful in describing the
observed statistics of Lagrangian acceleration, with a few simple
hypotheses and one fitting parameter, the one-dimensional Langevin
RIN models (\ref{P})-(\ref{FPconstant}) and
(\ref{Pnormal})-(\ref{LangevinExp}) show departures from the
experimental data and suffer from the lack of physical
interpretation in the context of the 3D Navier-Stokes turbulence.

%% 3.2
\subsection{Hnat-Chapman-Rowlands model}\label{Sec:HCRmodel}

A nonlinear Langevin and the associated Fokker-Planck equations
obtained by a direct requirement that the probability distribution
satisfies some model-independent scaling relation have been
recently proposed by Hnat, Chapman, and Rowlands,\cite{Hnat} to
describe the measured time series of the solar wind bulk plasma
parameters. It should be emphasized that this approach is related
to properties of Fokker-Planck equation rather than to those of
Langevin-type equation. Nevertheless, we find this result relevant
to fluid turbulence since it is based on a stochastic dynamical
framework and leads to the stationary distribution with
exponentially truncated power-law tails, similar to that implied
by the RIN models of Sec.~\ref{Sec:RINmodels}.

The Hnat-Chapman-Rowlands model is aimed to describe the observed
time series of the solar wind bulk plasma parameters and is based
on the construction of Fokker-Planck equation for which the
probability density function obeys the following one-parametric
model-independent rescaling:
\be\label{Pscaling}
P(x,t) = t^{-\alpha_0}P_s(xt^{-\alpha_0}).
\ee
Here, $x$ denotes fluctuating plasma parameter and $\alpha_0$ is
the scaling index. Particularly, the value $\alpha_0=1/2$
corresponds to a self-similar (Brownian) walk with Gaussian
probability density functions at all time scales. The fitted value
is different, $\alpha_0=0.41$, and corresponds to a single
non-Gaussian distribution $P_s(x_s)$, to which the observed
distributions of some four plasma parameters collapse under the
scaling $x_s=xt^{-\alpha_0}$.

The Langevin equation of this model assumes only additive noise,
and in such ansatz it was found to be
\be
\partial_t x = D_1(x)+D_2(x)\eta(t),
\ee
where $D_1$ and $D_2$ are of the form
\be
D_1(x) = \sqrt{\frac{b_0}{D_0}}x^{1-\alpha_0^{-1}/2},
\ee
\be
D_2(x) = [b_0(1-\alpha_0^{-1}/2)-a_0]x^{1-\alpha_0^{-1}},
\ee
$a_0$ and $b_0$ are free parameters and $2D_0$ is intensity of the
$\delta$-correlated Gaussian-white additive noise $\eta(t)$,
$\langle\eta(t)\rangle=0$. By construction, this specific form of
the dynamical equation ensures that the corresponding
Fokker-Planck equation
\be\label{FPHnat}
\partial_t P(x,t)
 = \partial_x[a_0 x^{1-\alpha_0^{-1}}P + b_0x^{2-\alpha_0^{-1}}\partial_xP]
\ee
has the general solution $P(x,t)$, which exhibits the scaling
(\ref{Pscaling}). The fitted values are $a_0/b_0=2$, $b_0=10$, and
$\alpha_0^{-1}=2.44$. The rescaled distribution $P_s(x_s)$
corresponding to Eq.~(\ref{FPHnat}) is characterized by power-law
tails truncated by stretched exponential. It provides a good fit
to the tails of the experimental distribution but {\em diverges}
at the origin $x_s\to 0$.

To sum up, we point out that this diffusion model uses the
generalized self-similarity principle resembling that used in the
Eulerian description of the energy cascade in the developed 3D
fluid turbulence and appears to be valid only asymptotically for
large values of the variable, with the fitted value of exponent
parameter being about $\alpha_0=0.41$.

%% 3.3
\subsection{Reynolds stochastic models of Lagrangian acceleration}
\label{Sec:Reynoldsmodels}

The second-order Lagrangian stochastic model of
Reynolds\cite{ReynoldsPF2003,ReynoldsPRL2003} which extends
Sawford model (\ref{LangevinSawford}) is prescribed by the
following equation for a component of acceleration:
\be
\label{Reynolds2}
 da =
 -(T_L^{-1}\!+\! \tau_\eta^{-1}
 \!-\!
 \sigma_{a|\epsilon}^{-1}\frac{d\sigma_{a|\epsilon}}{dt} )adt
 \!-\!
 T_L^{-1}\tau_\eta^{-1}udt
 \!+\!
 \sqrt{2\sigma_u^2(T_L^{-1} \!+\!\tau_\eta^{-1})T_L^{-1}\tau_\eta^{-1}} d\xi.
\ee
Here, $d\xi$ is an incremental Wiener process with zero mean and
variance $dt$, the energy-containing and dissipative time scales
of flow $T_L=2\sigma_u^2/(C_0\epsilon)$ and
$\tau_\eta=C_0\nu^{1/2}/(2a_0\epsilon^{1/2})$ are defined in terms
of the stochastic dissipation rate $\epsilon$, universal
Lagrangian structure constants $a_0$ and $C_0$, the kinematic
viscosity $\nu$, and the velocity variance $\sigma_u^2\equiv
\langle u^2\rangle \equiv {\bar u}^2$.

Model prediction for the Lagrangian acceleration PDF is due to
$P(a)=\int_{0}^{\infty} P(a|\epsilon)f(\epsilon)d\epsilon$, where
one assumes independent zero-mean Gaussian distributions for
velocity and acceleration with variances $\sigma_u^2$ and
$\sigma_{a|\epsilon}^2$ respectively. This model is featured by
accounting for fluctuations of $\epsilon$ and is consistent with
the log-normal model reviewed in
Sec.~\ref{Sec:LognormalDistribution}.

In order to calculate $P(a)$ one should determine distribution of
the turbulent energy dissipation rate $\epsilon$. Following Pope
and Chen,\cite{Pope1990} the evolution of logarithm of the
normalized dissipation rate $\chi= \ln(\epsilon/{\bar\epsilon})$
along a Lagrangian trajectory is governed by the stochastic
equation
\be\label{Pope}
d\chi = -(\chi - \langle\chi\rangle)T_\chi^{-1}dt
+\sqrt{2\sigma_\chi^2T_\chi^{-1}}d\xi',
\ee
where $d\xi'$ is independent incremental Wiener process with zero
mean and the time scale
$T_\chi=2\sigma_u^2/C_0\langle\chi\rangle$. The distribution of
$\chi$ is thus Gaussian, and its variance was approximated by
$\sigma_\chi^2=0.354+0.289\ln R_\lambda$, in accordance with K62
theory and DNS data by Yeung and Pope.\cite{Yeung1989} {{Thus,
Reynolds-number effects are incorporated into the model, which is
applicable to large time scale. It should be emphasized that the
introduction of fluctuating $\chi$ means that the model
(\ref{Reynolds2}) incorporates both additive and multiplicative
noises.}}

The resulting acceleration flatness factor behaves as $F=
1.35R_\lambda^{\,0.65}$, which is in agreement with the recent
pressure gradient DNS data by Kraichnan and
Gotoh\cite{Kraichnan0305040} and with lower bound on $F$ set by
the experiment.\cite{Bodenschatz} The obtained PDF $P(a)$ is in
agreement with the measured acceleration
distribution.\cite{Bodenschatz,Bodenschatz2} Also, the model
agrees well with the observed extended self-similarity of the
Lagrangian velocity structure functions, the exponential shape of
the Lagrangian velocity autocorrelation functions, and the
observed ultraslow correlation of the modulus of the
acceleration\cite{Mordant0206013,Mordant0103084} (see
Sec.~\ref{Sec:GaussianDistribution}).

Extension of this model to account for the observed dependence of
the acceleration variance on velocity\cite{ReynoldsNEXT2003} will
be reviewed in Sec.~\ref{Sec:ConditionalProbability} below.

One of the important dynamical quantities of a fluid particle
motion are trajectory rotations. Recently developed second-order
3D model by Reynolds and Veneziani\cite{ReynoldsPLA2004} shows
that non-zero mean trajectory-rotations are associated with
spiralling trajectories, oscillatory Lagrangian velocity
autocorrelation functions, suppressed rates of turbulent
dispersion for given turbulent kinetic energy and mean dissipation
rate, and skew diffusion; see also work by Borgas, Flesch, and
Sawford.\cite{BorgasJFM1997} The 3D stochastic model equation
includes terms of the form $\epsilon_{ijk}\Omega_ju_kdt$, and the
resulting stationary joint PDF
\be
\label{ReynoldsPDF}
 P(a,u) = (2\pi\sigma_a^2\sigma_u^2)^{-3}
 \exp\left[\frac{(a_i-\epsilon_{ijk}\Omega_ju_k)^2}{2\sigma_a^2}\right]
 \exp\left[\frac{u_i^2}{2\sigma_u^2}\right]
\ee
is of a Gaussian form with respect to $a_i$ and $u_i$. Here,
$\sigma_a^2=\sigma_u^2/(T_L\tau_\eta)$ is the conditional
acceleration variance. The conditional mean acceleration $\langle
a_i|u\rangle=\epsilon_{ijk}\Omega_ju_k$ entering the distribution
(\ref{ReynoldsPDF}) endows trajectories with a preferred sense of
rotation.

They showed that rotations of the Lagrangian velocity vector
produced by such model coincide closely with the intense rotations
measured in the recent seminal experiment by Zeff {\it et
al.}\cite{Zeff2003} and are described by Tsallis
statistics\cite{Tsallis,Aringazin0204359,Beck3} to a good
accuracy. Model predictions for the rotational statistics of the
North Atlantic Ocean are found to be in close agreement with
simulation data produced by the Miami Isopycnic-Coordinate Ocean
Model.

In a recent paper Reynolds\cite{ReynoldsPF2003b} constructed
phenomenological third-order Lagrangian stochastic model. The
model describes evolution of the material derivative of
acceleration, the so called hyper-acceleration ${\dot a}=da/dt$,
in analogy with the second-order model. The 1D stochastic equation
for ${\dot a}$ is
\begin{eqnarray}
\label{Reynolds3}
 d {\dot a} =
 -(T_L^{-1}\! +\! \tau_\eta^{-1} + t_3^{-1}){\dot a}dt
 -(T_L^{-1}\tau_\eta^{-1}\! +\! T_L^{-1}t_3^{-1} + \tau_\eta^{-1}t_3^{-1})adt
 \nonumber\\
 -T_L^{-1}\tau_\eta^{-1}t_3^{-1}udt
 + \sqrt{2\sigma_{{\dot a}|a\,u}^2(T_L^{-1}\! +\!
\tau_\eta^{-1}+t_3^{-1})} d\xi,
\end{eqnarray}
$da={\dot a}dt$, $du=adt$, two timescales $T_L$ and $\tau_\eta$
are defined as above, and $t_3$ is third timescale related to the
hypothesis of finite hyper-acceleration variance. The
hyper-acceleration is assumed to be autocorrelated exponentially
on the timescale $t_3 \ll \tau_\eta$, in a fully developed
turbulent flow. The hyper-acceleration variance $\sigma_{{\dot
a}|a\,u}^2= \sigma_{\dot a}^2-\sigma_{a}^4/\sigma_u^2$ is taken
conditional on both the acceleration $a$ and velocity $u$ through
their variances.

As in the case of the second-order model, this model applies to
stationary homogeneous and isotropic turbulence with Gaussian
velocity and acceleration statistics. The hyper-acceleration
statistics is Gaussian. The 1D toy model given by
Eq.~(\ref{Reynolds3}) is determined uniquely by the well-mixed
condition. The corresponding Lagrangian velocity autocorrelation
function is consistent with the inertial range and dissipation
range forms of Lagrangian velocity structure functions by
Kolmogorov similarity theory.

%%Second-order models imply divergent hyper-acceleration variance.
%%The finite-time autocorrelation allows more flexible description
%%than the zero-time one.

The model timescales are related to the variances of velocity,
acceleration, and hyper-acceleration by
\be
\sigma_a^2= \sigma_u^2(T_L \tau_\eta+T_Lt_3+\tau_\eta t_3)^{-1},
\quad \sigma_{\dot a}^2=\sigma_a^2(T_L^{-1}\tau_\eta^{-1}\! +\!
T_L^{-1}t_3^{-1} + \tau_\eta^{-1}t_3^{-1}).
\ee
For $t_3=0$ it follows particularly that
$\sigma_a^2=\sigma_u^2/(T_L\tau_\eta)=
a_0{\bar\epsilon}^{3/2}\nu^{-1/2}$ in an agreement with
Heisenberg-Yaglom scaling law (non-intermittent K41 picture) given
by Eq.~(\ref{HYscaling}); ${\bar \epsilon}={\bar u }^3/L$ and
$a_0$ is constant.

In fact, the third-order model goes beyond Kolmogorov
phenomenological theory by introducing third characteristic
time-scale $t_3$, in addition to the conventional time scales
$T_L$ (energy-containing) and $\tau_\eta$ (dissipative). For
$t_3\not=0$ one thus expects deviations from the K41 predictions.
Particularly, corrections due to intermittency are known to imply
dependence of the Kolmogorov constant $a_0$ on Reynolds number.
While for high-Re flows $a_0$ is found approximately constant
(however, weak deviations such as $a_0\simeq R_\lambda^{0.14}$ can
not be ruled out), for $R_\lambda<500$ there is a clear deviation
from the K41 scaling of the acceleration
variance.\cite{Bodenschatz}

Agreement between third-order model predictions and DNS data for
Lagrangian velocity structure function and Lagrangian acceleration
autocorrelation function is found significantly better than that
obtained with the second-order Lagrangian stochastic model
(\ref{Reynolds2}). The effects of third-order processes were found
comparable in magnitude to the effects of second-order processes.

This means that the third-order dynamics at the characteristic
time scale $t_3$ is essential for a better description of
homogeneous isotropic turbulence within the framework of such a
Lagrangian stochastic modeling approach. Physical interpretation
of such (and higher-order) processes in the context of 3D
Navier-Stokes equation and turbulence phenomenology is one of the
open problems.

Anisotropy effects in turbulence within the framework of
Lagrangian stochastic third-order model has been described by
Reynolds, Yeo, and Lee.\cite{ReynoldsPRE2004} The simplest 3D
stochastic equation for the hyper-acceleration in homogeneous
anisotropic turbulence is given by
\be
\label{ReynoldsAnisotropic}
 d{\dot a}_i =
  (a_{ij}+c_{ij}){\dot a}_jdt
 +(b_{ij}-c_{ik}a_{kj})a_jdt
 +c_{ik}b_{kj}u_jdt
 +\sqrt{-2(a_{ij}+c_{ij})\lambda_{jk}}d\xi_k,
\ee
with $da_i={\dot a}_idt$ and $du_i=a_idt$. Here,
 $a_{ij}=-C_0\epsilon\tau_{ij}-2a_0C_0^{-1}(\epsilon/\nu)^{1/2}\delta_{ij}$,
 $b_{ij}=-a_0(\epsilon^3/\nu)^{1/2}\tau_{ij}$,
 $c_{ij}(\chi_{jk}+b_{jl}\sigma_{lk})=-a_{ij}\chi_{jk}$,
 $\tau_{ij}=[\sigma^{-1}]_{ij}$, and
 $\sigma_{ij}=\langle u_iu_j\rangle$,
 $\chi_{ij}=\langle a_ia_j\rangle$, and
$\lambda_{ij}=\langle {\dot a}_i{\dot a}_j\rangle$ are Lagrangian
velocity, acceleration, and hyper-acceleration covariances
respectively; $i,j,\dots=1,2,3$ (summation over repeated indices
is assumed). When the inverse time scales $c_{ij}$ of third-order
processes tends to infinity the model (\ref{ReynoldsAnisotropic})
reduces to a second-order Lagrangian stochastic model.

Model predictions were compared with the results of DNS acquired
for a turbulent channel flow with low Reynolds number $R_\lambda
\simeq 30$. The third-order model is shown to account naturally
for the anisotropy of acceleration variances in low-Re turbulent
flows and for their dependency upon the energy-containing scales
of motion. The experimental values $C_0=6.0$ and $a_0=5.5$ were
taken and the hyper-acceleration variances were subsequently
chosen to guarantee consistency with the DNS data for the
acceleration variances.

As to the high-Re limit, it was argued that if $t_3/\tau_\eta$
tends to a finite value when $R_\lambda\to\infty$ then
hyper-acceleration statistics may account for the observed
anisotropy of acceleration variances in high-Re
flows.\cite{Bodenschatz} Alternatively, if $t_3/\tau_\eta\to0$ for
$R_\lambda\to\infty$ then one ends up with the isotropic scaling
prediction.
%%One may wonder how the small time scale $t_3\not=0$ could be
%%responsible for anisotropy effects in a high-Re flow which mainly
%%come from large scales.
It was argued that $t_3\not=0$ implies in general dependence of
the acceleration on the energy-containing scales of motion through
the dependence of Lagrangian acceleration variance $\sigma_a^2$
upon $\sigma_u^2$ for each component separately.

Non-universality of parameters in conventional Lagrangian
stochastic models, when one tries to fit them to experimental and
DNS data, was shown to be a consequence of truncation at either
first or second order and not an inherent deficiency of the
approach in general. Anisotropy of acceleration variance implies
``anisotropy'' of $a_0$ (different values of $a_0$ for different
components) when $t_3=0$. This is not necessarily the case for
$t_3\not=0$. The additional dynamical degree of freedom provided
by the third-order model, i.e. inclusion of third-order processes
to describe homogeneous turbulence, thus allows one to keep
universal character of some parameters when fitting model
predictions to the experiments and DNS.

%% 3.4
\subsection{Laval-Dubrulle-Nazarenko model of small-scale
turbulence} \label{Sec:LDNmodel}

The above one-dimensional Langevin toy models of Lagrangian
turbulence considered in Secs.~\ref{Sec:RINmodels},
\ref{Sec:HCRmodel}, and \ref{Sec:Reynoldsmodels} all suffer from
the lack of physical interpretation, {e.g.} of short-term
dynamics, or small- and large-scale contributions, in the context
of 3D Navier-Stokes equation.

The crucial point is to make a link between Langevin-type
equations and the 3D Navier-Stokes equation. This includes
determination of statistical properties of stochastic terms and
the functional form of deterministic terms, as well as their
dependence on the parameters entering the Navier-Stokes equation,
justified for fully developed turbulence. Also, some extension of
the stochastic equation may be required to account for dependence
of the parameters on Lagrangian velocity, in the spirit of RIN
approach of Sec.~\ref{Sec:RINmodels}, and in correspondence to the
Navier-Stokes equation as the pressure gradient term in the
Eulerian framework can be expressed in terms of the velocity owing
to the incompressibility condition. Strong and nonlocal character
of Lagrangian particle coupling as the result of pressure effects
makes it difficult to derive theoretically turbulence statistics
from the 3D Navier-Stokes equation. One is thus left with more or
less justified modeling approaches.

The Navier-Stokes equation based approach to describe statistical
properties of small scale velocity increments, both in the
Eulerian and Lagrangian frameworks, was developed in much detail
by Laval, Dubrulle, and Nazarenko;\cite{Laval0101036} see also
recent paper by Laval, Dubrulle, and McWilliams.\cite{Laval2}

This approach is based on featuring {\em nonlocal} interactions
between well separated large and small scales ---elongated
triads--- and is referred to as the {Rapid Distortion Theory}
(RDT) approach. Decomposition of velocities into large- and
small-scale parts was made by introducing a certain spatial filter
of a cutoff type. Within the framework of this approach, 3D
Langevin-type model of small-scale turbulence was proposed.

The main assumption of the Laval-Dubrulle-Nazarenko (LDN) approach
to the 3D Navier-Stokes turbulence is to introduce and separate
large-scale ($L$) and small-scale ($l$) parts in the 3D
Navier-Stokes equation and using the Gabor transform (Fourier
transform in windows)\cite{Laval0101036,Nazarenko}
\be
\label{Gabor}
 {\hat u}_i(x_n,k_m,t)= \int
 f(\varepsilon|x_j-x'_j|)e^{ik_m(x_m-x'_m)}u_i(x'_n,t)\mathrm{d}^3x'_s.
\ee
Here, the parameter $\varepsilon$ is such that
$2\pi/L\ll\varepsilon\ll 1$ and $f(x)$ is a function which
decreases rapidly in infinity. The ``window'' to which Fourier
transform applies is centered at the point $x_m$ and has the size
between small and large scales $l$ and $L$ respectively.

This allows to consider analytically small-scale turbulence
coupled to large-scale terms, i.e. account for an inter-scale
coupling. Such nonlocal interactions were argued to be important
in understanding intermittency in developed turbulent flows. The
other, large-scale, part of the equation can be treated separately
(and, in principle, solved numerically given the forcing and
boundary conditions) since the forcing is characterized by
presumably narrow range of small wave numbers, and the small
scales make little effect on it. Small-scale interactions are
modeled by a turbulent viscosity {and were shown numerically to
make small contribution to the anomalous scaling (intermittency)
in the decaying turbulence. Nevertheless, these are important when
fitting model distribution to the experimental data.} The 3D LDN
model of small scale turbulence was used to formulate surrogate 1D
LDN model, which was studied both in the Eulerian and Lagrangian
frames.\cite{Laval0101036}

The 1D LDN toy model of the Lagrangian turbulence implies a
nonlinear Langevin-type equation for a component of small-scale
velocity increments in time.\cite{Laval0101036} Such a toy model
can also be viewed as a passive scalar in a compressible 1D flow.
For sufficiently small time scale $\tau$ it corresponds to the
acceleration $a$ of fluid particle, $a=\delta_\tau u/\tau$, and is
written as\cite{Aringazin0305186,Aringazin0312415}
\be\label{LangevinLaval}
\partial_t a = (\xi - \nu_{\mathrm t}k^2)a + \sigma_\perp.
\ee
This equation corresponds to a Lagrangian description in the scale
space, along a wave-number packet, defined by the Gabor transform.
Here,
\be\label{turbviscosity}
\nu_{\mathrm t} = \sqrt{\nu_0^2+ B^2a^2/k^2}
\ee
stands for the turbulent viscosity introduced to describe
small-scale interactions, $\nu_0$ is the kinematic viscosity, $B$
is a free parameter, $k$ is the wave number [$\partial_tk = -k
\xi$, $k(0)=k_0$, to model the RDT stretching effect in 1D case],
$\xi$ and $\sigma_\perp$ are multiplicative and additive noises
associated with the velocity derivative tensor and forcing of
small scales by large scales (the energy transfer from large to
small scales), respectively. We refer to the model
(\ref{LangevinLaval}) as the 1D LDN-type model of Lagrangian
acceleration in turbulence.

The entities $\sigma_\perp$ and $\xi$ in Eq.~(\ref{LangevinLaval})
are surrogate versions of the 3D entities related to large- and
small-scale parts $U_i$ and $u_i$ of the velocity field as
follows:\cite{Laval0101036}
\be\label{xi}
\hat{\bm\xi} = \bm\nabla\left[\frac{2\bm{k}}{{\bm
k}^2}(\bm{k}\cdot\bm{U})-\bm{U}\right],
\ee
\be\label{sigmaperp}
\hat{\bm\sigma}_\perp = \hat{\bm\sigma} - \frac{\bm k}{{\bm
k}^2}(\bm{k}\cdot\hat{\bm\sigma}),
\ee
\be\label{sigmai}
\sigma_i=\partial_j(\overline{U_iU_j} -U_iU_j + \overline{u_jU_i}
-\overline{U_ju_i}),
\ee
where the hat denotes Gabor transform (\ref{Gabor}) and the
overline stands for certain spatial cutoff retaining a large-scale
part. Statistical properties of all the components of ${\bm\xi}$
and ${\bm\sigma}_\perp$ were studied numerically for decaying
turbulence and reveal rich and complex behavior.

One can see from Eqs.~(\ref{xi})-(\ref{sigmai}) that the noises
are related to the velocities and their derivatives, and the
additive noise $\bm \sigma_\perp$ is associated with interaction
terms between the large- and small-scale dynamics. One therefore
expects that this noise may exhibit both the short- and long-time
autocorrelations. Physically, this would correspond to vortical
structures dynamics of which is essentially characterized by two
well separated time scales.

This gives support to the idea that the intermittency is caused
also by some nonlocal interactions including the inertial-range
flow modes and not merely by small scales. We remark that one
would also like to know the role of the dissipative scale in this
integrated picture.\cite{Chevillard0310105}

Noisy character of the entities (\ref{xi}) and (\ref{sigmaperp})
may not be seen as a consequence of the Navier-Stokes equation,
which does not contain external random forces at the
characteristic time scale. In the RDT approach, $\xi$ and
$\sigma_\perp$ are treated as independent stochastic processes
entering the small-scale dynamics (\ref{LangevinLaval}) owing to
the fact that the large-scale dynamics is weakly affected by small
scales (which corresponds to a direct energy cascade in 3D flow)
and thus can be viewed, in the first approximation, as a given
force of a stochastic character with certain ascribed
autocorrelation along a particle trajectory.

Eqs.~(\ref{LangevinLaval}) and (\ref{xi})-(\ref{sigmai}) could be
used to trace back the origin of multiplicative and additive
noises entering various surrogate Langevin-type models of the
developed turbulence, and to provide important information on the
dynamics underlying the intermittency.

As a first step, in 1D case these noises were modeled in the
Lagrangian frame by coupled Gaussian-white
noises\cite{Laval0101036} inspired by the Kraichnan ensemble used
for turbulent passive scalar and the Kazantsev-Kraichnan model of
turbulent dynamo,
\begin{eqnarray}\label{noises}
\langle\xi(t)\rangle=0, \
\langle\xi(t)\xi(t')\rangle = 2D\delta(t-t'), \nonumber \\
\langle\sigma_\perp(t)\rangle = 0, \
\langle\sigma_\perp(t)\sigma_\perp(t')\rangle = 2\alpha\delta(t-t'), \\
\langle\xi(t)\sigma_\perp(t')\rangle = 2\lambda\delta(t-t'), \nonumber
\end{eqnarray}
where $D$, $\alpha$, and $\lambda$ are free parameters depending
on scale via $k_0$. The positive parameters $D$ and $\alpha$
measure intensities of the multiplicative and additive noises
respectively, while $\lambda$ (to be not confused with the Taylor
microscale) measures correlation between the noises.

The representation (\ref{noises}) puts an obvious limitation but
is partially justified by DNS.\cite{Laval0101036} The averaging in
Eq. (\ref{noises}) is made over ensemble realizations. Zero means
correspond to isotropy of the stochastic forces along a
trajectory. Physically, the small scales are thus assumed to be
essentially distorted in a stochastic way, as a combined effect of
much larger scales. The model (\ref{noises}) accounts for
short-time autocorrelations (approximated by zero-time
autocorrelations), with the parameters $D$, $\alpha$, and
$\lambda$ treated here as constants along a particle trajectory.

We stress that the correlation between the noises $\xi$ and
$\sigma_\perp$ defined by Eq. (\ref{noises}) is not {\it ad hoc}
assumption but a consequence of the structure of their 3D
counterparts (\ref{xi}) and (\ref{sigmaperp}) as they contain the
same large-scale velocity serving as a unifying agent between the
noises.

Partial support of the argument that large scales influence much
smaller scales is due to the recent study by Pearson {\it et
al.}\cite{Pearson0404114} of the so called ``zeroth law'' of
turbulence. Comparing Lumley's forward cascade model prediction
and DNS data they argue that some amount of energy is passed to
all higher wave numbers, not totally to the neighboring wave
numbers, at least for low $R_\lambda \sim O(10^2)$. Whether this
holds for higher-Re flows is however still an open question.

Stationary solution of the Fokker-Planck equation associated with
Eq.~(\ref{LangevinLaval}) and the noises (\ref{noises}),
%%\begin{eqnarray}
\be
 \label{FP}
\partial_t P(a,t) = \partial_a(\nu_{\mathrm t}k^2P)
 + D\partial_a (a \partial_a a P)
 - \lambda \partial_a (a\partial_a P)
%% \nonumber \\
 -\lambda \partial^2_a (a P)
 +\alpha \partial^2_a P,\
 \ee
%%\end{eqnarray}
is given by\cite{Laval0101036}
\be\label{StationaryLaval}
P(a) = C\exp\left[ \int_{0}^{a}dy \frac{-\nu_{\mathrm t}k^2y - Dy
+\lambda}{Dy^2 -2\lambda y +\alpha}\right],
\ee
where $C$ is a normalization constant and six parameters can be
used to fit the experimental data. This model specifies the
one-dimensional LDN model (\ref{LangevinLaval}), and we refer to
this model as the LDN model with $\delta$-correlated noises (dLDN
model).

Thus, one makes a closure by treating the combined effect of large
scales, for which one has a different dynamical LDN equation that
could be in principle solved numerically,\cite{Dubrulle0304035}
and nonlocal inter-scale coupling as a pair of given external
noises. The price of the simplification (\ref{noises}) is that one
introduces free parameters $\alpha$, $D$, and $\lambda$ to the
description. Matching large-scale dynamics to boundary conditions
deserves a separate study. Despite 3D turbulence is known to be
more sensitive to the large-scale forcing or boundary conditions,
as compared to the 2D one, the used simplification (\ref{noises})
is relevant for high-Re flows to some
extent\cite{Laval0101036,Dubrulle0304035} and allows one to
advance in analytical treatment of the problem.

It should be stressed that the 1D LDN toy model
(\ref{LangevinLaval}) as well as its particular case, dLDN model,
have several limitations related to the LDN separation of small
and large scales allowing to study exclusively nonlocal effects
associated with the linear process of distortions of small scales
by a strain produced by large scales; the use of model turbulent
viscosity; and one-dimensionality. {{Applicability of the model at
Lagrangian integral time scale is limited due to the used
separation of scales. Formulation of the model to include
description at the integral time scale in a consistent way is of
much interest. Our remark is that one of the possible formal ways
to account for Reynolds-number effects is to assume that
intensities of the external noises, which represent large scales,
depend on Reynolds number.}}

Langevin-type equation containing both the Gaussian-white
multiplicative and additive noises was studied in detail by
Nakao.\cite{Nakao9802030} The associated Fokker-Planck equation
was also analyzed. The dLDN model (\ref{LangevinLaval}) extends
Nakao's set up by incorporating two new features: (i) the
nonlinearity controlled by $B$ in Eq.~(\ref{turbviscosity}) and
(ii) the coupling of the noises controlled by $\lambda$ in
Eq.~(\ref{noises}).

It is interesting to note that the RDT approach qualitatively
resembles the model studied by Kuramoto and
Nakao,\cite{KuramotoPRL1996} a system of spatially distributed
chaotic elements driven by a field produced by nonlocal coupling,
which is spatially long-wave and temporally irregular. Such
systems, in which the multiplicative noise is the local Lyapunov
exponent fluctuating randomly due to the chaotic motion of the
elements, show power-law correlations, intermittency, and
structure functions similar to that of the developed fluid
turbulence.

Finally, we note that different strategy to obtain single-particle
PDF of Lagrangian velocity increments from the 3D Navier-Stokes
equation, without referring to Langevin-type equations, is due to
the Lagrangian PDF method\cite{FriedrichPRL2003} outlined in
Introduction.

Before turning to a comparison of the LDN and RIN models which
will be made in Sec.~\ref{Sec:Comparison} below, it is worthwhile
to outline recent results on RDT approach to small-scale
turbulence.

Within the framework of RDT approach, new physical measures of
intermittency such as the mean polarization and the spectral
flatness have been introduced recently by Dubrulle, Laval,
Nazarenko, and Zaboronski\cite{Dubrulle0304035} by using the Gabor
transform of separated large- and small-scale velocities and a
method previously developed for the kinematic dynamo
problem.\cite{Nazarenko2003} The resulting equation for the
small-scale Gabor transformed velocity ${\hat u}_m={\hat
u}_m(k,x(t),t)$, for the fluid-particle trajectory determined by
$\partial_t x_i(t)=U_i$, has been found in the following form:
\be\label{SDT}
\partial_t {\hat u}_m
 =\sigma_{ij}k_i\frac{\partial {\hat u}_m}{\partial k_j}
 -\sigma_{mi}{\hat u}_i
 +\frac{2}{k^2}k_m\sigma_{ij}k_i{\hat u}_j - \nu
k^2{\hat u}_m,
\ee
where $\sigma_{ij}(t)=\nabla_jU_i$ is the large-scale strain. The
nonlinear advection term entering the 3D Navier-Stokes equation
has been neglected. Hence local interactions among small scales
are not considered. Nevertheless, Eq.~(\ref{SDT}) is of much
interest since one can investigate contribution of nonlocal
interactions.

Gaussian white-noise processes for large-scale strain matrix
components (rapid stochastic distortions of small scales), the
form of which ensures statistical isotropy near a fluid-particle
path and the incompressibility,
\be
 \label{strain}
%% \sigma_{ij}(t) = \Omega(A_{ij}(t)-d^{-1}A_{nn}(t)\delta_{ij}),
%% \quad
 \langle \sigma_{ij}(\tau)\sigma_{kl}(0)\rangle =
 \Omega(\delta_{ik}\delta_{jl}-d^{-1}\delta_{ij}\delta_{kl})\delta(\tau),
 \ee
have been taken as a model representation; $i,j,\dots=1,2,\dots,
d$. This representation in Eq.~(\ref{SDT}) was used to derive
time-evolution equation for the generating function
\be \label{Z}
Z(k,t) = \langle\exp[\lambda_1|{\hat u}_m|^2 +\lambda_2{\hat
u}_m^2 +\lambda_3{\hat u}_m^{*2}]\rangle,
\ee
where $\lambda_{1,2,3}$ are auxiliary parameters, $\Omega$ is
constant, and $d=3$ for three dimensions. Due to the isotropy the
final equation depends only on the module of wave number, $k=|\bm
k|$. The model (\ref{SDT})-(\ref{strain}) is referred to as
Stochastic Distortion Theory (SDT) model.

From the equation for $Z$ an evolution equation for the
Gabor-velocity correlators of even order can be extracted in a
standard way. Inviscid regime and dissipative regime
approximations of the obtained correlators have been presented.

Fourth-order wave-number space correlators are shown to carry
essential information on the turbulence statistics and
intermittency which is not available from the two-point spatial
correlators widely used in studying turbulence. Particularly, the
mean polarization carries information on amplitudes and phase
difference of Gabor modes. All turbulence wavepackets are shown
eventually to become plane polarized, $\langle|{\hat
u}_i|^4\rangle \simeq \langle|{\hat u}_i^2|^2\rangle$, i.e. the
turbulence tends to be strongly non-Gaussian. Also, the spectral
flatness, in the inviscid regime, increases as $k^{3/2}$ with the
increase of wave number $k$ that indicates presence of small-scale
intermittency. The effects of dissipation have been quantified.

It was also argued that the log-normal character of turbulence
statistics (nonlinearity in the order of correlator appears to be
of a square form in the exponent) arises because the strain in
SDT model is a multiplicative noise for the velocity, which
becomes nearly one-dimensional, and the time integrated strain
tends to become a Gaussian process.

We note that the used model statistics of strain (\ref{strain}) is
characterized by $\delta$-correlations and does not contain
long-time fluctuating terms. Such fluctuations could be accounted
for by taking, for example, $\Omega$ to be long-time correlated
stochastic process, which corresponds to a slow-varying random
intensity of the noisy strain along a particle trajectory.

In the following Section, we make a comparison of the RIN model
(\ref{P})-(\ref{FPconstant}) with the LDN-type model
(\ref{LangevinLaval}), as well as its particular case, the dLDN
model (\ref{StationaryLaval}).\cite{Aringazin0305186}

%% 4
\section{Comparison of the simple RIN and LDN-type models}
\label{Sec:Comparison}

\subsection{A qualitative comparison}\label{Sec:QualitativeComparison}

A direct comparison of the Langevin equations (\ref{Langevin}) and
(\ref{LangevinLaval}) of the two models suggests the following
evident identifications:
\be\label{Comparison}
 \gamma F(a) = (\xi - \nu_{\mathrm t}k^2)a, \quad
 \sigma L = \sigma_\perp.
\ee
Hence the additive noises can be made identical to each other by
putting $\sigma^2=\alpha$. Further, in the case of a linear drift
force, $F(a)=-a$, and constant viscosity, $\nu_{\mathrm t}=\nu_0$,
we can identify the remaining parameters, $\gamma=\nu_0k^2-\xi$,
so that we get
\be\label{beta}
\beta \equiv \gamma/\sigma^2=(\nu_0k^2-\xi)/\alpha.
\ee
This relation implies that the parameter $\beta$ can be viewed as
a stochastic variable with a nonzero mean due to the stochastic
nature of $\xi$ assumed in the LDN model. This is in agreement
with the simple RIN model, the defining feature of which is just
that the fluctuating part of $\beta$ follows some statistical
distribution. Such a general procedure for obtaining Tsallis-type
statistics was suggested in Ref.~\refcite{Aringazin0204359}.

In the dLDN model (\ref{LangevinLaval})-(\ref{StationaryLaval}),
both the additive and multiplicative noises are taken {\em
$\delta$-correlated} due to Eq.~(\ref{noises}). This is in a sharp
contrast to the assumption that $\beta$ can be taken constant to
derive the stationary solution (\ref{PGauss}) which is the
foundation of the simple RIN model. More precisely, the solution
in the form (\ref{PGauss}) can be obtained as the lowest-order
approximation if $\beta$ is slow varying in time as compared to a
typical time scale associated with the additive noise $L(t)$ (the
adiabatic approximation). This suggests that the multiplicative
noise $\xi$ should be taken as a sufficiently slow varying
stochastic variable, to meet the ansatz used in RIN models.

The detailed numerical analysis of the noises\cite{Laval0101036}
for the turbulent flow at relatively low Reynolds numbers,
$57<R_\lambda<80$, shows that the autocorrelation of the
multiplicative noise $\xi$ decays much slower (by about one order
of magnitude) than that of the additive noise $\sigma_\perp$.
Hence the typical time scale $\tau_\xi$ at which $\xi$ varies is
considerably bigger than that $\tau_\sigma$ of $\sigma_\perp$.
   %%%% We expect that this time scale difference becomes
   %%%% even bigger for higher Reynolds numbers.
Also, the cross-correlation between the two noises was found to be
rather weak, i.e. $\lambda\ll D$ and $\lambda\ll \alpha$, by about
two orders of magnitude in the longitudinal case, and $\lambda=0$
in the transverse case. Altogether this allows one to introduce
the time-scale hierarchy $\tau_\xi \gg \tau_\sigma$ and to
decouple the noises, {i.e.} to put $\lambda=0$, which justifies
the adiabatic approximation and gives support to one-dimensional
RIN models.

The presence of the long-time correlated amplitude $e^{\omega(t)}$
and the short-time correlated directional part $L(t)$ of the
stochastic driving force in the Langevin-type equation considered
by Mordant {\it et al.}\cite{Mordant0206013} also supports the
above adiabatic approximation (two well separated time scales in
the single additive stochastic force, in the Lagrangian
framework). Also, as established by Hill\cite{Hill-PRL2002} for
locally isotropic turbulence, fluid-particle acceleration
correlation is governed by two length scales: one arises from the
pressure gradient, the other from the viscous force.

As usual, the $\delta$-correlated noise originates from taking the
limit of zero correlation time in a system with the smallest
finite correlation time of the noise.

On the contrary, in the dLDN model one assumes the approximation
of comparable time scales, $\tau_\xi \simeq \tau_\sigma$, and
retains the coupling parameter $\lambda$ (which relates
small-scale stretching with vorticity in the 3D case, and is
responsible for the skewness generation along the scale).

The use of the constant turbulent viscosity $\nu_{\mathrm
t}=\nu_0$ makes a good approximation in describing intermittency
corrections since both the constant and turbulent viscosities were
found to produce corrections which are of the same level as the
DNS result.\cite{Laval0101036} In the physical context, this means
that the small scale interactions are not of much important in the
dynamics underlying the intermittency. This justifies the use of
the approximation of linear forcing $F(a)=-a$ in simple RIN
models. We note that this is also in an agreement with both the
experimental results for the Lagrangian velocity autocorrelation
function by Mordant, Metz, Michel, and
Pinton,\cite{Mordant0103084} and the recent experimental Eulerian
results for the spatial velocity increments by Renner, Peinke, and
Friedrich.\cite{Friedrich}

Alternatively, one can consider a {more general} RIN model
characterized by the presence of Gaussian-white additive and {\em
multiplicative} noises and fluctuating intensities of both the
noises. This will lead to a model similar to the dLDN model
(\ref{LangevinLaval}) in which the noise intensities $D$ and
$\alpha$, and the coupling parameter $\lambda$ are assumed to
fluctuate at large time scale.

In summary, we found that the one-dimensional RIN model
(\ref{P})-(\ref{FPconstant}) can be viewed as a particular case of
the one-dimensional LDN-type model (\ref{LangevinLaval}) of
turbulence which is based on the RDT approach by Laval, Dubrulle,
and Nazarenko.\cite{Laval0101036} It should be stressed that while
both the toy models assume introduction of some external
statistics ---the correlator of $L(t)$ and the distribution
$f(\beta)$ in Eq.~(\ref{P}) and the correlators of $\xi$ and
$\sigma_\perp$ in Eq.~(\ref{LangevinLaval})--- the LDN-type model
is characterized by a solid foundation and reveals a rich
structure as compared to RIN models.

In the first approximation, {i.e.} $\lambda=0$, $\nu_{\mathrm
t}=\nu_0$, and $\tau_\xi \gg \tau_\sigma$, the class of RIN models
is in a quite good qualitative correspondence with the LDN-type
model (\ref{LangevinLaval}) and differs from the specific dLDN
model (\ref{LangevinLaval})-(\ref{StationaryLaval}) by the only
fact that in the latter one assumes $\tau_\xi \simeq \tau_\sigma$
and introduces a $\delta$-correlated multiplicative noise. Hence
the different resulting probability density functions for the
acceleration of fluid particle in the developed turbulent flow,
Eqs.~(\ref{PAringazin})-(\ref{PBeck}) and (\ref{StationaryLaval}),
respectively.

%% 4.1
\subsection{A quantitative comparison}
\label{Sec:QuantitativeComparison}

With the above result of qualitative comparison, we are led to
make a more detailed, quantitative comparison of the dLDN model
(\ref{LangevinLaval})-(\ref{StationaryLaval}) and the simple RIN
model (\ref{P})-(\ref{FPconstant}) with the underlying $\chi^2$ or
log-normal distribution of $\beta$, in order to determine which
approximation, $\tau_\xi \simeq \tau_\sigma$ or $\tau_\xi \gg
\tau_\sigma$, is better when used to describe the Lagrangian
statistical properties of the developed turbulent flow. We take
the recent high precision Lagrangian experimental
data\cite{Bodenschatz,Bodenschatz2} on statistics of fluid
particle acceleration in the developed turbulent flow as a
testbed. Actually we follow the remark made in
Ref.~\refcite{Laval0101036} that the $\delta$-approximation of
$\xi$ is debatable and the performance of such a model should be
further examined in the future.

In Ref.~\refcite{Laval0101036}, explicit analytic evaluation of
the distribution (\ref{StationaryLaval}) is given for the
particular case $\nu_{\mathrm t} =\nu_0$, while the general case
is treated in terms of $d\ln P(a)/da$ when fitting to the
numerical RDT data. %% ($-4 \leq a \leq 4$).
 In order to make fits to the experimental probability density
function $P(a)$ and to the contribution to the fourth-order
moment, $a^4P(a)$, covering wide range of the normalized
acceleration, $-60\leq a\leq 60$, one needs in an analytic or
numerical evaluation of the rhs of Eq.~(\ref{StationaryLaval}). To
this end, we have calculated {exactly} the integral appearing in
the expression (\ref{StationaryLaval}). Despite the integral may
look simple the resulting expression is rather complicated. The
exact result is presented in Appendix A.\cite{Aringazin0305186}

The $\chi^2$ and log-normal distribution-based probability density
functions (\ref{PAringazin}) and (\ref{PBeck}) are both
realizations of the RIN model and contain one fitting parameter,
$a_c$ and $s$, respectively. The result of comparison of fitting
qualities of these functions,\cite{Aringazin0301040} with
$a_c=39.0$ and $s=3.0$, is that the probability density function
(\ref{PBeck}) provides a better fit to the experimental
data\cite{Bodenschatz2} on low-probability tails and on the
contribution to the kurtosis, which summarizes peakedness of the
distribution. However, since the integral in Eq.~(\ref{PBeck})
cannot be evaluated analytically we will use the
distribution~(\ref{PAringazin}), which provides a better fit to
the central part of the experimental distribution, when dealing
with analytic expressions.

The dLDN probability density function (\ref{StationaryLaval})
contains six parameters which can be used for a fitting: the
multiplicative noise intensity $D$, the additive noise intensity
$\alpha$, the coupling $\lambda$ between the multiplicative and
additive noises, the turbulent viscosity parameter $B$, the
parameter $\nu_0$, and the wave number parameter $k$.

{\em The parameter $k$}. For the fitting, we can put $k=1$ without
loss of generality since it can be absorbed by the following
redefinition of the parameters $\nu_0$ and $B$:
\be
\nu_0k^2 \to \nu_0, \quad Bk \to B.
\ee

{\em The parameter $\alpha$}.  The structure of the rhs of
Eq.~(\ref{StationaryLaval}) is such that only four parameters out
of five can be used for the fitting. For example, one can put
$\alpha=1$ without loss of generality by using the following
redefinitions:
\be\label{redef}
\nu_0/\alpha \to \nu_0,\ B/\alpha \to B,\ D/\alpha \to D,\
\lambda/\alpha \to \lambda.
\ee
Alternatively, one can put $D=1$ provided the following
redefinitions:
\be\label{redefD}
\nu_0/D \to \nu_0,\ B/D \to B,\ \alpha/D \to \alpha,\
\lambda/D \to \lambda.
\ee

{\em The parameter $\lambda$}. Due to Eq.~(\ref{A5}) we have
$c=-i\sqrt{D\alpha-\lambda^2}$, which is imaginary for
$D\alpha>\lambda^2$ and real for $D\alpha<\lambda^2$. For $c=0$,
{i.e.} $D\alpha=\lambda^2$, the integral (\ref{LavalIntegral3}) is
finite since divergent terms in $F(c)$ and $F(-c)$ defined by
Eq.~(\ref{A4}) cancel each other. Since the parameter $\lambda$
measuring the coupling between the noises is assumed to be much
smaller than both the noise intensities $D$ and
$\alpha$,\cite{Laval0101036} we put $D\alpha>\lambda^2$ in our
subsequent analysis. Moreover, the parameter $\lambda$ responsible
for the skewness can be set to zero since we will be interested,
as a first step, in approximately isotropic and homogeneous
turbulent flows, for which the experimental distribution $P(a)$
exhibits very small or zero skewness.\cite{Bodenschatz}

Thus, we can use three redefined free parameters $\nu_0$, $B$, and
$D$ for the fitting, by putting $k=1$, $\alpha=1$, and
$\lambda=0$. However, we shall keep $k$ and $\alpha$ in an
explicit way in the formulas below, to provide a general
representation.

We start by considering two important particular cases of the dLDN
probability density function (\ref{StationaryLaval}): the constant
viscosity $\nu_{\mathrm t} = \nu_0$ and the dominating turbulent
viscosity $\nu_{\mathrm t} = B|a|/k$.

%% 4.1.1
\subsubsection{Constant viscosity}\label{Sec:Constant}

At $\lambda=0$ (symmetric case) and $B=0$, {i.e.} constant
viscosity $\nu_{\mathrm t} = \nu_0$, using
Eq.~(\ref{LavalIntegral1}) in Eq.~(\ref{StationaryLaval}) we get
(cf. Ref.~\refcite{Laval0101036})
\be\label{LavalPowerLaw}
P(a) = C(Da^2+\alpha)^{-(1+\nu_0k^2/D)/2},
\ee
where $C$ is normalization constant. This distribution is of a
power-law type and one can compare it with the result
(\ref{PAringazin}), which contains a Gaussian truncation of
similar power-law tails.

We note that with the identifications,
\be
 D/\alpha = 2(q-1), \quad
(1 + \nu_0k^2/D)/2 = 1/(q-1),
\ee
the distribution (\ref{LavalPowerLaw}) reproduces that obtained in
the context of generalized statistics with the underlying $\chi^2$
distribution.\cite{Beck} Particularly, for $q=3/2$ (i.e. $n=3$ and
$\beta_0=3$) used there, it follows that $D/\alpha=1$ and
$\nu_0k^2=3$. These values can be used as estimates.

It is highly remarkable to note that the two different approaches
yield stationary distribution of exactly the same power-law form,
with certain identification of the parameters; namely, the
Gaussian-white multiplicative and additive noises with constant
intensities and a linear drift term imply $P(a)$ of the {same}
form as that obtained in the RIN model with $\chi^2$ distributed
$\beta$, the ratio of the drift coefficient to the intensity of
the Gaussian-white additive noise. It follows that the effect of
$\chi^2$ distributed $\beta$ mimics the presence of the
multiplicative noise, and {\it vice versa}, in this particular
case.

The power-law distribution (\ref{LavalPowerLaw}) can be used to
get a good fit of the Lagrangian experimental $P(a)$ data for
small accelerations, {e.g.} with the normalized values
%%$a/\langle a^2\rangle^{1/2}$
ranging from $-10$ to $10$, but in contrast to the Gaussian
truncated one (\ref{PAringazin}) it exhibits strong deviations for
large acceleration magnitudes, and for $(1 + \nu_0k^2/D)/2\leq 2$
leads to a divergent fourth-order moment, which is known to be
finite.\cite{Bodenschatz,Aringazin0212642,Aringazin0301040}

Introducing the noise intensity ratio parameter
\be
b=\sqrt{D/\alpha}
\ee
and denoting
\be
\kappa = -(1 + \nu_0k^2/D)/2,
\ee
we can rewrite the normalized distribution (\ref{LavalPowerLaw})
as follows (cf. Ref.~\refcite{Nakao9802030})
\be\label{PNakao}
P(a) = \frac{(1+b^2a^2)^\kappa}
            {2 {_2\!}F_1(-\kappa;\frac{1}{2};\frac{3}{2};-b^2)},
\ee
where ${_2\!}F_1$ is the hypergeometric function. In accord to the
analysis made by Nakao,\cite{Nakao9802030} for {\em small}
additive noise intensity, {i.e.} at $b \gg 1$, this distribution
exhibits a pronounced plateau near the origin, and the $n$th order
moments, {\em truncated} by reflective walls at some fixed $|a|$,
behave as a power of $b$,
\begin{eqnarray}
\langle a^n\rangle \sim b^{-\nu_0k^2/D} {\textrm{ for } }
n>\nu_0k^2/D,\\
\langle a^n\rangle \sim b^{-n} {\textrm{ for } } n<\nu_0k^2/D,
\end{eqnarray}
where $n>0$. Thus, the truncated moments behave as
\be\label{anNakao}
\langle a^n\rangle \sim G_0 + G_1b^{-H(n)},
\ee
where $G_{0,1}$ are some constants and the function $H(n)$ is zero
at $\nu_0k^2=0$ and monotonically increases to saturate at $n$,
for large values of $\nu_0k^2$. It should be stressed that such a
behavior of the moments for small additive noise intensity is not
specific to the distribution (\ref{PNakao}) since it gives
divergent moments but arises after some truncation of it, for
example, by means of reflective walls or nonlinearity.
Particularly, a truncation of the power-law tails of the
distribution naturally arises when accounting for the turbulent
viscosity to which we turn below.

%% 4.1.2
\subsubsection{Dominating turbulent viscosity}\label{Sec:Dominating}

At $\lambda=0$ (symmetric case), for the case of dominating
turbulent viscosity, $\nu_{\mathrm t} = B|a|/k$, using
Eq.~(\ref{LavalIntegral2}) we get for positive (upper sign) and
negative (lower sign) values of $a$:
\be\label{LavalExpLaw}
P(a) = \frac{C\exp\left\{\mp Bka/D \pm {Bk\alpha^{1/2}}{D^{-3/2}}
 \mathrm{arctan}[(D/\alpha)^{1/2}a]\right\}}
 {(Da^2+\alpha)^{1/2}},
\ee
where $C$ is normalization constant. One can see that, as
expected, the power-law dependence is of a similar form as in
Eq.~(\ref{LavalPowerLaw}) but it is exponentially truncated at
large $|a|$ owing to the turbulent viscosity term controlled by
the parameter $B$. This distribution is similar to the Gaussian
truncated one (\ref{PAringazin}) but the truncation is of an
exponential type and there is some symmetric enhancement of the
tails supplied by the $\mathrm{arctan}$ term.

Now we turn to the general symmetric case, which provides a link
between the two particular cases $\nu_{\mathrm t} = \nu_0$ and
$\nu_{\mathrm t} = B|a|/k$ considered above.

%% 4.1.3
\subsubsection{The general symmetric case}\label{Sec:GeneralCase}

At $\lambda=0$ (symmetric case), from  Eqs.~(\ref{A5})-(\ref{A7})
we have
\be\label{c12}
 c=-id_2,\
 c_1=id_1^2d_2,\
 c_2=kd_1,\
\ee
where we have denoted
\be\label{d12}
 d_1=\sqrt{D(Dk^2\nu_0^2-B^2\alpha)},\quad
 d_2=\sqrt{D\alpha}.
\ee
Note that $c$ is imaginary and the rhs of
Eq.~(\ref{LavalIntegral3}) is much simplified yielding a symmetric
distribution with respect to $a\to-a$. The complex entity $c_2$
defined by Eq.~(\ref{A7}) may be either real (for
$Dk^2\nu_0^2>B^2\alpha$) or imaginary (for
$Dk^2\nu_0^2<B^2\alpha$). In particular, for the case of constant
viscosity, $\nu_0^2 \gg B^2$, it is real while for the case of
dominating turbulent part of the viscosity, $B^2 \gg \nu_0^2$, it
is imaginary, provided that the intensities of noises $D$ and
$\alpha$ are of the same order of magnitude. These two particular
cases lead to different final expressions for the distribution,
(\ref{LavalPowerLaw}) and (\ref{LavalExpLaw}), respectively,
obtained above.

In the general case, using Eqs.~(\ref{c12}) and (\ref{d12}) in
Eq.~(\ref{LavalIntegral3}) after some algebra we obtain the
following expression for the dLDN probability density function
(\ref{StationaryLaval}), at $\lambda=0$:
\be\label{LavalGenLaw} P(a) =
\frac{Ce^{-\nu_{\mathrm t}k^2/D}}
 {(Da^2+\alpha)^{1/2}}%%\nonumber \\
 %%\times
 \left[
 \frac{4D^5(B^4D\alpha a^2 +k^2(Dk\nu_0^2 +d_1\nu_{\mathrm t})^2)}
 {d_1^6k^2(Da^2+\alpha)}
 \right]^{kd_1/(2D^2)},
\ee
where $C$ is normalization constant, $d_1$ is given in
Eq.~(\ref{d12}), and $\nu_{\mathrm t}$ is defined by
Eq.~(\ref{turbviscosity}).

It can be easily checked that Eq.~(\ref{LavalGenLaw}) reduces to
Eq.~(\ref{LavalPowerLaw}) at $B=0$, while to verify that it
reduces to Eq.~(\ref{LavalExpLaw}) at $\nu_{\mathrm t} = B|a|/k$
it is required to account for that $d_1$ becomes imaginary,
returning back to the logarithmic representation due to
Eq.~(\ref{A4}), and the use of the identity~(\ref{A8}).

The distribution (\ref{LavalGenLaw}) is characterized by the
power-law tails, which are (i) exponentially truncated and (ii)
enhanced by the power-law part of the numerator, with both the
effects being solely related to the nonzero turbulent viscosity
coefficient $B$ responsible for the small-scale dynamics.

We conclude that to provide an acceptable fit of the dLDN model
prediction to the Lagrangian experimental data\cite{Bodenschatz2}
small-scale interactions encoded in the turbulent viscosity
$\nu_{\mathrm t}$ are essential.

%% FIGURE 8
%%%%%%%%%%%%%%%%%%%%%%%%%%%%%%%%%%%%%%%%%%%%%%%%%%%%%%%
\begin{figure}[tbp!]
\begin{center}
\includegraphics[width=0.75\textwidth]{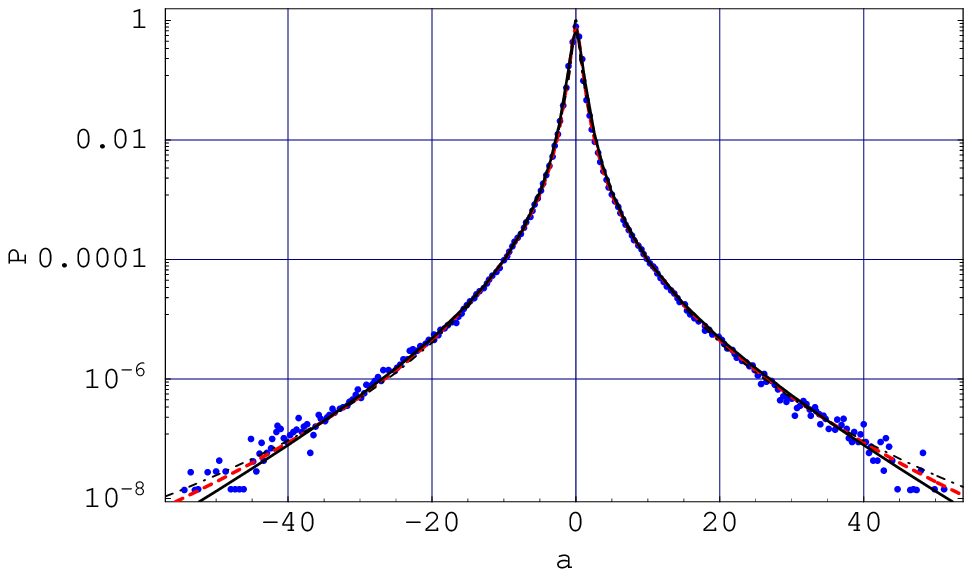}
\includegraphics[width=0.75\textwidth]{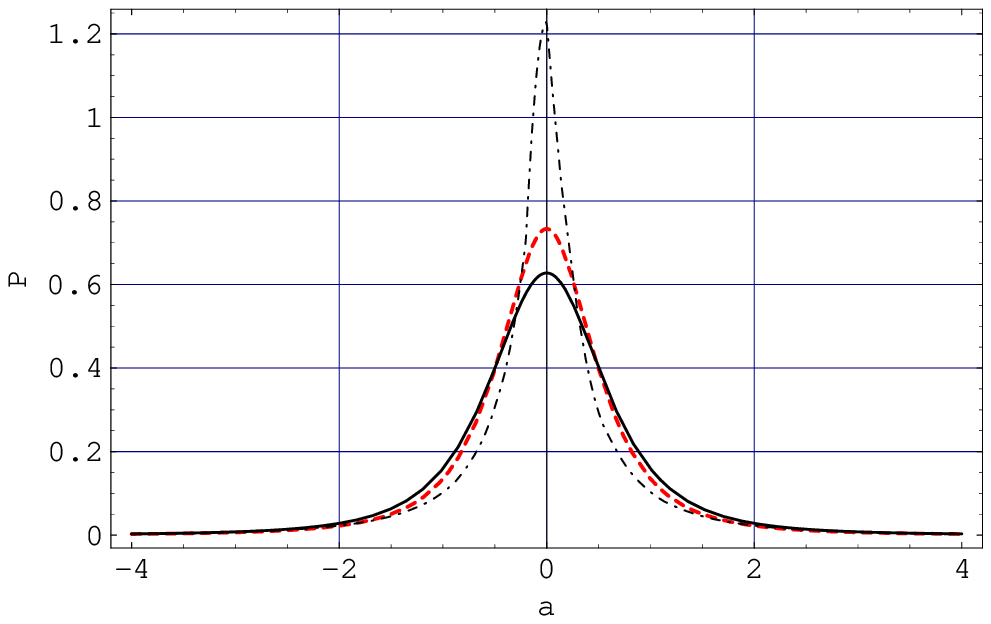}
\end{center}
\caption{ \label{Fig7} Lagrangian acceleration probability density
function $P(a)$. Dots: experimental data for the $R_\lambda=690$
flow by Crawford, Mordant, Bodenschatz, and Reynolds.$^2$
%%\cite{Bodenschatz2}.
Dashed line: the stretched exponential fit (\ref{Pexper}).
%%$b_1=0.513$, $b_2= 0.563$, $b_3= 1.600$, $C=0.733$.
Dot-dashed line: Beck log-normal model (\ref{PBeck}), $s=3.0$.
Solid line: LDN-type model (\ref{LavalGenLaw}), $k = 1$,
$\alpha=1$, $D = 1.130$, $B = 0.163$, $\nu_0 = 2.631$, $C=1.805$.
$a$ denotes acceleration normalized to unit variance.}
\end{figure}
%%%%%%%%%%%%%%%%%%%%%%%%%%%%%%%%%%%%%%%%%%%%%%%%%%%%%%%

%% FIGURE 9
%%%%%%%%%%%%%%%%%%%%%%%%%%%%%%%%%%%%%%%%%%%%%%%%%%%%%%%
\begin{figure}[tbp!]
\begin{center}
\includegraphics[width=0.75\textwidth]{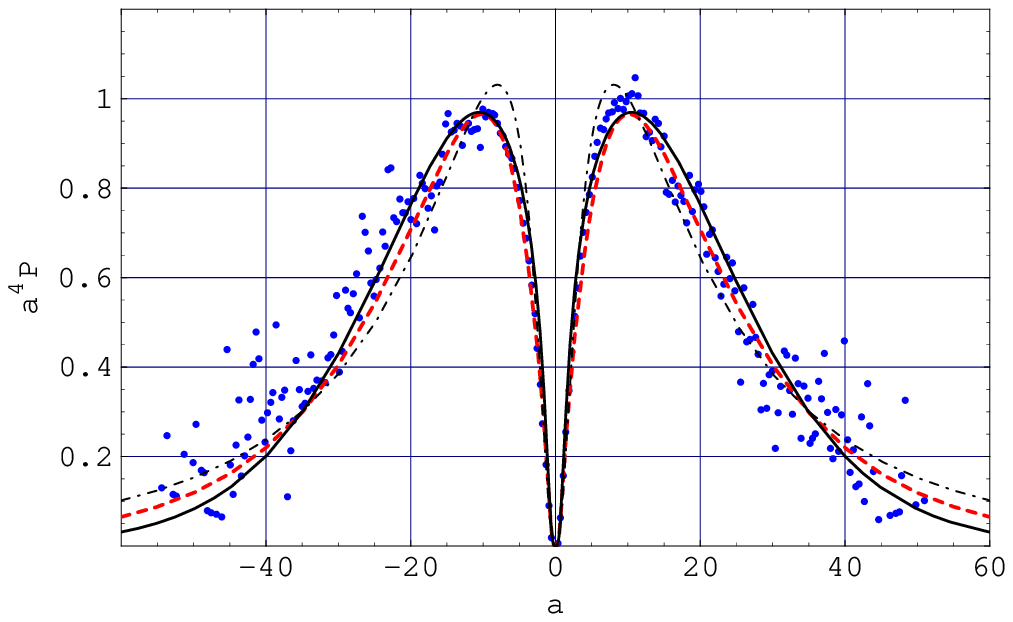}
\end{center}
\caption{ \label{Fig8} The contribution to fourth-order moment
$a^4P(a)$. Same notation as in Fig.~\ref{Fig7}.}
\end{figure}
%%%%%%%%%%%%%%%%%%%%%%%%%%%%%%%%%%%%%%%%%%%%%%%%%%%%%%%

Sample fit of the dLDN probability density function $P(a)$ given
by Eq.~(\ref{LavalGenLaw}) and the corresponding contribution to
fourth-order moment are shown in Figs.~\ref{Fig7} and \ref{Fig8},
respectively. In the numerical fit, we have put, in accordance to
the redefinitions (\ref{redef}), the wave number parameter $k=1$
and the additive noise intensity parameter $\alpha=1$ in
Eq.~(\ref{LavalGenLaw}) and fitted the remaining three parameters
$\nu_0$, $D$, and $B$. One can observe a good agreement with the
experimental data. Particularly, the dLDN contribution to the
kurtosis $a^4P(a)$ plotted in Fig.~\ref{Fig8} does peak at the
same points as the experimental curve (positions of the peaks
depend mainly on $D$). The core of the dLDN distribution shown in
Fig.~\ref{Fig7} fits the experimental data to a higher accuracy as
compared with the log-normal model (\ref{PBeck}) but yet
underestimates that of the experimental curve. This departure
could be attributed to the approximation of $\delta$-correlated
multiplicative noise used in the dLDN model (see discussion in
Sec.~\ref{Sec:QualitativeComparison} above).

Having the general symmetric form of the dLDN distribution
evaluated explicitly, Eq.~(\ref{LavalGenLaw}), one can derive the
acceleration moments $\langle a^{n}\rangle$, $n=2,4,\dots$ The
associated integrals are not analytically tractable and can be
evaluated numerically. We will consider these below in
Sec.~\ref{Sec:ConditionalProbability}.

In the most general case ($\lambda\not=0$) the resulting $P(a)$ is
given due to an exponential of the exact result
(\ref{LavalIntegral3}) which will be considered below in
Sec.~\ref{Sec:ConditionalProbability}.

To sum up, we have made an important step forward with the dLDN
model by having calculated $P(a)$ exactly (see Appendix). We have
shown that the dLDN model is capable to reproduce the recent
Lagrangian experimental data on the acceleration statistics to a
good accuracy. Particularly, we found that the predicted
fourth-order moment density function does peak at the same value
of acceleration, $|a|/\langle a^2\rangle^{1/2}\simeq 10.2$, as the
experimental curve, in contrast to the predictions of the other
considered stochastic models. The presence of the
$\delta$-correlated multiplicative noise and the nonlinearity
(turbulent viscosity) in the Langevin-type equation was found to
be of much importance. The considered RIN models provide less but
yet acceptable accuracy of the low-probability tails although they
employ only one free parameter, which can be fixed by certain
phenomenological arguments, as compared to the dLDN model, which
contains four free parameters. However, we stress that in contrast
to the LDN model the considered RIN models have a meager support
from the turbulence dynamics.

%% 5
\section{Lagrangian acceleration statistics conditional on the velocity}
 \label{Sec:ConditionalProbability}

In a recent paper Mordant, Crawford, and
Bodenschatz\cite{Mordant0303003} reported experimental data on the
probability density function $P(a|u)$ of the transverse
acceleration conditional on the same component of Lagrangian
velocity.

Sawford, Yeung, Borgas, Vedula, Porta, Crawford, and
Bodenschatz\cite{SawfordPF2003} have studied acceleration
statistics from laboratory measurements and direct numerical
simulations in 3D turbulence at $R_\lambda$ ranging from 38 to
1000. For large $|u|$, the conditional acceleration variance was
argued to behave as
\be\label{a2u}
\langle a^2|u\rangle \sim u^6,
\ee
obtained to a leading order in the same component $u$ (to be not
confused with the {rms} velocity $\bar u \equiv \langle
u^2\rangle^{1/2}$). This is qualitatively consistent with the
stretched exponential tails of the unconditional acceleration PDF.
The conditional mean rate of change of the acceleration derived
from the data has been shown consistent with the drift term in
second-order Lagrangian stochastic models of turbulent transport.
The correlation between the square of the acceleration and the
square of the velocity has been shown small but not negligible.

In a recent paper Reynolds\cite{ReynoldsNEXT2003} developed a
self-consistent second-order stochastic model with additive noise
which accounts for dependence of the Lagrangian acceleration
covariance matrix $Q_{ij}=\langle a_ia_j|u\rangle$ on Lagrangian
velocity $u$:
\be
Q_{ij} = [f(|u|)-g(|u|)]\frac{u_iu_j}{u^2} + g(|u|)\delta_{ij}.
\ee
Here, isotropy of acceleration variances is assumed and the
longitudinal and transverse functions $f(|u|)$ and $g(|u|)$
describe covariance of components in acceleration parallel and
orthogonal to the velocity vector. This model extends the
second-order Lagrangian stochastic model reviewed in
Sec.~\ref{Sec:Reynoldsmodels} to account for the observed
dependency of Lagrangian acceleration statistics on
velocity:\cite{Mordant0303003,SawfordPF2003}
\begin{eqnarray}
\label{Reynolds2u}
 da_i = -(T_L^{-1}\! +\! \tau_\eta^{-1})a_idt
 + \left(\frac{1}{2}\frac{\partial Q_{ij}}{\partial u_j}
 -\sigma_u^{-2}Q_{ij}u_j\right)dt
 - \frac{1}{2}Q_{jk}^{-1}\frac{\partial Q_{ik}}{\partial t}a_idt \nonumber\\
 +\frac{1}{2}Q_{im}\frac{\partial Q_{km}}{\partial u_j}a_ja_kdt
 + \sqrt{2\sigma_u^2(T_L^{-1}\! +\! \tau_\eta^{-1})T_L^{-1}\tau_\eta^{-1}} d\xi.
\end{eqnarray}

Fitting to the DNS data has been made by the polynomials
$f=f_0+f_1u^2+f_2u^4+f_3u^6$ and
$g=g_0+g_1u^2+g_2u^4+g_3u^6$.\cite{SawfordPF2003} The observed
dependence of the conditional acceleration variance $\langle
a^2|u\rangle$ on $u$ was partially understood in terms of
Lagrangian accelerations induced by vortex tubes within which the
vorticity is constant and outside which the vorticity vanishes.
Dimensional arguments were invoked to derive the above third-order
polynomial structure of the isotropic covariance matrix as a
function of squared velocity $u^2$.\cite{SawfordPF2003} The
parameters $f_1$, $f_2$, $g_1$, and $g_2$ are argued to be
constants, and $f_0$ and $g_0$ are taken positive. The tails of
the resulting $P(a_i)$ can not be expressed in analytical
functions but nevertheless one can show that the tails of $P(|{\bm
a}|)$ have a log-normal form.

The inclusion of such conditional acceleration covariances in the
model resulted in a significant reduction of the predicted
occurrence of small accelerations that meets experimental and DNS
data for unconditional distributions. The cores of the resulting
conditional acceleration distributions were found to broaden with
increasing $|u|$, in a qualitative agreement with the experiment,
and in general they have almost the same shape as the
unconditional distribution, in accordance with the experimental
data.\cite{Mordant0303003}

During the course of derivation of the generalized Fokker-Planck
for single-particle Lagrangian PDF of velocity increments from the
3D Navier-Stokes equation, Friedrich\cite{FriedrichPRL2003}
considered the two-point two-time acceleration autocorrelation
conditional on Lagrangian velocity and position, $\langle\bm a(\bm
r, t)\bm a(\bm r-\bm l, t')|\bm u(\bm x_0,t'), \bm x(\bm
x_0,t')\rangle_{\bm l = \bm v(t-t')}$. Here, $\bm x_0$ is the
initial position of a fluid particle. This approach and the used
closure scheme were outlined in Introduction. The conditional
dependence was ignored in order to get simple approximation to the
diffusion term. This is consistent with K41 theory. It is of much
interest to account for such a conditional dependence within the
framework of this constitutive approach since experimental data
and DNS show essential dependence of the Lagrangian acceleration
variance on the velocity.

Recently, the multifractal approach has been used by Biferale {\it
et al.}\cite{Biferale0403020} to obtain acceleration moments
conditional on the velocity. Particularly, the multifractal
prediction
\be\label{varianceBiferale}
\langle a^2|u\rangle \sim u^{4.57}
\ee
agrees well with the DNS data for large velocity magnitudes. The
exponent 4.57 differs from the value 6 predicted by Sawford {\it
et al.}\cite{SawfordPF2003} and is very close to the
Heisenberg-Yaglom scaling exponent value 9/2 entering
Eq.~(\ref{HYscaling}). This indicates that the averaging of the
conditional acceleration variance (\ref{varianceBiferale}) over
Gaussian distributed velocity $u$ is consistent with
Heisenberg-Yaglom scaling law (see Eq.~(68) and remark in
Ref.~\refcite{Aringazin0305186}).

%%We will consider the Lagrangian acceleration statistics
%%conditional on the velocity in more details below in
%%Sec.~\ref{Sec:Conditional}.

The experimental data reveal highly non-Gaussian, stretched
exponential character of $P(a|u)$, very similar to that of $P(a)$,
for fixed $u$ ranging from zero up to about three {rms}
velocity\cite{Mordant0303003} as opposed to the theoretical result
that $P(a|u)$ is a Gaussian in $a$ due to the simple RIN
model~(\ref{PGauss}) for arbitrary $\beta=\beta(u)$, or due to the
more general RIN model~(\ref{gu}). Similarity between the
experimental $P(a|u)$ and $P(a)$ suggests that they share the
process underlying the fluctuations.

Below, we consider this important problem within the framework of
the general RIN approach.\cite{Aringazin0305186,Aringazin0312415}

The idea is that stretched exponential form of the tails of
observed conditional distribution $P(a|u)$ could be assigned
solely to small time scales, while the marginal probability
distribution $P(a)$ is developed from $P(a|u)$ at large time
scales, in accord to the two-time-scale dynamics.

This requires some modification in simple RIN models reviewed in
Sec.~\ref{Sec:RINmodels}. The sole use of the Gaussian-white
additive noise, with fluctuating intensity depending on $u$, and a
linear force $F(a)=-a$ with fluctuating $\gamma=\gamma(u)$, is
{\em not} capable to explain the stretching in the observed
$P(a|u)$, as it implies only Gaussian conditional probability
density function $P(a|u)$, for any fixed $u$. However, it is known
that accounting for the {\em multiplicative} Gaussian-white noise
in the drift term of Langevin-type equation implies stretched
exponential tails.

Hence one can simply follow the dLDN ansatz as a constitutive
model (see Sec.~\ref{Sec:Comparison}) using the assumption that
the additive noise intensity $\alpha$ appearing in the stationary
probability distribution $P(a|D,\alpha,B,\nu_0)$ given by
Eq.~(\ref{LavalGenLaw}) depends on $u$. Also, we will generalize
consideration by taking the parameter $\lambda$ to be nonzero and
depending on $u$ as well.

The stationary acceleration PDF stemming from the stochastic model
(\ref{LangevinLaval})-(\ref{noises}) has been calculated
exactly\cite{Aringazin0305186} and due to
Eq.~(\ref{LavalIntegral3}) is given by
\be
%%\begin{eqnarray}
\label{PLaval} P(a)
 =
 %%C\exp\left[\int_{0}^{a} \!\!\! dx
%% \frac{-k^2x\sqrt{\nu_0^2+ B^2x^2/k^2} \!-\! Dx \!+\! \lambda}{Dx^2
%%-2\lambda x +\alpha}\right]\nonumber\\
  \frac{C \exp[-{\nu_{\mathrm
t}k^2}/{D}+F(c,a)+F(-c,a)]} {(Da^2\!-\!2\lambda a
\!+\!\alpha)^{1/2}(2Bka+\nu_{\mathrm t}k^2)^{{2B\lambda
k}/{D^2}}},
%%\end{eqnarray}
\ee
for constant parameters. Here, $C$ is normalization constant and
$\nu_{\mathrm t}=\nu_{\mathrm t}(a)$, $F(c,a)$ and $c$ are given
by Eqs.~(\ref{A4})-(\ref{A7}).
 %%
%%and we have denoted
%%\begin{eqnarray}
%%\label{Fc} F(c)
%% = \frac{c_1k^2}{2c_2D^2c}\ln[\frac{2D^3}{c_1c_2(c-Da+\lambda)}
%%   \nonumber \\
%%   \times(
%%   B^2(\lambda^2 + c\lambda-D\alpha)a
%%   + c(D\nu_{\mathrm t}^2k^2+c_2\nu_{\mathrm t})
%%    )
%%   ],\\
%%c=-i\sqrt{D\alpha-\lambda^2},\\
%%c_1 = B^2(4\lambda^3\!+\!4c\lambda^2\! -\!
%%3D\alpha\lambda-cD\alpha)
%%    \!+\! D^2(c\!+\!\lambda)\nu_0^2k^2,\\
%%c_2 = \sqrt{B^2(2\lambda^2 + 2c\lambda-D\alpha)k^2 +
%%D^2\nu_0^2k^4}.
%%\end{eqnarray}
The distribution (\ref{PLaval}) is characterized by the presence
of exponential cutoff, complicated power-law dependence, and terms
responsible for a skewness of the distribution (asymmetry with
respect to $a\to-a$). While the skewness generation can not be
directly applied to our one-dimensional case, we do not set
$\lambda$ to zero for some reasons to be explained below.

One way when comparing the model predictions with experiment is to
make a direct fit of the obtained PDF (\ref{PLaval}) to the
experimental data on unconditional acceleration distribution by
assuming all the parameters and wave number to be constant (see
Sec.~\ref{Sec:Comparison}).

Particularly, this implies a reduction of the original 1D LDN
model since wave number is taken to be fixed so that the
artificial 1D compressibility aimed to model RDT stretching effect
in 1D case is not considered. We note that the Lagrangian
acceleration is usually associated with the dissipative scale, and
in the present paper we do not study dependence of the parameters
on the wave number. Such a dependence for velocity increments was
analyzed in Ref.~\refcite{Laval0101036} with the expected result
that for larger scales the velocity increment PDF tends to a
Gaussian form. The Gaussian form is reproduced also when $D \to 0$
and $B \to 0$, i.e. the process becomes purely additive with a
linear drift term.

Without loss of generality one can put, in a numerical study,
$k=1$ and the additive noise intensity $\alpha=1$ by rescaling the
multiplicative noise intensity $D>0$, the turbulent viscosity
parameter $B>0$, the kinematic viscosity $\nu_0>0$, and the cross
correlation parameter $\lambda$. The particular cases $B=0$ and
$\nu_0=0$ at $\lambda=0$, and the general case at $\lambda=0$ were
studied in detail in Sec.~\ref{Sec:Comparison}. Nonzero $\lambda$
is responsible for an asymmetry of the PDF (\ref{PLaval}) and in
3D picture corresponds to a correlation between stretching and
vorticity (the energy cascade). Particularly, in the Eulerian
framework the third-order moment of spatial velocity increment
$\langle (\delta_l u)^3\rangle$ was found to be proportional to
{cross-correlation parameter}, in accord to a kind of generalized
K\'arm\'an-Howarth relationship.\cite{Laval0101036}
%%The third-order Eulerian velocity structure function is known to
%%be generally very small as compared to even-order ones. This
%%indicates that the parameter $\lambda$ is about zero, and the
%%distribution is almost symmetrical.

However, the approximation based on constant parameters does not
allow one to consider both the conditional and unconditional
acceleration statistics within the same model.

In the following Section, we will consider an extension of the
model (\ref{LangevinLaval})-(\ref{StationaryLaval}) with the
solution (\ref{PLaval}) by assuming certain model parameters in
the obtained stationary acceleration PDF to be dependent on random
velocity.\cite{Aringazin0312415}

This extension is compatible with the 3D LDN approach as $\xi$ and
$\sigma_\perp$ depend on velocity and contain large-scale
quantities due to their definitions (\ref{xi}) and
(\ref{sigmaperp}). Such a functional dependence and the associated
longtime fluctuations have been ignored when making the
simplification (\ref{noises}) with constant parameters $D$,
$\alpha$, and $\lambda$. We partially restore them. {This is the
main point of our subsequent consideration, and the functional
form of the distribution is thus due to Eq.~(\ref{PLaval}) with
certain parameters being now treated as functions of stochastic
velocity $u$.} Observations are that the acceleration variance
does depend on the same component of velocity.

We point out that characteristic time of variation of the
parameters should be sufficiently large to justify the
approximation that the resulting PDF (\ref{PLaval}) is used with
independent randomized parameters, $P(a|\textrm{Parameters})$. Two
well separated timescales in the Lagrangian velocity increment
autocorrelation have been established both by experiments and
DNS.\cite{Mordant0206013} The large timescale has been found of
the order of the Lagrangian integral scale and corresponds to a
magnitude part that is in accord to our assumption that the
intensity of noise along the trajectory is longtime fluctuating.

%% 5.1
\subsection{LDN-type model of the conditional acceleration
statistics} \label{Sec:Conditional}

%% FIGURE 10
%%%%%%%%%%%%%%%%%%%%%%%%%%%%%%%%%%%%%%%%%%%%%%%%%%%%%%%
\begin{figure}[tbp!]
\begin{center}
\includegraphics[width=0.85\textwidth]{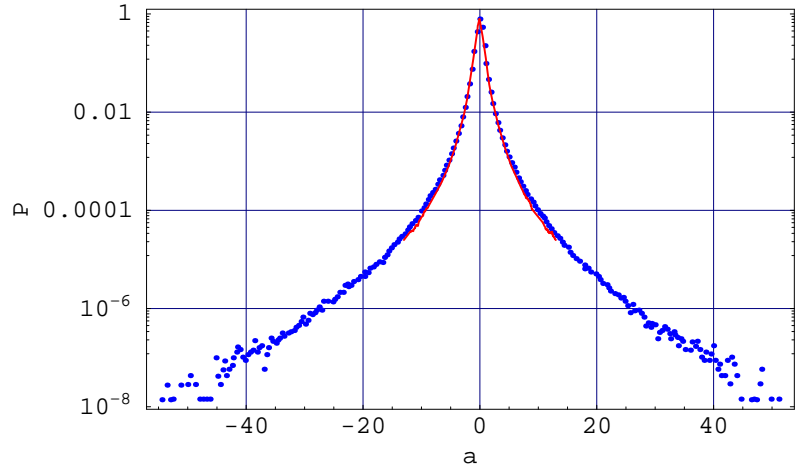}
\caption{\label{FigM1} A comparison of  the experimental
unconditional Lagrangian acceleration PDF (dots) by Crawford,
Mordant, Bodenschatz, and Reynolds$^2$ and the experimental
conditional Lagrangian acceleration PDF at Lagrangian velocity
$u=0$ (line) by Mordant, Crawford, and Bodenschatz.$^3$ $a$
denotes acceleration normalized to unit variance.}
\end{center}
\end{figure}
%%%%%%%%%%%%%%%%%%%%%%%%%%%%%%%%%%%%%%%%%%%%%%%%%%%%%%%

%% 5.1.1
\subsubsection{Conditional and unconditional
Lagrangian acceleration distributions}
\label{Sec:ConditionalUnconditional}

The experimental conditional and unconditional distributions,
which we denote for brevity by $P_{\mathrm{expt}}(a|u)$ and
$P_{\mathrm{expt}}(a)$ respectively, were found to be
approximately of the same stretched exponential form and both
reveal a strong Lagrangian turbulence intermittency. In
Fig.~\ref{FigM1} we compare $P_{\mathrm{expt}}(a|0)$ and
$P_{\mathrm{expt}}(a)$.\cite{Mordant0303003} This similarity
indicates that they share the same process underlying the
intermittency.

Accordingly, in our previous
studies\cite{Aringazin0305186,Aringazin0305459,Aringazin0306022,Aringazin0311098}
reviewed in Sec.~\ref{Sec:Comparison} we used the result of a
direct fit of the PDF (\ref{PLaval}) to $P_{\mathrm{expt}}(a)$,
which was measured with a high precision: 3\% relative uncertainty
for $|a|/\langle a^2\rangle^{1/2}\leq
10$.\cite{Bodenschatz2,Mordant0303003} We assumed that the
parameters $\alpha$ and $\lambda$ entering Eq.~(\ref{PLaval})
depend on the amplitude of Lagrangian velocity $u$, while $D$,
$B$, and $\nu_0$ are taken to be fixed at the fitted values
($k=1$). Theoretically, only $\alpha$ and $\lambda$ may depend on
velocity due to Eqs.~(\ref{xi}) and (\ref{sigmaperp}), while the
other parameters not.

An exponential form of $\alpha(u)$ has been proposed in
Ref.~\refcite{Aringazin0305186} and was found to be relevant from
both K62 phenomenological and experimental points of view.
Particularly, such a form leads to the log-normal RIN model when
$u$ is independent zero-mean Gaussian
distributed,\cite{Aringazin0301245} and yields the acceleration
PDF whose low-probability tails are in a good agreement with
experiments.\cite{Beck4,Aringazin0305186} Also, we used an
exponential form of $\lambda(u)$ so that the conditional
acceleration PDF (\ref{PLaval}) takes the form
\be
P(a|u)=P(a|\alpha(u),\lambda(u)).
\ee
Such a form was found to provide good fits of (i) the {conditional
probability density function} $P(a|u)$ to
$P_{\mathrm{expt}}(a|u)$; (ii) the {conditional acceleration
variance} $\langle a^2|u\rangle$; and (iii) the {conditional mean
acceleration}\cite{Aringazin0306022} $\langle a|u\rangle$ at
various $u$ that meet the experimental data.\cite{Mordant0303003}
A brief report on these results is presented in
Ref.~\refcite{Aringazin0311098}.

%% FIGURE 11
%%%%%%%%%%%%%%%%%%%%%%%%%%%%%%%%%%%%%%%%%%%%%%%%%%%%%%%
\begin{figure}[tbp!]
\begin{center}
\includegraphics[width=0.75\textwidth]{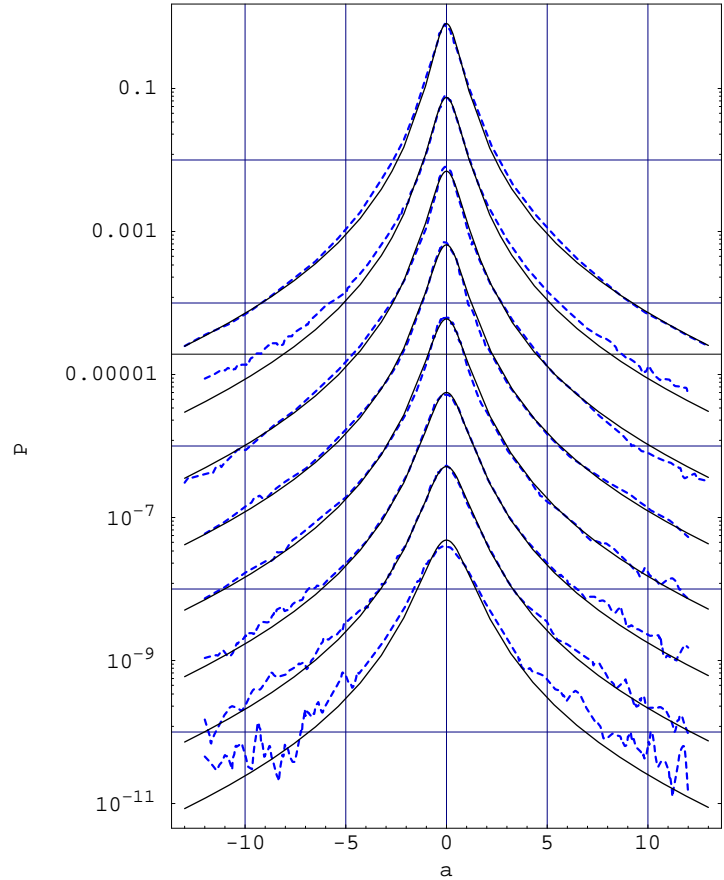}
\caption{\label{FigM2} Theoretical conditional Lagrangian
acceleration PDF $P(a|u)$ (line) and the experimental conditional
Lagrangian acceleration PDF (dashed line) at velocities $u=$ 0,
0.45, 0.89, 1.3, 1.8, 2.2, 2.7, 3.1 (from top to bottom, shifted
by repeated factor 0.1 for clarity) by Mordant, Crawford, and
Bodenschatz.$^3$ Lagrangian acceleration and velocity components
$a$ and $u$ are normalized to unit variances.}
\end{center}
\end{figure}
%%%%%%%%%%%%%%%%%%%%%%%%%%%%%%%%%%%%%%%%%%%%%%%%%%%%%%%

However, self-consistent consideration of the model assumes
different strategy:\cite{Aringazin0312415} $P(a|u)$ should be
fitted to $P_{\mathrm{expt}}(a|u)$ and the marginal PDF computed
due to
\be \label{Pmarginal}
P_{\mathrm{m}}(a)=\int_{-\infty}^{\infty}P(a|u)g(u)du,
\ee
where $g(u)$ is PDF of independent random velocity, should
reproduce $P_{\mathrm{expt}}(a)$. The marginal distribution
(\ref{Pmarginal}) corresponds to a convolution of the stationary
acceleration statistics with (independent) random velocity.

The task is thus to fit a variety of the experimental data, {both
on the conditional and unconditional statistics of acceleration,}
with a single set of fit parameters. For this purpose we use the
following natural steps.

First we fit $P(a|u)=P(a|\alpha(u),\lambda(u))$ given by
Eq.~(\ref{PLaval}) to $P_{\mathrm{expt}}(a|u)$ assuming that the
parameters depend on $u$ in an exponential way,
\be\label{alphalambda}
\alpha(u) = \alpha_0\exp[{|u|/u_\alpha}],\quad
\lambda(u)=\lambda_0\exp[{|u|/u_\lambda}].
\ee
Hereafter, we use acceleration $a$ and velocity $u$ normalized to
unit variances. The fit parameter set is $D>0$, $\nu_0>0$, $B>0$,
$\lambda_0$, $u_\alpha>0$, and $u_\lambda>0$ ($\alpha_0=1$,
$k=1$). The relations in Eq.~(\ref{alphalambda}) mean that the
additive noise intensity and the correlation between the noises
become higher for increasing amplitude of velocity.

We fit $P(a|0)$ to $P_{\mathrm{expt}}(a|0)$, that excludes the
free parameters $u_\alpha$ and $u_\lambda$ from consideration, by
varying $D$, $\nu_0$, and $B$ at $\alpha_0=1$ and
$\lambda_0=-0.005$. We notice that the available conditional
statistics $P_{\mathrm{expt}}(a|u)$ is low for high velocities,
the presented acceleration range is small, $-14 <a<14$, so that a
rather big uncertainty remains when determining fit values of the
parameters. Changes in shape of $P_{\mathrm{expt}}(a|u)$ with $u$
increasing from $u=0$ to $u=3.1$ are captured independently by the
fit parameters $u_\alpha$ and $u_\lambda$. The result is shown in
Fig.~\ref{FigM2}. {Good overlapping of each theoretical curve with
data points at all fixed magnitudes of $u$ has been achieved.}

%% FIGURE 12
%%%%%%%%%%%%%%%%%%%%%%%%%%%%%%%%%%%%%%%%%%%%%%%%%%%%%%%
\begin{figure}[tbp!]
\begin{center}
\includegraphics[width=0.75\textwidth]{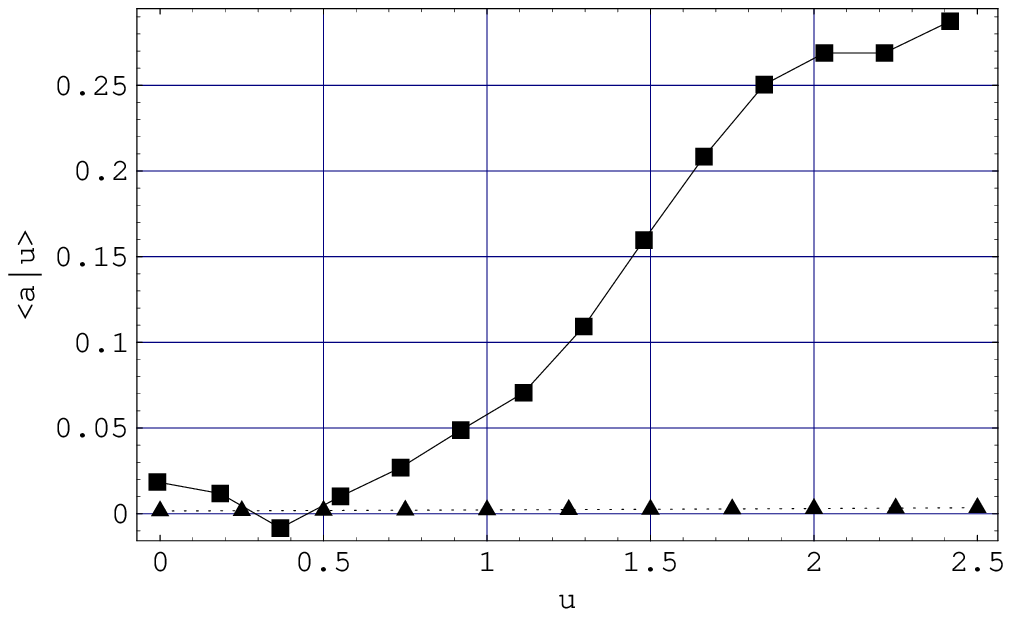}
\caption{\label{FigM3} Theoretical conditional Lagrangian
acceleration mean $\langle a|u\rangle$ (triangles) and the
experimental conditional Lagrangian acceleration mean  by Mordant,
Crawford, and Bodenschatz$^3$ (squares) as functions of the
Lagrangian velocity $u$ normalized to unit variance.}
\end{center}
\end{figure}
%%%%%%%%%%%%%%%%%%%%%%%%%%%%%%%%%%%%%%%%%%%%%%%%%%%%%%%

%% FIGURE 13
%%%%%%%%%%%%%%%%%%%%%%%%%%%%%%%%%%%%%%%%%%%%%%%%%%%%%%%
\begin{figure}[tbp!]
\begin{center}
\includegraphics[width=0.75\textwidth]{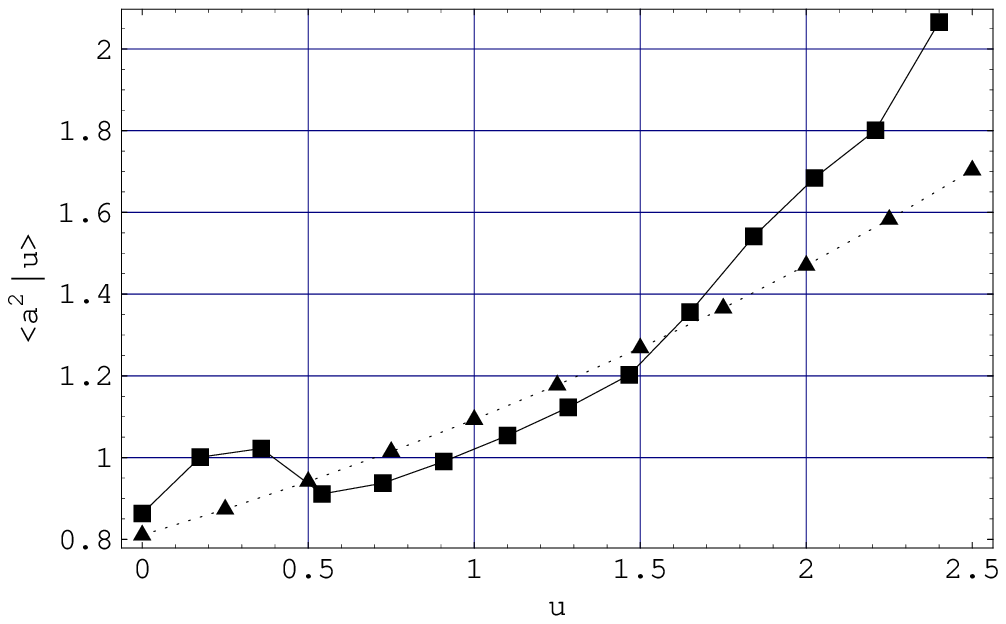}
\caption{\label{FigM4} Theoretical conditional Lagrangian
acceleration variance $\langle a^2|u\rangle$ (triangles) and the
experimental conditional Lagrangian acceleration variance by
Mordant, Crawford, and Bodenschatz$^3$ (squares) as functions of
the Lagrangian velocity normalized to unit variance.}
\end{center}
\end{figure}
%%%%%%%%%%%%%%%%%%%%%%%%%%%%%%%%%%%%%%%%%%%%%%%%%%%%%%%

Second we calculate the conditional mean $\langle a|u\rangle$ and
the conditional variance $\langle a^2|u\rangle$ and compare them
with the experimental data. This decreases uncertainty in fit
parameter values. The results are shown in Figs.~\ref{FigM3} and
\ref{FigM4}. Note that the predicted conditional mean $\langle
a|u\rangle$ as a function of $u$ appears to be very small and does
not match the experiment when the variance is fitted. We will
discuss this large discrepancy in Sec.~\ref{Sec:ConditionalMean}
below.

%% FIGURE 14
%%%%%%%%%%%%%%%%%%%%%%%%%%%%%%%%%%%%%%%%%%%%%%%%%%%%%%%
\begin{figure}[tbp!]
\begin{center}
\includegraphics[width=0.75\textwidth]{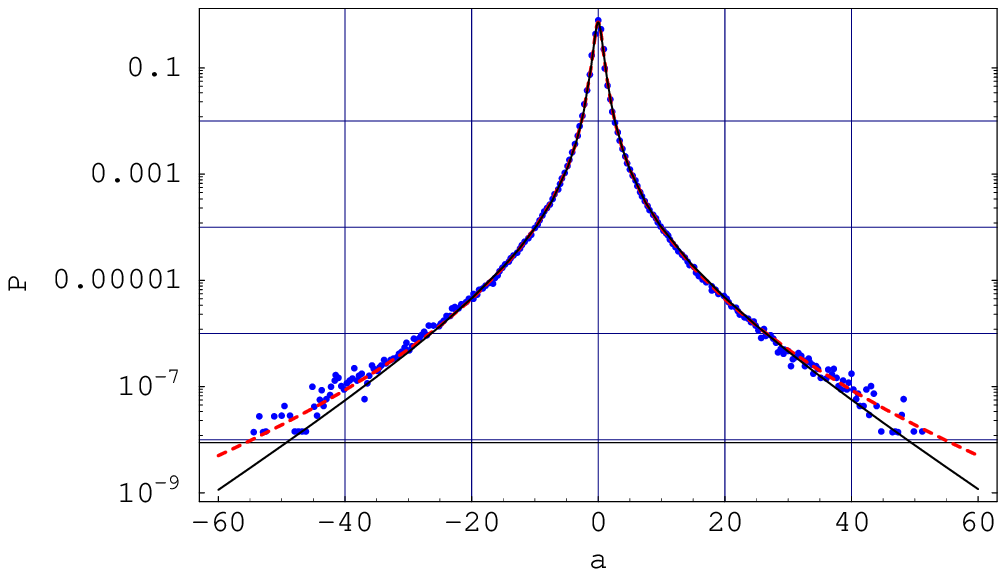}
\caption{\label{FigM5} Theoretical marginal Lagrangian
acceleration probability density function (\ref{Pmarginal}) with
Gaussian distributed Lagrangian velocity (line), experimental data
for the $R_\lambda=690$ flow (dots) by Crawford, Mordant,
Bodenschatz, and Reynolds,$^2$ and the stretched exponential fit
(\ref{Pexper}) (dashed line).}
\end{center}
\end{figure}
%%%%%%%%%%%%%%%%%%%%%%%%%%%%%%%%%%%%%%%%%%%%%%%%%%%%%%%

%% FIGURE 15
%%%%%%%%%%%%%%%%%%%%%%%%%%%%%%%%%%%%%%%%%%%%%%%%%%%%%%%
\begin{figure}[tbp!]
\begin{center}
\includegraphics[width=0.75\textwidth]{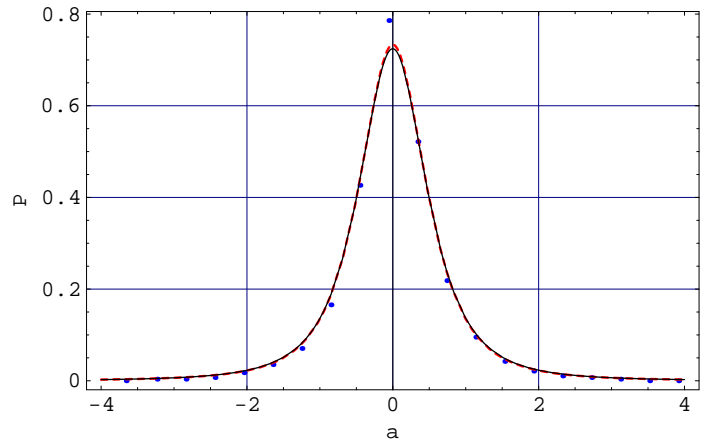}
\caption{\label{FigM6} Lin-lin plot of the central part of the
curves of Fig.~\ref{FigM5}. Same notation as in Fig.~\ref{FigM5}.}
\end{center}
\end{figure}
%%%%%%%%%%%%%%%%%%%%%%%%%%%%%%%%%%%%%%%%%%%%%%%%%%%%%%%

%% FIGURE 16
%%%%%%%%%%%%%%%%%%%%%%%%%%%%%%%%%%%%%%%%%%%%%%%%%%%%%%%
\begin{figure}[tbp!]
\begin{center}
\includegraphics[width=0.75\textwidth]{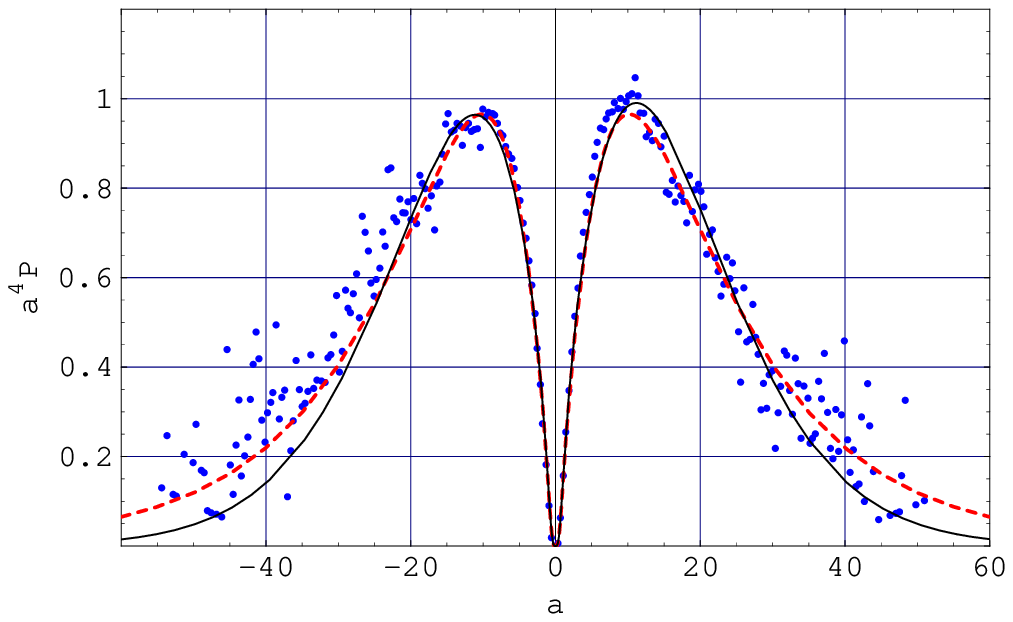}
\caption{\label{FigM7} The contribution to fourth-order moment,
$a^4P(a)$. Same notation as in Fig.~\ref{FigM5}.}
\end{center}
\end{figure}
%%%%%%%%%%%%%%%%%%%%%%%%%%%%%%%%%%%%%%%%%%%%%%%%%%%%%%%

Finally we calculate  the marginal distribution
$P_{\mathrm{m}}(a)$ defined by Eq.~(\ref{Pmarginal}) by using the
conditional PDF $P(a|\alpha(u),\lambda(u))$ and Gaussian
distribution of velocity,
\be\label{Gaussianu}
g(u)=\frac{1}{\sqrt{2\pi}}\exp\left[-\frac{u^2}{2}\right],
\ee
at fixed $a$. In the numerical calculation we use $a$ ranging from
$-100$ to 100 with the step 0.1. Then we make an interpolation of
the obtained points and fit the resulting curve to
$P_{\mathrm{expt}}(a)$. A noticeable effect of the integration
over $u$ with Gaussian $g(u)$ is a widening of tails of the
distribution that meets the experimental data shown in
Fig.~\ref{FigM1}. The used integration range for velocity is
finite, $-20\leq u\leq 20$.

The fit of $P_{\mathrm{m}}(a)$ to $P_{\mathrm{expt}}(a)$ strongly
decreases the uncertainty but the most strict determination of fit
values comes due to a comparison of the theoretical contribution
to fourth-order moment, $a^4P(a)$, with the experimental data. The
results are shown in Figs.~\ref{FigM5}, \ref{FigM6}, and
\ref{FigM7}. Quality of these sample fits is better than in the
other recent stochastic models reviewed in
Ref.~\refcite{Aringazin0305186}. In particular, the core of the
unconditional distribution reproduces very well that given by the
stretched exponential (\ref{Pexper}) as one can see from
Fig.~\ref{FigM6}. However, both curves a bit underestimate the
height at $a=0$. This means that there is small overestimation in
the range of medium accelerations, from $a=10$ to 30.

The value $\lambda_0=-0.005$ has been obtained by adjusting the
theoretical curve to slightly different heights of the peaks of
the observed $a^4P(a)$ shown in Fig.~\ref{FigM7}. Note that the
model does not assume the use of {\it ad hoc} skewness of the
forcing. Nonzero cross correlation parameter $\lambda$ naturally
results not only in nonzero mean acceleration but also in a
skewness of both the theoretical distributions $P(a|u)$ and
$P_{\mathrm{m}}(a)$. One should extract acceleration components
parallel to the velocity vector and transverse to it, to verify
whether there is skewness in their distributions.

This skewness could be associated with the Eulerian downscale
skewness generation, which despite of being small for homogeneous
flows is known to be of a fundamental character in the inertial
range (Kolmogorov four-fifth law), since the Eulerian velocity
structure function $\langle(\delta_l u)^3\rangle$ was found to be
related to cross-correlation between two noises. However, we
stress that the observed very small skewness of acceleration
distribution is attributed to the effect of anisotropy of the
studied flow. How the large-scale anisotropy affects smallest
scales of the flow is an interesting problem. Our fit made by
using nonzero $\lambda$ is of an illustrative character, to verify
whether it can explain the observed increase of the conditional
mean acceleration with increasing velocity shown in
Fig.~\ref{FigM3}. This issue requires a separate study and will be
discussed further in Sec.~\ref{Sec:ConditionalMean}.

Anisotropic aspects of turbulent statistical fluctuations by using
irreducible representations of SO(3) group, which represents a
rotational symmetry of the forceless 3D Navier-Stokes equation,
have been reviewed in a recent paper by Biferale and
Procaccia.\cite{Biferale0404014} They stressed that the anisotropy
associated with anisotropic large-scale forcing decays upon going
to smaller and smaller scales of a high-Re flow, and the
conflicting experimental measurements on the decay of anisotropy
are explained and systematized, in an agreement with the SO(3)
decomposition theory.

The following remarks are in order. Our finding is that the
condition $u_\alpha \leq u_\lambda$ provides a convergence of
$P_{\mathrm{m}}(a)$. Also, $u_\lambda$ should not be small to
provide assumed condition $\lambda \ll \alpha$ at arbitrary $u$
(the cross correlation is small as compared to both noise
intensities $\alpha$ and $D$).\cite{Laval0101036,Aringazin0305186}
We used these criteria when making the fits.

The resulting sample fit values are given by
\begin{eqnarray}\label{fits}
D=2.1, \ \nu_0=5.0, \ B=0.35,\nonumber\\
 \lambda_0=-0.005, \ u_\alpha=3.0,\ u_\lambda=3.0,
\end{eqnarray}
with $\alpha_0=1$ and $k=1$. The theoretical curves in
Figs.~\ref{FigM2}--\ref{FigM7} are shown for this sample set of
values, which require a further fine tuning. Such a small value of
$|\lambda|$ as compared to $\alpha$ or $D$ is in agreement with
that obtained in the LDN direct numerical simulations. The
calculated flatness factor $F=49.3$ of $P_{\mathrm{m}}(a)$ is in
agreement with the experimental value (\ref{flatness}).

To summarize, the considered Navier-Stokes equation based 1D toy
model (\ref{PLaval})-(\ref{alphalambda}) is capable to fit all the
available high-precision experimental data on the conditional and
unconditional Lagrangian acceleration
statistics\cite{Bodenschatz,Bodenschatz2,Mordant0303003} with the
single set of parameters (\ref{fits}) to a good accuracy, with an
exception being only the conditional mean acceleration, which we
will consider below.

%% 5.1.2
\subsubsection{Conditional mean acceleration}
\label{Sec:ConditionalMean}

As one can see from Fig.~\ref{FigM3}, at the values of fit
parameters (\ref{fits}) the predicted conditional mean
acceleration $\langle a|u\rangle$ qualitatively is in agreement
but greatly deviates from the experimental data. Namely, it is
nonzero due to nonzero $\lambda$ and increases with the increase
of $|u|$ but remains to be much smaller than the experimental data
even at high values of $|u|$. The conditional mean acceleration is
evidently zero for a symmetrical distribution (i.e. at
$\lambda=0$) and should be zero for statistically homogeneous
isotropic turbulence when one studies acceleration components
aligned to fixed directions. The observed departure from zero is
thought to reflect large-scale anisotropy of the studied
$R_\lambda=690$ flow. Remarkably, the DNS of (approximately)
homogeneous isotropic turbulence\cite{Mordant0303003} also reveals
slightly nonzero mean. In
Refs.~\refcite{Biferale0402032,Biferale0403020} {\it a priori}
statistical equivalence between all the acceleration components in
the laboratory frame of reference was used and the subsequent
averaging over the three components gives zero-mean acceleration
to a good accuracy. The latter procedure is partially justified by
the high-Re effect of anisotropy decay when going to smaller
scales.\cite{Biferale0404014}

To reduce the discrepancy, one can try the value $u_\lambda=1.0$
instead of $u_\lambda=3.0$ to provide faster increase of
$|\lambda|$ for higher $|u|$. This implies a good fit to the
experimental conditional mean acceleration (see, e.g. Fig.~2 in
Ref.~\refcite{Aringazin0311098}) but we found an excess asymmetry
of $P(a|u)$ at high $|u|$, with big departure from observations,
and divergencies when calculating
$P_{\mathrm{m}}(a)$.\cite{Aringazin0312415} The reason of the
divergency is that $\lambda(u)$ at $u_\lambda=1.0$ grows faster
than $\alpha(u)$ at $u_\alpha=3.0$ so that $\lambda$ becomes
comparable or bigger than $\alpha$ with increasing $u$, and when
$\lambda^2 \to D\alpha$ the function $F(c)$ defined by
Eq.~(\ref{A4}) undergoes unbound growth. Thus we conclude that the
observed nonzero conditional mean acceleration may be due to a
persistent effect of the flow anisotropy rather than some
intrinsic dynamical (cascade) mechanism associated with the
developed turbulence. It should be emphasized however that some
aspects of anisotropic fluctuations in high-Re flows were found to
be universal.\cite{Biferale0404014}

In general one observes a rather small relative increase of the
experimental conditional mean acceleration for higher $|u|$ that
eventually reflects a coupling of the acceleration to large scales
of the studied flow.\cite{Chevillard0310105,Laval0101036} This
coupling could be accounted for particularly by introducing a
correlation between the Lagrangian acceleration and velocity. This
possibility is of much interest to explore as it may yield the
deficient increase of $\langle a|u\rangle$ but it is beyond the
scope of the present formalism, which assumes an independent
velocity statistics. We note also that in contrast to the
experimental data on the variance $\langle a^2|u\rangle$ the
experimental $\langle a|u\rangle$ exhibits small asymmetry with
respect to $u\to -u$ (not shown in
Fig.~\ref{FigM3}).\cite{Mordant0303003}

The multiplicative noise intensity $D$ was taken to be independent
on the velocity $u$. The effect of variation of $D$ has been
considered in Ref.~\refcite{Aringazin0305186} with the qualitative
result that it does not provide the specific change in shape of
$P(a|u)$ observed in experiments as shown in Fig.~\ref{FigM2}.
However, a weak dependence of $D$ on $u$ can not be ruled out.

%% 5.1.3
\subsubsection{Summary}
\label{Sec:SummaryConditional}

In summary, the presented 1D LDN-type stochastic toy model with
the velocity-dependent additive noise intensity and cross
correlation parameter is shown to capture main features of the
observed conditional {\it and} unconditional Lagrangian
acceleration statistics to a good accuracy except for the
discrepancy in the conditional mean acceleration which can be
attributed to certain coupling of the acceleration to large scales
of the studied $R_\lambda=690$ flow.\cite{Bodenschatz}
%%(finite Reynolds number effect).

One of the technical advantages of this model is that one obtains
the conditional acceleration distribution in an explicit
analytical form, Eq.~(\ref{PLaval}), that enables to trace back
all the produced effects and perform integration over velocity.

The main result is of course not only good sample fits of a
variety of experimental data which are important to test
performance of the model but also certain advance in understanding
of the mechanism of Lagrangian intermittency provided by the
dynamical Laval-Dubrulle-Nazarenko approach to small-scale
turbulence.

The central point is that the LDN toy model has a deductive
support from Navier-Stokes turbulence. The obtained exact analytic
result for the conditional acceleration distribution and the use
of recent high-precision Lagrangian experimental data on
conditional and unconditional acceleration statistics allow one to
make a detailed analysis of the mechanism within the adopted
framework.

Effects of large scales and turbulent viscosity have been found of
much importance in steady-state Lagrangian acceleration
statistics. The detailed study of conditional acceleration
statistics reveals a specific model structure of the external
large-scale dynamics and nonlocal inter-scale coupling for
homogeneous high-Re flows. Namely, the additive noise
$\sigma_\perp$ associated with the downscale energy transfer
mechanism encodes the main contribution to velocity dependence of
the acceleration statistics.

In general, a cross correlation between model additive and
multiplicative noises, associated with a correlation between
stretching and vorticity in the 3D case, naturally provides a
skewness of distributions and a nonzero mean. Weakness of this
correlation measured by the parameter $\lambda$ is a theoretical
requirement that meets the Lagrangian and Eulerian experiments and
DNS of homogeneous isotropic turbulence. It was shown that the
observed conditional mean acceleration is not related to non-zero
$\lambda$ and may be attributed to flow anisotropy effects. In the
Eulerian frame, the cross correlation between noises is related to
the four-fifth Kolmogorov law but the effect of skewness generated
by $\lambda\not=0$ is negligibly small as the result of relatively
large intensity of the additive noise, which tends to symmetrize
acceleration distributions. This is a dynamical evidence implied
by the model rather than a result of {\it a priori} local isotropy
in the spirit of K41 theory.

The use of exponential dependence of certain noise parameters on
statistically independent Gaussian distributed Lagrangian velocity
has been found appropriate to cover new experimental data on
conditional statistics and subsequently to transfer from the
conditional to unconditional acceleration distribution both
exhibiting a strong Lagrangian intermittency of the flow. Such a
dependence is also compatible with the log-normal statistics
assumed by K62 theory and obtained in the SDT
model.\cite{Dubrulle0304035} Gaussian-white multiplicative noise
and a longtime correlated intensity of Gaussian-white additive
noise were both found to make essential contributions to
intermittent bursts of acceleration.

It is worthwhile to mention new aspect of stochastic dynamics
emerging from fluctuating character of the intensity $\alpha$ of
additive noise in the presence of the multiplicative noise. The
ratio between intensities of these noises, $b^2=D/\alpha$,
controls the character of dynamics of fluid particle. For $b\ll 1$
or $b\gg 1$ the acceleration evolution is dominated by the
influence of additive or multiplicative noise respectively. Since
$b$ is longtime correlated stochastic parameter, the dynamics can
be of a Brownian-like ($b\ll 1$), burst ($b\gg 1$), or mixed
($b\simeq 1$) character. The regimes of dynamics alternate
randomly due to random $\alpha$ and are characterized by different
chances of occurrence of bursts of acceleration. Relative weight
of each of the regimes in a statistically stationary consideration
is controlled by the ascribed distribution of $\alpha$.

The additive sector of stochastic forces entering the
Langevin-type equation (\ref{LangevinLaval}) for the acceleration
is of the form $e^{u(t)}L(t)$, where $L(t)$ is Gaussian-white
noise and $u(t)$ is an independent stationary process with
independent increments associated with Lagrangian velocity. The
stochastic process related to such representation has been
recently considered by Muzy and Bacry\cite{Muzy0206202} in the
context of multifractal random walk\cite{MuzyEJP2000,BacryPRE2001}
with uncorrelated increments. We will discuss this interesting
observation\cite{Aringazin0301245} in Sec.~\ref{Sec:MRW} below.

%% 6
\section{Multifractal random walk and Lagrangian velocity}
\label{Sec:MRW}

In our previous
studies\cite{Aringazin0301245,Aringazin0305186,Aringazin0312415}
we took Lagrangian velocity component $u$ in a stationary
turbulent flow to be Gaussian distributed with zero mean, i.e.
$u(t)$ is understood as an independent stationary Gaussian
process. For Gaussian distributed $u$ the characteristic parameter
is the rms velocity ${\bar u}$, which completely determines
statistical properties of the zero-mean random $u$. The latter is
an agreement with approximately Gaussian form of the experimental
curve for both of two measured components of Lagrangian velocity
in developed turbulent flows.\cite{Bodenschatz,Paulain0306005}

In this Section, we go beyond the Gaussian modeling of the 1D
process $u(t)$ by adopting the following assumption. Lagrangian
velocity $u(t)$ is an independent stationary stochastically
continuous process with {\em independent increments}, i.e. the
increments $u(t_1)-u(t_0)$, $u(t_2)-u(t_1)$, $u(t_3)-u(t_2)$,
$\dots$, for $t_{k+1}>t_k$ are uncorrelated and the joint
distribution of $u(t_0+t)$, $u(t_1+t)$, $u(t_2+t)$, $\dots$, for
any $t_k$ ($k=0,1,2,\dots, n$) and any $n$ does not depend on time
$t$.

In essence, this assumption restates {universality of the
Lagrangian velocity statistics} in the inertial range, in accord
to K41 similarity hypotheses formulated in the Lagrangian
framework, and strictly specifies {statistical properties of the
stochastic process} $u(t)$.

In general, stationary stochastic processes with independent
increments, the simplest example of which is the usual Brownian
motion, is a wide class characterized by strong Markovian property
(no memory effects) and the so called infinite divisibility. The
most important feature of such processes for the present
consideration is that one can get a general analytic form of the
characteristic function of the increments which allows one to
determine all statistical properties of the process for a given
stochastic measure. This enables one to use a broader class of
models (as compared to the Gaussian modeling) yet having well
known analytical tools developed by
Kolmogorov,\cite{Kolmogorov1932} Levy,\cite{Levy1934} and
Khinchin.\cite{Khinchin1936}

By definition, for stationary infinitely divisible processes the
characteristic function $\varphi(z)$ of the distribution of
$u(s+t)-u(s)$
%% p.40.
can always be represented as $\varphi(z)=e^{g(z)}$, where $g(z)$
is determined by the celebrated Levy-Khinchin formula; see
Eq.~(\ref{LevyKhinchin}) below. For {infinitely divisible
processes} there always exists some function $\varphi_n(z)$ such
that the relation
\be\label{id}
\varphi(z)=[\varphi_n(z)]^n
\ee
holds for any positive integer $n$. Indeed, one can use the
representation $\varphi_n(z)=\exp[\frac{1}{n}\ln\varphi(z)]$,
where we fix arg$\,\varphi(0)=0$ to provide uniqueness of the
representation.

We start by considering simple example, when the Lagrangian
velocity is viewed as a stationary {\em continuous} process $u(t)$
with independent increments. In many cases, without loss of
generality one can put the initial value $u(0)=0$. This process
corresponds to the Brownian motion, i.e. $u(t)$ is Gaussian
process with the mean $\langle u(t)\rangle=0$. We stress that this
motion is not due to a molecular structure of the medium but is
attributed to turbulent fluctuations. This process is
stochastically continuous (of course, this does not mean that
realizations of the process are continuous) and infinitely
divisible. Indeed, it can be proven that in this case the
increments $u(s+t)-u(s)$ are a stationary Gaussian process, i.e.
$u$ is normally distributed, and the homogeneity implies
independence on the value of $s$. Particularly, for the
statistical law $\ln \epsilon \simeq u$ we conclude that the
stochastic energy dissipation rate $\epsilon$ is log-normally
distributed. This meets K62 log-normal model, in which log-normal
distribution of $\epsilon$ is postulated.

Let $u(t)$ be a stationary {\em jump} process with independent
increments, i.e.
\be
u(t)=\mathrm{const}, \textrm{  for } t\in [0,t_1), (t_1,t_2),
\dots, (t_n,T),
\ee
where 0$<$$t_1$$<$$\cdots$$<$$t_n$$<$$T$ and $n$ is finite number.
The velocity is constant at each of the finite-time intervals.
Again, this process is stochastically continuous and infinitely
divisible. It can be proven that the jump process is strongly
Markovian and the number of jumps, $j(t)$, is itself a
stochastically continuous stationary Poisson jump process with
independent increments. With the use of $j(t)$ the process $u(t)$
can be represented as
\be
\label{iid} u(t)=\sum_{k=1}^{j(t)}\eta_k,
\ee
where $\eta_k$ are independent identically distributed (i.i.d)
random variables, and is referred to as the generalized Poisson
process.\cite{Khinchin1936}

In particular, if the increments $u(s+t)-u(s)$ are a stationary
Poisson process with independent increments (the characteristic
function is $\varphi(z)=\exp[\lambda(e^{iz}-1)]$), we conclude
that for $\ln\epsilon \simeq u$ the energy dissipation rate
$\epsilon$ is log-Poisson distributed. This meets
She-Leveque\cite{She94} cascade model and extended self-similarity
inspired model by Dubrulle\cite{Dubrulle94}, which are in good
agreement with the experimental data on scaling exponents of the
Eulerian velocity structure functions.

In general case, a stationary stochastically continuous process
with independent increments can always be {\em decomposed} into
the continuous and the jump process parts.\cite{Levy1934}
Therefore, the log-Poisson distribution of $\epsilon$ stemming
from the jump part of the process can receive ``corrections''
coming from the continuous part of the process (log-normal
distribution), and vice versa. Hence, the distribution of
$\epsilon$ arising from the general consideration of the
stationary stochastically continuous process $u(t)$ with
independent increments and the statistical law $\ln\epsilon \simeq
u$ can be defined as a {\em weighted superposition} of log-Poisson
and log-normal distributions. The weight parameter may depend on
the scale.

Muzy and Bacry\cite{Muzy0206202} has recently studied a class of
log-infinitely divisible multifractal random processes of the
following form:
\be
X_\tau(t) = \int_{0}^{t} e^{\omega_\tau(t')}dw(t'),
\ee
where $X_\tau(t)=X(t+\tau)-X(t)$ is the increment of stochastic
process $X(t)$, $dw(t)$ is Gaussian white-in-time noise, i.e.
$w(t)$ is Wiener process, and $\omega_\tau(t)=\ln W_{\tau/T}(t)$.
Here, $W_{\tau/T}(t)$ is a continuous version of independent
identically distributed (i.i.d.) random variables of the
well-known Mandelbrot's discrete multiplicative
cascade,\cite{Mandelbrot1967}
$M_\tau(t)=W_{\tau/\tau'}(t)M_{\tau'}(t)$, in the sense that the
scale $\tau$ takes continuous values in replace of the discrete
$\tau_n$. This corresponds to a continuous extension of the
discrete cascade.

The stochastic process $\omega_\tau(t)$ can then be represented as
a sum of arbitrary number of i.i.d. random variables so that by
definition $\omega_\tau(t)$ is an infinitely divisible random
process. Note that this process is defined on the scale-time
$(\tau,t)$-plane, with the upper value $\tau=T$. The process
$\omega_\tau(t)$ has the following remarkable scaling property:
\be
\omega_{\lambda \tau}(\lambda t)= \omega_\tau(t)+\Omega_\tau,
\ee
for $\lambda<1$, where $\Omega_\tau$ is an infinitely divisible
random variable.

The multifractal random measure is defined as the small-scale
limit, $M(dt)=\lim_{\tau\to 0}M_\tau(dt)$, of the stochastic
measure $M_\tau(dt)=e^{\omega_\tau(t)}dt$. The process
$M_\tau(t)=\int_0^t e^{\omega_\tau(t')}dt'$ is a jump process for
certain choice of properties of the characteristic function of
$\omega_\tau(t)$, namely, when the Levy measure has no Gaussian
component.

The small-scale limit, $X(t)=\lim_{\tau\to0}X_\tau(t) \equiv
B(M(t))$, converges and is referred to as the Mandelbrot-Taylor
subordinated Brownian process.\cite{Mandelbrot1967} The process
$X(t)$ represents an example of the multifractal random walk
(MRW).

The MRW can thus be thought of as the usual Brownian motion
defined on the ``multifractal'' time
$\tau=M(t)$.\cite{Muzy0206202} Such a viewpoint follows in general
from the representation of the process $X(t)$ as the stochastic
integral of a {\em random} function $f(t)=e^{\omega(t)}$ over
Wiener process $w(t)$. In the case the function $f(t)$ is taken to
be deterministic one ends up with the usual continuous stationary
Gaussian process for $X(t)$, i.e. the Brownian motion $B(t)$
defined on usual time $t$.

The theory of one-dimensional infinitely divisible stochastic
processes developed by Kolmogorov, Levy, and Khinchin was applied
to the above setup. The result is that the multifractal scaling
property of the absolute moments of $X(t)$ can be found exactly
with the use of the Levy-Khinchin formula for the characteristic
function $\varphi(z)$ of the increments of $X(t)$,
\be
\label{LevyKhinchinExp} \varphi(z)=e^{g(z)},
\ee
where
\be\label{LevyKhinchin}
 g(z) = imz + \int (e^{izx}-1 - iz\sin
x)\frac{\nu(dx)}{x^2},
\ee
$\nu(dx)$ is the canonical Levy measure and $m=\langle
X(t)\rangle$ is the mean. Namely, for the case of stationary
multifractal measure $M(dt)$ (and finite second-order moment of
the process) one has\cite{Muzy0206202}
\be
\label{MRWscaling} \langle |X(t)|^p\rangle =
\frac{2^{p/2}\sigma^p\Gamma(\frac{p+1}{2})}{\Gamma(\frac{1}{2})}
K_{p/2}t^{\zeta(p)},
\ee
where $K_p=T^{-\zeta (p)}\langle M^p\rangle$, the temporal scaling
exponent is
\be
%%\zeta(p)=p -\psi(p),
\zeta(p)=\frac{p}{2} -\psi\left(\frac{p}{2}\right),
\ee
and the real convex cumulant-generating function $\psi(z)$ is
defined by $\psi(z)=g(-iz)$, for which without loss of generality
one can put $\psi(1)=0$. The intermittency parameter measures
nonlinearity of $\zeta(p)$ in $p$ and is defined by
\be
\label{lambda}
 -\frac{\partial^2\zeta(p)}{\partial p^2}{|_{p=0}}.
\ee
Note also that in Eq.~(\ref{MRWscaling}) one gets the prefactor in
analytically exact form.

Mordant {\it et al.}\cite{Mordant0206013} have recently
demonstrated that the above MRW formalism in constructing of the
additive noise in the form $e^{\omega(t)}L(t)$, which models the
stochastic force having uncorrelated direction (i.e. $L(t)$ is
Gaussian-white) and longtime correlation of its magnitude [i.e.
$\omega(t)$ is Gaussian and ultraslow autocorrelated due to
Eq.~(\ref{ultraslow})], leads to the Lagrangian scaling exponent
$\zeta(p)$ of the form
\be
\label{zetaLogNormal}
\zeta(p)=\left(\frac{1}{2}+\lambda^2\right)p-\frac{\lambda^2}{2}p^2,
\ee
$\zeta(2)=1$. This corresponds to the log-normal MRW with the
generating function $\psi(p)=mp+\mu^2p^2/2$, where $m$ and $\mu$
are parameters. Also, the autocorrelation functions of logarithm
of the amplitude of infinitesimal Lagrangian velocity increments
in time are in agreement with the experimental data on tracer
particle in developed turbulent flow. The fit value of the
intermittency parameter was found to be $\lambda^2=0.115\pm0.01$.
We note that this approach does not include a multiplicative noise
sector, such as that represented by the first term in
Eq.~(\ref{LangevinLaval}). While the ultraslow autocorrelated
$\omega(t)$ mimics the presence of multiplicative noise, this
phenomenological approach may result in essential departures from
the experimental data since the multiplicative noise, which
particularly arises naturally in the LDN approach reviewed in
Sec.~\ref{Sec:LDNmodel}, is not present explicitly.

We note that within the same MRW framework one can consider,
alternatively, log-Poisson MRW,
$\psi(p)=[m-\sin(\ln\delta)]p-\gamma(1-\delta^p)$,  which yields a
different prediction for the Lagrangian scaling exponent
\be
\label{zetaLogPoisson}
 \zeta(p)=m' p+\gamma(1-\delta^{p}),
\ee
where $m'$, $\gamma$, and $\delta$ are parameters. For $\delta\to
1$ and $\gamma(\ln\delta)^2\to\lambda^2$ one recovers the
log-normal case. The form of this scaling exponent is similar to
those of She-Leveque\cite{She94} and Dubrulle\cite{Dubrulle94}
cascade models corresponding to log-Poisson distribution of the
normalized energy dissipation rate at spatial scale $l$,
\be
\label{zetaLogPoisson2}
 \zeta^E(p)=(1-\Delta)\frac{p}{3} + \frac{\Delta(1-\beta^{p/3})}{1-\beta},
\ee
where $\Delta$ and $\beta$ are parameters, $\Delta=\beta=2/3$
corresponds to She-Leveque model and $\zeta^E(3)=1$; see also work
by L'vov and Procaccia.\cite{LvovPRE2000}

The predicted Lagrangian scaling exponents can be fitted to
experimental data to a good accuracy. The experimental data by
Mordant {\it et al.}\cite{Mordant0103084} on the relative
Lagrangian scaling exponents are ($\zeta(2)=1$):
\be
 \label{zetaexp}
 \zeta(p)/\zeta(2)= 0.56\pm0.01, \ 1,\  1.34\pm0.02,\
1.56\pm0.06, \ 1.8\pm0.2
\ee
for $p=1,2,3,4,5$ respectively.
%% Eulerian exponents
%% $\zeta(p)/\zeta_3$ = (0.42, 0.75, 1, 1.17, 1.28),
%% $p=1,2,3,4,5$
%% \cite{Mordant0103084}.

Recent DNS results by Biferale {\it et al.}\cite{Biferale0402032}
for $R_\lambda=280$ flow yield slightly different values:
$\zeta(p)/\zeta(2)$ = 1.7$\pm$0.05, 2.0$\pm$0.05, 2.2$\pm$0.07 for
$p=4,5,6$. They mention that the range of time delays over which
relative scaling occurs is 10$\tau_\eta$ to 70$\tau_\eta$, and in
this range anisotropic contributions induced by the large-scale
flow appear to influence the scaling properties. In the range from
$\tau_\eta$ to 10$\tau_\eta$ all local slopes were found to
converge around the value 2. This effect is most likely due to the
capture by very intense small-scale vortical structures, the
relative contribution of which to the scaling properties is high
in the Lagrangian frame, as compared to that in the Eulerian one.
The influence of dissipative range $\tau<\tau_\eta$ can not be
associated with these corrections since it tends to increase the
value of the local slope rather than to decrease it.  It was
emphasized that the two-time-scale dynamics, $\tau \in [\tau_\eta,
10\tau_\eta]$ and $\tau \in [10\tau_\eta, T_L]$, is a feature of
velocity fluctuations along Lagrangian trajectories. How to
incorporate such dynamical processes in the Lagrangian
multifractal description and stochastic modelization of particle
diffusion was mentioned as one of the most challenging open
problems arising from their analysis.

In general, the above situation resembles that of fitting to the
experimental Eulerian scaling exponents for which the K62
log-normal model shows an increasing departure for large orders
$p$ while the She-Leveque and Dubrulle hierarchic models are in a
good agreement for all experimentally accessible orders $p$. This
indicates that {log-Poisson} models give qualitatively acceptable
description of higher inhomogeneities of fully turbulent flow in
the Eulerian framework.

This suggests that within the present model for the longtime
behavior the {jump} process approximation is appropriate for
qualitative and quantitative description of the random walk of the
Lagrangian velocity in turbulent flow as compared with the
{continuous} process approximation. This can be understood as the
fundamental effect of finite size and finite discrete cascade
mechanism of the developed turbulence.

However, both types of processes may contribute to the dynamics,
and our main conclusion is that one can construct a model which
includes them on equal footing by adopting the viewpoint that the
process $u(t)$ is infinitely divisible with independent
increments. Such a process includes both the continuous (Gaussian)
and jump (Poisson) parts.

This approach could be used to address the problem pointed out
recently by Chevillard {\it et al.}\cite{Chevillard0310105} that
the Lagrangian (L) and Eulerian (E) singularity spectra $D(h)$ can
not be both log-normal (parabolic), namely,
\be
D^L(h)= -h+(1+h)D^E(h/(1+h)).
\ee
Whereas the left-hand side of the measured and DNS singularity
spectrum curves for Lagrangian velocity increments are fitted, the
right-hand side of them is reproduced by neither the log-normal
nor the log-Poisson statistics based models mapped to the
Lagrangian domain.

Here we note that this is expected result since intense and weak
increments are driven by different processes related to the
``burst'' and ``diffusive'' dynamical regimes. The burst regime
corresponds to capturing by very intense vortical structures while
the diffusive regime describes motion in their incoherent
surround. An example illustrating the difference is given by the
log-normal model\cite{Beck4} which reproduces low-probability
tails of the experimental acceleration distribution to a very good
accuracy but greatly deviates (overestimates) at the central part
of distribution which is characterized by low accelerations.

Within the framework of the 1D LDN model it seems to be reasonable
to take the multiplicative noise $\xi$ to be Poisson distributed
and use Gaussian approximation for the additive noise
$\sigma_\perp$. We expect however robustness of the model due to
the numerical results on the noisy on-off intermittency
model\cite{Nakao9802030} for which three quite different types of
the multiplicative noise (Gauss-Markov noise produced by
Ornstein-Uhlenbeck process, dichotomous noise, and chaotic noise
produced by Lorenz model) were shown to produce the same power-law
dependence of the moments for the model with reflective walls.

The above MRW approach gives an additional support to the
representation $e^{u(t)}L(t)$ and thus to the choice $\alpha=e^u$,
where $u$ is the magnitude of Lagrangian velocity, considered in
Sec.~\ref{Sec:ConditionalProbability}. Leaving aside the choice of
stochastic process for $u(t)$ (Gaussian or Poisson, or the
weighted superposition of them), the difference between MRW
approach by Mordant {\it et al.}\cite{Mordant0206013} and RIN
approach to 1D LDN-type
model\cite{Aringazin0305186,Aringazin0312415} is that, in addition
to the above additive noise incorporating two well separated time
scales, the latter model accounts for (i) local interactions via
the turbulent viscosity, (ii) the multiplicative noise effects,
(iii) the coupling of additive and multiplicative noises, and (iv)
the velocity-dependence of acceleration statistics.

%%\section*{Acknowledgements}
%%
%%The authors are grateful to L.~Biferale, E.~Bodenschatz,
%%L.~Chevillard, B.~Devenish, R.~Friedrich, A.~M.~Reynolds, and
%%F.~Toschi for the interest to our work and comments.

\appendix

\section{Exact integrals}\label{Sec:ExactIntegrals}

Exact indefinite integrals, up to a constant term which does not
depend on $a$, used in calculating the definite integral entering
the probability density function (\ref{StationaryLaval}) are given
below.

At $\nu_{\mathrm t} = \nu_0$,
%%(A1)
\begin{eqnarray}\label{LavalIntegral1}
\int\!\!\! da \frac{-\nu_0k^2a - Da +\lambda}{Da^2 -2\lambda a
+\alpha}
 = -\frac{D+\nu_0k^2}{2D} \ln[Da^2\!-\!2\lambda a \!+\!\alpha] \nonumber
 \\
 + \frac{\lambda\nu_0k^2}{D\sqrt{D\alpha-\lambda^2}}\
 \mathrm{arctan}\frac{Da-\lambda}{\sqrt{D\alpha-\lambda^2}}.\quad
\end{eqnarray}

At $\nu_{\mathrm t} = B|a|/k$, for positive (upper sign) and
negative (lower sign) $a$, respectively,
%%(A2)
\begin{eqnarray}\label{LavalIntegral2}
\int da \frac{\mp Bka^2 - Da +\lambda}{Da^2 -2\lambda a +\alpha}
 = \mp\frac{B k a}{D} \nonumber \\
 -\frac{D^2 \pm 2B\lambda k}{2D^2} \ln[Da^2-2\lambda a +\alpha]
 \nonumber \\
 \pm \frac{B(D\alpha-2\lambda^2)k}{D^2\sqrt{D\alpha-\lambda^2}}\
 \mathrm{arctan}\frac{Da-\lambda}{\sqrt{D\alpha-\lambda^2}},
\end{eqnarray}
%%and for negative $a$,
%%(A2')
%%\begin{eqnarray}\label{LavalIntegral2'}
%%\int da \frac{Bka^2 - Da +\lambda}{Da^2 -2\lambda a +\alpha}
%% = \frac{Bka}{D} \nonumber \\
%% -\frac{D^2-2B\lambda k}{2D^2} \ln[Da^2-2\lambda a +\alpha]
%% \nonumber \\
%% - \frac{B(D\alpha-2\lambda^2)k}{D^2\sqrt{D\alpha-\lambda^2}}\
%% \mathrm{arctan}\frac{Da-\lambda}{\sqrt{D\alpha-\lambda^2}}.
%%\end{eqnarray}

In the general case, we have obtained a cumbersome expression
%%(A3)
\begin{eqnarray}\label{LavalIntegral3}
\int \!\! da \frac{-\nu_{\mathrm t}k^2a - Da +\lambda}{Da^2
-2\lambda a +\alpha}
 = -\frac{\nu_{\mathrm t}k^2}{D}
   -\frac{1}{2}\ln[Da^2\!-\!2\lambda a \!+\!\alpha] \nonumber \\
   -\frac{2B\lambda k}{D^2}\ln[2Bka+\nu_{\mathrm t}k^2]
   +F(c) + F(-c), \quad
\end{eqnarray}
where we have denoted, for brevity,
%%(A4)
\begin{eqnarray}\label{A4}
F(c,a)
 = \frac{c_1k^2}{2c_2D^2c}\ln\left[\frac{2D^3}{c_1c_2(c-Da+\lambda)}\times
   \right.\nonumber \\
   \left.
   \times(
   B^2(\lambda^2 + c\lambda-D\alpha)a
   + c(D\nu_{\mathrm t}^2k^2+c_2\nu_{\mathrm t})
    )
   \right],
\end{eqnarray}
%%(A5)
\be\label{A5}
c=-i\sqrt{D\alpha-\lambda^2}, \quad \nu_{\mathrm t} =
\sqrt{\nu_0^2+ B^2a^2/k^2},
\ee
%%(A6)
\be\label{A6}
c_1 = B^2(4\lambda^3+4c\lambda^2 - 3D\alpha\lambda-cD\alpha)
    + D^2(c +\lambda)\nu_0^2k^2,
\ee
%%(A7)
\be\label{A7}
c_2 = \sqrt{B^2(2\lambda^2 + 2c\lambda-D\alpha)k^2 +
D^2\nu_0^2k^4}.
\ee
Some useful formulas used in verifying the limits $B\to 0$ and
$D\to 0$ are:
%%(A8)
\be\label{A8}
\mathrm{arctan}\, x = \frac{i}{2}(\ln(1-ix)- \ln(1+ix)),
\ee
%%(A9)
\be\label{A9}
\lim_{D\to 0}\frac{1}{D}\ln[1+Da^2] = a^2.
\ee

\end{document}